\documentclass[aps,superscriptaddress,twocolumn,showpacs,floatfix,notitlepage]{revtex4-1}

\usepackage{amsmath}
\usepackage[makeroom]{cancel}
\usepackage{amsfonts}
\usepackage{amssymb}
\usepackage{graphicx}
\usepackage{color}
\usepackage[svgnames]{xcolor}
\usepackage{xspace}
\usepackage{dsfont}
\usepackage{physics}
\usepackage[mathscr]{euscript}

\DeclareFontFamily{OT1}{pzc}{}
\DeclareFontShape{OT1}{pzc}{m}{it}%
{<-> s * [1.15] pzcmi7t}{}
\DeclareMathAlphabet{\mathpzc}{OT1}{pzc}{m}{it}

\newcommand{\macro}[1]{\texttt{\textbackslash#1}}
\newcommand{\m}[1]{\macro{#1}}

\newcommand{\ud}[1]{{#1^{\dagger}}}

\newcommand{\Vuckovic}{Vu\u{c}kovi\'c\xspace}
\newcommand{\g}[1]{g^{(#1)}}
\newcommand{\av}[1]{\langle  #1 \rangle}
\newcommand{\pop}[1]{\langle \ud{#1} #1 \rangle}
\newcommand{\coh}[1]{\langle  #1 \rangle}

\newcommand{\corr}[3]{\langle #1^{\dagger #2} #1^{#3} \rangle}

\usepackage[colorlinks=true, linkcolor=blue, citecolor=magenta]{hyperref}

\newcommand{\mean}[1]{\langle#1\rangle}
\allowdisplaybreaks

\begin{document} 
\title{Conventional and unconventional photon statistics}

\author{Eduardo {Zubizarreta Casalengua}}
\affiliation{Departamento de F\'isica Te\'orica de la Materia
Condensada, Universidad Aut\'onoma de Madrid, 28049 Madrid,
Spain}

\author{{Juan Camilo} {L\'{o}pez~Carre\~no}}
\affiliation{Departamento de F\'isica Te\'orica de la Materia
  Condensada, Universidad Aut\'onoma de Madrid, 28049 Madrid, Spain}
\affiliation{Faculty of Science and Engineering, University of
  Wolverhampton, Wulfruna St, Wolverhampton WV1 1LY, UK}

\author{Fabrice P. Laussy}
\affiliation{Faculty of Science and Engineering,
  University of Wolverhampton, Wulfruna St, Wolverhampton WV1 1LY, UK}
\affiliation{Russian Quantum Center, Novaya 100, 143025 Skolkovo,
  Moscow Region, Russia}

\author{Elena {del Valle}}
\email{elena.delvalle.reboul@gmail.com}
\affiliation{Departamento de F\'isica Te\'orica de la Materia
Condensada, Universidad Aut\'onoma de Madrid, 28049 Madrid,
Spain}

\date{\today}

\begin{abstract}
  We show how the photon statistics emitted by a large variety of
  light-matter systems under weak coherent driving can be understood,
  to lowest order in the driving, in the framework of an admixture of
  (or interference between) a squeezed state and a coherent state,
  with the resulting state accounting for all bunching and
  antibunching features. One can further identify two mechanisms that
  produce resonances for the photon correlations: i)
  \emph{conventional statistics} describes cases that involve a
  particular quantum level or set of levels in the excitation/emission
  processes with interferences occurring to all orders in the photon
  numbers, while \emph{unconventional} statistics describes cases
  where the driving laser is far from resonance with any level and the
  interference occurs for a particular number of photons only,
  yielding stronger correlations but only for a definite number of
  photons. Such an understanding and classification allows for a
  comprehensive and transparent description of the photon statistics
  from a wide range of disparate systems, where optimum conditions for
  various types of photon correlations can be found and realized.
\end{abstract}

\maketitle
\tableofcontents

\section{Introduction}

Quantum optics was born with the study of photon
statistics~\cite{glauber06a}. Following Hanbury Brown's discovery of
photon bunching~\cite{hanburybrown52a,hanburybrown56c} and Kimble
\emph{et al.}'s observation of antibunching~\cite{kimble77a}, there
has been a burgeoning activity of tracking how pairs of detected
photons are related to each other. From the various mechanisms that
generate correlated photons, one that turns an uncorrelated stream
into antibunched photons has attracted much attention across different
platforms for its practicality of operation (with a laser) and
appealing underlying
mechanism~\cite{imamoglu94a,werner99a,kim99a,michler00a,birnbaum05a,faraon08a,dayan08a,lang11a,hoffman11a}.
This so-called ``blockade'' effect describes how the occupation of an
energy level by a particle forbids another particle to occupy the same
level. At such, it is reminiscent of Pauli's exclusion
principle~\cite{kaplan_book16a,massimi_book05a} and indeed the first
type of blockade involved electron in the so-called ``Coulomb
blockade''~\cite{averin86a,fulton87a,kastner08a}. However, while
Pauli's principle relies on the antisymmetry of fermionic
wavefunctions, one can also implement a ``bosonic blockade'' with
photons exciting an anharmonic system~\cite{carusotto01a}, based on
nonlinearities in the energy levels due to self-interactions. The idea
is simple: when photons are resonant with the frequency of the
oscillator, a first photon can excite the system, but due to
photon-photon interactions, a subsequent photon is now detuned from
the oscillator's frequency. If its energy is not sufficient to climb
the ladder of states, it cannot excite the system, that remains with
one photon.  In this way, one can turn a coherent---that is,
uncorrelated---stream of photons, with its characteristic Poissonian
fluctuations, into a more ordered stream of separated photons,
effectively acting as a ``photon turnstile''. The quality of such a
suppression of the Poisson bursts can be measured at the two-photon
level with Glauber's second-order correlation
function~$g^{(2)}(\tau)$, that compares the coincidences in time to
those expected from a random process of same intensity. Correlations
decrease from~1, with no blockade, towards~0 as the nonlinearity~$U$
increases. In the limit where~$U$ becomes infinite, putting the second
excited state arbitrarily far and realizing a two-level system (2LS),
a second photon is strictly forbidden and $g^{(2)}$ becomes perfectly
antibunched. For open bosonic systems, the ratio of the interaction to
the decay rate is an important variable for the blocking to be
effective. The ``blockade'' regime is reached when interactions start
to dominate the dissipation. This can be marked as the onset of
antibunching: one photon starts to suppress the next
one~\cite{sanvitto19a}.

The driven damped anharmonic system is an important model, not least
because it is one of the few cases to enjoy an exact analytical
solution~\cite{drummond80a}. While much of the mechanism is contained
in this particular case, compound systems---where the anharmonic
system is coupled to a single-mode cavity---have also attracted
considerable attention.  This describes for instance interacting
quantum-well excitons coupled to a microcavity
mode~\cite{kavokin_book17a}. The effect is then known as
``polariton-blockade'', after the eponymous light-matter particles
that constitute the elementary excitations of such systems.  This
configuration was first addressed theoretically by Verger \emph{et
  al.}~\cite{verger06a} who studied the response of the cavity around
the lower-polariton resonance, predicting antibunching indeed,
although of too small magnitude with the parameters of typical systems
to be observed easily. This spurred interest in polariton boxes and
other ways of confining polaritons to enhance their
interactions~\cite{eldaif06a}. A few years later, Liew and Savona
computed a much stronger antibunching from a seemingly distinct
compound system with the same order of nonlinearity, namely, coupled
cavities~\cite{liew10a}. This so-called ``unconventional polariton
blockade'' was quickly understood as originating not from the
particular configuration of coupled cavities with weak Kerr
nonlinearities but from a subtler type of blocking, due to destructive
interferences between probability amplitudes whenever there are two
paths that can reach the excited state with two
photons~\cite{bamba11a}. This result has generated considerable
attention, although it was later remarked~\cite{lemonde14a} that it
was a known effect~\cite{carmichael85a,carmichael91b}, observed
decades earlier~\cite{foster00a} where it received a much smaller
followup. Besides, Lemonde \emph{et al.}~\cite{lemonde14a} further
clarified how unconventional photon blockade is connected to squeezing
rather than single-photon states, which had been presented as one of
the main interest of the effect.  Recently, both
conventional~\cite{arXiv_delteil18a,arXiv_munozmatutano17a} and
unconventional~\cite{vaneph18a,snijders18a} blockades have been
reported in solid-state systems, where the 2010 revival of the idea
had triggered intense activity.

In this text, we provide a unifying picture of the two types of
polariton blockades. We show how they typically sit next to each other
in interacting coupled light-matter systems along with other
phenomenologies, that produce superbunching instead of the blockade
antibunching.  In particular, we show that they are both rooted in the
single-component system, either a 2LS or an anharmonic oscillator,
with strong photon correlations produced by interfering the emitter's
incoherent signal with a coherent fraction.  We will nevertheless
highlight how the two blockades are intrinsically different mechanisms
with different characteristics. Most importantly, the conventional
blockade, based on dressed-state blocking, yields photon antibunching
at all orders in the number of photons, i.e., $g^{(N)}\rightarrow 0$
for all~$N\ge 2$, while the unconventional blockade can only target
one~$N$ in isolation, producing bunching for the others. Another
apparent similarity is that both types of blockades produce the same
state in what concerns the population and the two-photon
correlation~$g^{(2)}$ at the lowest order in the driving, but
differences occur at higher orders, namely, at the second-order
for~$g^{(2)}$ and at the third-order for the population. Differences
exist already at the lowest order in the driving for $g^{(3)}$ and
higher-order photon correlations, making it clear that the two
mechanisms differ substantially and produce different states, despite
strong resemblances in the quantities of easiest experimental reach.
The state produced by both types of blockade at the two-photon level
results from a simple interference between a squeezed state and a
coherent state. While the squeezing is typically produced by the
emitter, the coherent fraction can be either brought from outside,
idoneously, as a fraction of the driving laser itself---a technique
known as ``homodyning''---or be produced internally by the driven
system itself, a concept introduced in the literature under the apt
qualification of ``self-homodyning''~\cite{carmichael85a}.  We will
thus highlight that, to this order, essentially the same physics---of
tailoring two-photon statistics by admixing squeezed and coherent
light, discussed in Section~\ref{sec:ThuFeb15174655CET2018}---takes
place in a variety of platforms, overviewed in
Section~\ref{sec:vieene25080609GMT2019}.  We will further synthesize
this picture by unifying the cases where the nonlinearity is i)
strong, namely, provided by a 2LS
(Section~\ref{sec:FriFeb23095933CET2018}) or on the contrary ii) weak,
namely, provided by an anharmonic oscillator
(Section~\ref{sec:jueene10122434CET2019}), and how these are further
generalized in presence of a cavity where self-homodyning becomes a
compelling picture since a cavity is an ideal receptacle for coherent
states. This brings the 2LS into the Jaynes-Cummings model
(Section~\ref{sec:FriFeb23100104CET2018}) and the anharmonic
oscillator into microcavity polaritons
(Section~\ref{sec:FriFeb23100219CET2018}), respectively. There are
many variations in between all these configurations, that the
literature has touched upon in many forms, as we briefly overview in
the next Section. For our own discussion, while we have tried to
retain as much generality as possible for the variables that play a
significant role, we do not include for the sake of brevity all the
possible combinations, which could of course be done would the need
arise for a given platform. Section~\ref{sec:jueene24231247GMT2019}
summarizes and concludes.

\section{A short review of blockades and related effects}
\label{sec:vieene25080609GMT2019}

We will keep this overview very short since a thorough review would
require a full work on its own. A good review is found in
Ref.~\cite{flayac17b}. This Section will also allow us to introduce
the formalism and notations. For details of the microscopic
derivation, we refer to Ref.~\cite{verger06a}. 

A fairly general type of photon blockade is described by the
Hamiltonian
\begin{subequations}
  \label{eq:Mon5Jun145532BST2017}
  \begin{align}
    H={}&\hbar\omega_a\ud{a}a+\hbar\omega_b\ud{b}b+\hbar g(\ud{a}b+a\ud{b})\label{eq:Mon5Jun150144BST2017}\\
        &+\frac{U_a}{2}\ud{a}\ud{a}aa+\frac{U_b}{2}\ud{b}\ud{b}bb\\
        &+\Omega_a e^{i\varpi_a t}a+\Omega_b e^{i\varpi_b t}b\\
&+\mathrm{h.c.}\nonumber
  \end{align}
\end{subequations}
where~$\hbar\omega_c$ is the free energy of the modes~$c=a,b$, both
bosonic, $\hbar g$ describes their Rabi coupling, giving rise to
polaritons as eigenstates of line~(\ref{eq:Mon5Jun150144BST2017}),
$U_c$ are the nonlinearities of the respective modes, here again
for~$c=a,b$, and~$\Omega_c$ describes resonant excitation at the
energy~$\varpi_c$. This is brought to the dissipative regime through
the standard techniques of open quantum systems, namely, with a master
equation in the Lindblad form
$\partial_t \rho = i \left[ \rho ,H \right] + \sum_{c=a,b}
(\gamma_c/2) \mathcal{L}_{c} \rho$, where the superoperator
$\mathcal{L}_{c} \rho \equiv 2 c \rho \ud{c} - \ud{c}c \rho - \rho
\ud{c}c$ describes a decay rate of mode~$c$ at rate~$\gamma_c$ (we do
not include incoherent excitation nor dephasing, which are other
parameters one could easily account for in the following analysis from
a mere technical point of view).

Particular cases or variations of Eq.~(\ref{eq:Mon5Jun145532BST2017})
have been studied in a myriad of works, even when restricting to those
with a focus on the emitted photon statistics. This ranges from cases
retaining one mode only~\cite{ferretti12a} to the most general form of
Eq.~(\ref{eq:Mon5Jun145532BST2017})~\cite{ferretti10a, ferretti13a,
  flayac13a, flayac16a, flayac17a, arXiv_liang18a} with further
variations (such as using pulsed excitation~\cite{flayac15a}). A first
consideration of the effect of field admixture on the photon
statistics was made by Flayac and Savona~\cite{flayac13a}, who found
that that the conditions for strong correlations are shifted rather
than hampered. This touches upon, in the framework of input/output
theory, the mechanisms of mixing fields that we will highlight in the
following, where we will show that beyond being altered, correlations
can be drastically optimized (becoming exactly zero to first order for
antibunching and infinite for bunching). In a later
work~\cite{flayac16a}, they further progressed towards fully
exploiting homodyning by including a ``dissipative, one-directional
coupling'' term, which allowed them to achieve a considerable
improvement of the photon correlations, especially in time, with
suppression of oscillations and the emergence of a plateau at small
time delays. This is due to the same mechanism than the one we used
with identical consequences but in a different
context~\cite{lopezcarreno18b} (a two-level system admixed to an
external laser).  Emphasis should also be given to the bulk of work
devoted to related ideas of mixing fields by the \Vuckovic group,
starting with their use of self-homodyning to study the Mollow triplet
in a dynamical setting~\cite{fischer16a}. Initially used as a
suppression technique to access the quantum emitter's dynamics by
cancelling out the scattered coherent component from their driving
laser~\cite{muller16a,dory17a}, they later appreciated the widespread
application of their effect and its natural occurence in other
systems~\cite{fischer18a}, where it had passed unnoticed, as well as
the benefits of a tunability of the interfering
component~\cite{fischer17a}, which they proposed in the form of a
partially transmitting element in an on-chip integrated architecture
that combines a waveguide with a quantum-dot/photonic-crystal cavity
QED platform. As such, they have made a series of pioneering
contributions in the effect of homodyning for quantum engineering and
optimization, which is promised to a great followup~\cite{li18a}. In
the following, we will provide a unified theory of the mixing of
coherent and quantum light that has been developed and implemented
throughout the recent years by Fischer \emph{et
  al.}~\cite{fischer16a,fischer18a}.  The possibility and benefits of
an external laser to optimize photon correlations also appeared in a
work by Van Regemortel \emph{et al.}~\cite{vanregemortel18a}, with a
foothold in the same ideas. The effect of tuning two types of driving
was emphasized by Xu and Li~\cite{xu14a}, who reported among other
notable results how changing their ratio can bring the system from
strong antibunching to superbunching, an idea which we will revisit
from the point of view of interfering fields through (controlled)
homodyning or (self-consistent) self-homodyning.  Similar ideas have
then been explored and extended several times in many variations of
the problem~\cite{zhang14e,tang15a,shen15a,shen15b, li15c, xu13a,
  xu16a, wang16b, wang17a, cheng17a,
  deng17a,zhou16a,liu16a,kryuchkyan16a,zhou16b,yu17a} which all fit
nicely in the wider picture that we will present.  The
microcavity-polariton configuration with interactions in one mode only
(describing quantum well excitons, the other being a cavity mode) has
been studied mainly from the (conventional) polariton blockade point
of view~\cite{verger06a,gerace09a}, in which case, the (much stronger)
unconventional antibunching has been typically overlooked.  We will
focus in the following on this case rather than on the possibly more
popular two weakly-interacting sites. First, because this allows a
direct comparison with the Jaynes--Cumming limit, second, because this
configuration became timely following the recent experimental
breakthrough with polariton
blockade~\cite{arXiv_delteil18a,arXiv_munozmatutano17a}. Whatever the
platform, it needs be also emphasized that while many works have
focused on single-photon emission as the spotlight for the effect
(which is dubious when antibunching is produced from the
unconventional route), others have also stressed different
applications or suggested different contextualization, such as
phase-transitions~\cite{ferretti10a} or
entanglement~\cite{casteels17a}, and there is certainly much to
exploit from one perspective or another.

Among other configurations that cannot be accommodated by
Eq.~(\ref{eq:Mon5Jun145532BST2017}) as they add even more components,
one could mention examples from works that involve additional modes
(three in Refs.~\cite{majumdar12e,kyriienko14c,kyriienko14a}),
different types of nonlinearity, e.g., $a^2\ud{b}$ in
Refs.~\cite{majumdar13a,gerace14a,zhou15a}, a four-level system in
Ref.~\cite{bajcsy13a}), two two-level system in a
  cavity in Ref.~\cite{radulaski17a} and up to the general
  Tavvis--Cummings model~\cite{arXiv_trivedi19a}, pulsed coherent
  control of a two-level system in Ref.~\cite{arXiv_loredo18a}, two
coupled cavities each containing a two-level
system~\cite{knap11a,schwendimann12a} up to a complete
array~\cite{grujic13a}. It seems however clear to us that the
phenomenology reported in each of these particular cases would fall
within the classification that we will establish in the remaining of
the text, i.e., they can be understood as an homodyning effect of some
sort.

\section{Homodyne and self-homodyne interferences}
\label{sec:ThuFeb15174655CET2018}

We will return in the rest of this text to such systems as those
discussed in the previous Section---all a particular case or a
variation of Eq.~(\ref{eq:Mon5Jun145532BST2017})---to show that the
two-photon statistics of their emission can be described to lowest
order in the driving by a simple process: the mixing of a squeezed and
a coherent state. In this Section, we therefore study this
configuration in details.

\begin{figure}[t]
  \centering
  \includegraphics[width=\linewidth]{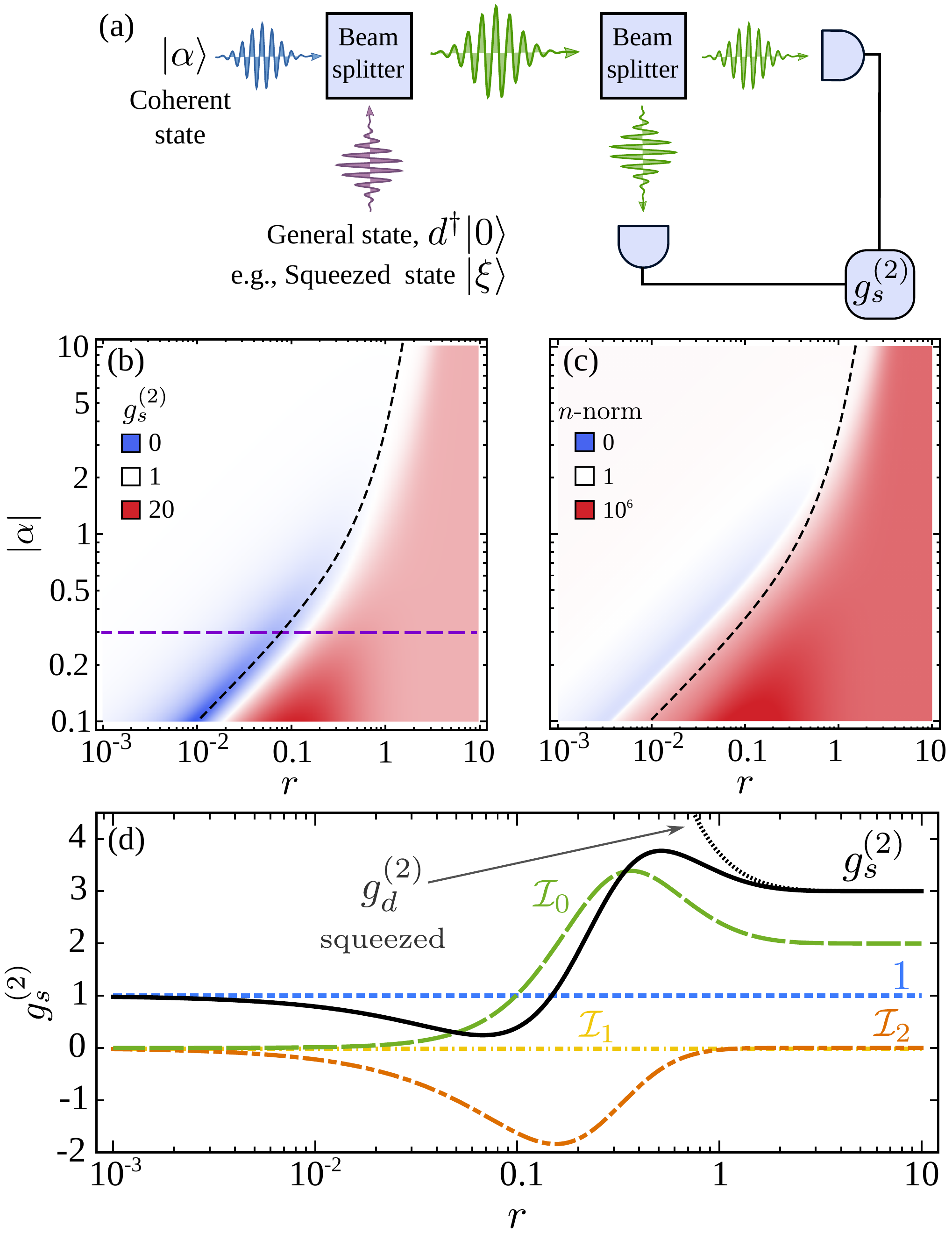}
  \caption{(Color online) Second-order coherence function $\g{2}_s$
    for the interference between a coherent and a quadrature-squeezed
    field, as in the setup shown in panel~(a). The resulting $\g{2}_s$
    is shown in panel~(b) as a function of the coherent field
    $|\alpha|$ and squeezing parameeter~$r$ (colour map in log
    scale). The relative phase between~$\phi$ between the squeezed and
    coherence state is the one that optimises antibunching, i.e.,
    $\phi = \theta/2$.  (c) The $n$-norm (up to $\g{6}$, i.e.,
    $n = 5$) as defined in Eq.~(\ref{eq:crispsnorm}).  Dashed black
    lines in both panels mark the minimum of $\g{2}_s$, showing that
    the best antibunching is no guarantee of good two-photon emission.
    (d) Cut of $\g{2}_s$ along the horizontal dashed line in panel~(b)
    ($|\alpha| =0.3$) and its decomposition given by
    Eqs.~(\ref{eq:g2mixdecomposition})
    and~(\ref{eq:ThuFeb22121240CET2018}). The black dotted line shows
    $\g{2}_d$ for the squeezed state only.}
  \label{fig:1}
\end{figure}

We first consider the mixture of any two fields as obtained in one of
the output arms of a balanced beam splitter (cf.~Fig.~\ref{fig:1}a),
which is fed by a coherent state on one of its arms, with a complex
amplitude~$\alpha=|\alpha|e^{i\phi}$, and another field of a general
nature, described with annihilation operator $d$, on the other
arm. The field that leaves the beam splitter is a mixture of the two
input fields, whose annihilation operator can be written
as~$s=\alpha+d$, where we are leaving out the normalization
($1/\sqrt{2}$) and $\frac{\pi}{2}$-phase shift in the reflected light
as reasoned in Appendix~\ref{app:1}. Within this description, any
normally-ordered correlator of the resulting field can be expressed in
terms of the inputs as
\begin{equation}
\label{eq:homodynecorrelators}
\corr{s}{n}{m} = \sum_{p=0}^{n} \sum_{q=0}^{m}
\binom{n}{p}\binom{m}{q} \alpha^{* p} \alpha^{q} \corr{d}{n-p}{m-q} \,.
\end{equation}
From this expression, we can compute any relevant observable of the
mixture. For instance, the total population is
\begin{equation}
	\mean{n_s} \equiv \pop{s} = |\alpha|^2 + \mean{n_d} + 2 \Re
        [ \alpha^* \coh{d} ] \,,
\end{equation}
with $n_d\equiv d^\dagger d$. Apart from the sum of both input
intensities, there is a contribution (last term) from the first-order
interference between the coherent components of each of the fields or
\emph{mean fields}. Similarly, the second-order coherence function,
which is defined as
\begin{subequations}
  \label{eq:MonNov27154304CET2017}
  \begin{align}
    \label{eq:MonNov27154304CET2017a}
    g_s^{(2)}(\tau)&=\lim_{t\rightarrow \infty}\frac{\mean{s^\dagger
                     (t) (s^\dagger s )(t+\tau) s
                     (t)}}{[\mean{s^\dagger s}(t)]^2}\,,\\
    \label{eq:MonNov27154304CET2017b}
                   & = \frac{\mean{s^\dagger (s^\dagger s )(\tau)
                     s}}{\mean{n_s}^2} \,, 
  \end{align}
\end{subequations}
can be readily obtained from the correlators in
Eq.~(\ref{eq:homodynecorrelators}). In this text, we omit the time~$t$
in all expressions, as it is considered to be large enough for the
system to have reached the steady state under the continuous drive,
and will also set the delay~$\tau=0$, thus focusing on
coincidences. This simplifies the notation
$g_s^{(2)}=g_s^{(2)}(t\rightarrow\infty,\tau=0)$. We will also
consider $N$-th order coherence functions, also at zero delay:
$g_s^{(N)}\equiv\mean{s^{\dagger N} s^N}/\mean{s^\dagger s}^N$. 

These correlators can always be written as a polynomial series of
powers of the amplitude of the coherent field~$\alpha$:
\begin{equation}
  \label{eq:ThuFeb22113232CET2018}
  g^{(N)}_s=\frac{\sum_{k=0}^{2N} c_k(\phi) |\alpha|^k}{\mean{n_s}^{2N}}\,,
\end{equation}
where~$c_k(\phi)$ are coefficients that depend on the phase of the
coherent field~$\phi$, and mean values of the
type~$\mean{d^{\dagger\mu} d^\nu}$ with $\mu+\nu\leq N^2$. In
particular, the 2nd-order correlation function,
Eq.~(\ref{eq:MonNov27154304CET2017b}), can be rearranged
as
\begin{equation}
\label{eq:g2mixdecomposition}
g^{(2)}_{s} = 1 + \mathcal{I}_0 + \mathcal{I}_1 + \mathcal{I}_2\,,
\end{equation}
with
$\mathcal{I}_m \sim
|\alpha|^m$~\cite{mandel82a,carmichael85a,vogel91a,vogel95a}, where 1
represents the coherent contribution of the total signal, and the
incoherent contributions read
\begin{subequations}
  \label{eq:ThuFeb22121240CET2018}
  \begin{align}
      \label{eq:ThuFeb22121240CET2018a}
	\mathcal{I}_0 &= \frac{\corr{d}{2}{2} -
                        \pop{d}^2}{\mean{n_s}^2}\,,\\
      \label{eq:ThuFeb22121240CET2018b}
	\mathcal{I}_1 &=4\frac{ \Re[\alpha^{*} (\av{\ud{d}  d^2}-
                        \pop{d} \av{d})]}{\mean{n_s}^2}\,,\\
      \label{eq:ThuFeb22121240CET2018c}
      \mathcal{I}_2 &= 2\frac{\Re[ {\alpha^*}^{2}
          \av{d^2}] - 2\Re[\alpha^*
                      \av{d}]^2  + |\alpha|^2 \pop{d}}{\mean{n_s}^2}\\
                      &
                        =4 \Big[
                        |\alpha|^2 
                        \left( \cos^2 \phi \ \av{{:}X_{d}^2{:}} + \sin^2 \phi
                        \ \av{{:}Y_{d}^2{:}} +{} \right. \nonumber \\
    & \quad \left. {}+ \cos \phi \sin \phi
    \ \av{\lbrace X_d , Y_d \rbrace} \right) - \Re[ \alpha^*
    \av{d}]^2 \Big]/\mean{n_s}^2\,.\nonumber
  \end{align}
\end{subequations}
Here, the notation ``$::$'' indicates normal ordering,
$\lbrace X_d,Y_d \rbrace = X_dY_d+Y_dX_d$, and
$X_d = \frac{1}{2} \left(\ud{d}+d\right) $,
$Y_d = \frac{i}{2} \left(\ud{d}- d \right)$ are the quadratures of the
field described with the annihilation operator~$d$. Note that there
are no explicit terms $\mathcal{I}_3$ and $\mathcal{I}_4$ because
through simplifications these get absorbed in the
term~$\mathcal{I}_1$.

The decomposition was, to the best of our knowledge, first introduced
by Carmichael~\cite{carmichael85a} and in fact precisely to show that
the same quantum-optical phenomenology observed in different systems
had the same origin, namely, to root nonclassical effects observed in
optical bistability with many atoms in a cavity, to the physics of a
single atom coherently driven, i.e., resonance fluorescence. This is
at this occasion of unifying squeezing and antibunching from two
seemingly unrelated platforms under the same umbrella of
self-homodyning that he introduced this terminology. These concepts
have thus been invoked and explored to some extent from the earliest
days of the field, but it is only recently that they seem to start
being fully understood and
exploited~\cite{fischer16a,flayac16a,fischer18a,vanregemortel18a,lopezcarreno18b}. Using
such interferences to analyse the squeezing properties of a signal of
interest (here, the field with annihilation operator~$d$) through a
controlled variation of a local oscillator (here, the coherent
field~$\alpha$), was first suggested by W.~Vogel~\cite{vogel91a,
  vogel95a}, subsequently implemented to show squeezing in resonance
fluorescence~\cite{schulte15a}, and recently impulsed in a series of
works from the \Vuckovic group, as previously discussed.  The physical
interpretation of the contributions to the decomposition are as
follows:

\begin{itemize}
\item The numerator of~$\mathcal{I}_0$ is the normally-ordered
  variance of the signal intensity, that is, $\mean{{:}(\Delta n_d)^2
    {:}}=\mean{{:}n_d^2{:}}-\mean{n_d}^2$ with $n_d=d^\dagger d$ and $\Delta
  n_d=n_d-\mean{n_d}$. Therefore, $\mathcal{I}_0<0$ indicates that the
  field~$d$ has sub-Poissonian statistics, which in turn contributes to
  the sub-Poissonian statistics of the total field~$s$.

\item The numerator of~$\mathcal{I}_1$ represents the normally-ordered
  correlation between the fluctuation-field strength and intensity,
  $\av{\ud{d} d^2}- \pop{d} \av{d}=\mean{{:}\Delta d \, \Delta n_d{:}}$,
  which have been referred to as \emph{anomalous
    moments}~\cite{vogel91a,vogel95a} and been recently
  measured~\cite{kuhn17a}. A squeezed-coherent state has such
  correlations. 

\item The numerator of the last component, $\mathcal{I}_2$, can also
  be written in terms of the quadratures of the field~$d$.  Having
  $\mathcal{I}_2<0$ necessarily implies that the state of light has a
  squeezing component. This can be proved by noting that
  ${:}X_d^2{:} = X_d^2 - 1/4$ (the same for ${:}Y_d^2{:}$). Then,
  rearranging the numerator of Eq.~(\ref{eq:ThuFeb22121240CET2018c})
  leads to:
\begin{equation}
\label{eq:I2transformation}
 \av{X_{d,\phi}^2}- \av{X_{d_,\phi}}^2 
 = \left( \Delta
 X_{d,\phi} \right)^2 
\,,
\end{equation}
where $X_{d,\phi} = (e^{i \phi} \ud{d} + e^{-i \phi} d)/2$ is the
quadrature with the same phase of the coherent field (given by the
angle $\phi$).  If $\mathcal{I}_2 < 0$, the dispersion of $X_{d,\phi}$
must be less than $1/2$, but since $X_{d,\phi}$ and its orthogonal
quadrature, $X_{d,\phi+\pi/2}$, must fulfil the Heisenberg uncertainty
relation, $\Delta X_{d,\phi} \Delta X_{d,\phi+ \pi/2} \geq 1/4$, then
$\Delta X_{d,\phi+ \pi/2} > 1/2$. This necessarily implies that there
is a certain degree of squeezing in $d$. Nevertheless, the opposite
statement is not true. A state with a non-zero degree of squeezing can
have $\mathcal{I}_2 \geq 0$, for instance, if the relative direction
between the coherent and squeezing contributions fulfils
$\theta-2\phi = \pi/2$ (a straightforward example is provided by the
displaced squeezed state).  Furthermore, if $\mean{d}=0$, the
numerator of $\mathcal{I}_2$ simplifies to
$4 |\alpha|^2 ( \av{{:}X_{d}^2{:}} - |\alpha|^2 )$.
\end{itemize}

An analogous procedure allows us to decompose the third-order
coherence function, $\g{3}_s= 1+\sum_{m=0}^{4} \mathcal{J}_m$ (the
expressions for $\mathcal{J}_m $ are given in
Appendix~\ref{app:g3decomp}). Naturally, higher-order-correlator
decompositions follow the same rules.

As an illustration which will be relevant in what follows, let us
consider the interference between a coherent and a squeezed state, as
shown schematically in Fig.~\ref{fig:1}(a). The coherent state can be
written as~$\ket{\alpha}=D_a(\alpha)\ket{0}$,
where~$\alpha = |\alpha|e^{i\phi}$ is the amplitude of the coherent
state as before, and $D_a(\alpha)=\exp(\alpha \ud{a}-\alpha^\ast a)$
is the displacement operator of the field with annihilation
operator~$a$. Similarly, the squeezed state may be written
as~$\ket{\xi}=S_d(\xi)\ket{0}$, where~$\xi=re^{i\theta}$ is the
\emph{squeezing parameter} and
$S_d(\xi)=\exp(\xi d^{\dagger\,2}-\xi^\ast d^2)$ is the squeezing
operator of the field with annihilation operator~$d$. Thus, the state
that feeds the beam splitter
is~$\ket{\psi_\mathrm{in}}=\ket{\alpha\,,\xi}$. The interference at
the beam splitter mixes these two states, and the state of the light
that leaves the beam splitter is a two-mode squeezed state that is
further squeezed and displaced (the detailed transformation is given
in Appendix~\ref{app:1}). Since we are only interested in the output
of one of the arms of the beam splitter, we take the partial trace
over the other arm and we end up with a \emph{displaced squeezed
  thermal state}~\cite{lemonde14a},
\begin{equation}
  \label{eq:ThuFeb22151812CET2018}
  \rho_{s} = \mathcal{D}_s (\alpha) \mathcal{S}_s (\xi)
  \rho_{\mathrm{th}} \left(\mean{n_\mathrm{th}}\right) 
  \ud{\mathcal{S}_s} \left(\xi \right) \ud{\mathcal{D}_s} \left(\alpha
  \right)\,,
\end{equation}
where now the displacement and squeezing operators correspond to the
operator~$s=a+d$, and $\mean{n_\mathrm{th}}$, the thermal population,
can be obtained from the population of the squeezed
state,~$\mean{n_d}$, which for a balanced beam splitter follows the
relation~$\mean{n_\mathrm{th}} = (\sqrt{1+\mean{n_d}}-1)/2$.  The
second-order correlation for the total signal is computed as in
Eq.~(\ref{eq:MonNov27154304CET2017}), taking the averages using the
state in Eq.~(\ref{eq:ThuFeb22151812CET2018}), namely
$ g^{(2)}_s = \Tr[\rho_s s^{\dagger\,2}s^2]/\Tr[\rho_s s^\dagger
s]^2$,
which can be decomposed as in Eq.~(\ref{eq:g2mixdecomposition}) into
\begin{subequations}
    \label{eq:FriOct20184657CEST2017}
    \begin{align}
        \label{eq:FriOct20184657CEST2017a}
       \mathcal{I}_0&= \frac{ \sinh[4](r)}{\av{n_s}^2}
               [1+\coth(r)^2]\,,\\
        \label{eq:FriOct20184657CEST2017b}
       \mathcal{I}_1&=0\,,\\
        \label{eq:FriOct20184657CEST2017c}
       \mathcal{I}_2&=\frac{2 |\alpha|^2 \sinh^2(r)}{\av{n_s}^2} \left[ 1 - \cos(\theta -
           2 \phi) \coth(r) \right] \,,
  \end{align}
\end{subequations}
where~$\mean{n_s}=|\alpha|^2 + \sinh^2(r)$. Here, $\mean{d}=0$ but
also Eq.~(\ref{eq:FriOct20184657CEST2017b}) is exactly zero because,
for a squeezed state, the correlators~$\corr{d}{\mu}{\nu}$ vanish
when~$\mu+\nu$ is an odd number. Useful expression for the
decomposition of $\g{2}_s$ and $\g{3}_s$ in terms of the incoherent
component and the two-photon coherence are given in
Appendix~\ref{app:1}.

Inspection of Eqs.~(\ref{eq:FriOct20184657CEST2017}) shows that the
only way in which $\g{2}_s<1$, that is, the statistics of the total
signal can be sub-Poissonian, regardless of the value of the squeezing
parameter, is for $\mathcal{I}_2$ to be negative, which implies that
the phases of the displacement and the squeezing must be related
by~$|\theta-2\phi|<\pi/2$. We take for simplicity the minimizing
alignment, $\theta=2\phi$, which means that the coherent and squeezed
excitations are driven with the same phase, since the phase of the
squeezed state is $\theta/2$. Using such a relation, the interference
yields the correlation map shown in Fig.~\ref{fig:1}(b) as a function
of the amplitude of the coherent~$|\alpha|$ and squeezing~$r$
intensities. The black dashed line shows the optimum amplitude of the
coherent state that minimizes~$\g{2}_s$ for a given squeezing, which
is given by
\begin{equation}
  \label{eq:FriOct20154005CEST2017}
  |\alpha|_\mathrm{min}=e^{r}\sqrt{\cosh(r)\sinh(r)}\,.
\end{equation}
Replacing this condition in Eqs.~(\ref{eq:FriOct20184657CEST2017}) we
obtain the minimum possible value of~$\g{2}_s$,
\begin{equation}
  \label{eq:FriOct20172135CEST2017}
  g_{s,\,\mathrm{min}}^{(2)}=1-\frac{e^{-2r}}{1+\sinh(2r)}\leq 1\,.
\end{equation}
This goes to zero although at the same time as the population goes to
zero.  Figure~\ref{fig:1}(d) shows a transverse cut of the correlation
map in~(b) along the purple long-dashed line, which corresponds
to~$|\alpha|=0.3$. The decomposition and total~$g_s^{(2)}$ are shown
as a function of the squeezing parameter, with
minimum~$g_s^{(2)}=0.26$ at $r\approx 0.078$.  Without the
interference with the coherent state, the squeezed state can never
have sub-Poissonian statistics. In fact, in such a case the
correlations become independent of the phase of the squeezing
parameter:
\begin{equation}
  \label{eq:ThuFeb22170058CET2018}
  g^{(2)}_{d}=g_s^{(2)}|_{\alpha\rightarrow 0}=
  2+\coth^2(r)\geq 3\,,
\end{equation}
which diverges at vanishing squeezing $r\rightarrow 0$ (with also
vanishing signal $\av{n_d}=\sinh^2(r)$), and is minimum when squeezing
is infinite~$r\rightarrow \infty$.

There is a great tunability from such a simple admixture since
$g^{(2)}_s$ of the light at the output of the beam splitter can be
varied between 0 and $\infty$ simply by adjusting the magnitudes of
the coherent field and the squeezing parameter. In particular, the
most sub-Poissonian statistics occurs when coherent light interferes
with a small amount of squeezing~$r<|\alpha_\mathrm{min}|$, in the
right intensity proportion, given by
Eq.~(\ref{eq:FriOct20154005CEST2017}). Counter-intuitively,
$g^{(2)}_s\ll 1$ occurs when the squeezed light itself is, on the
opposite, super-Poissonian (even super-chaotic
$g_d^{(2)}> 2$)~\footnote{In fact, in order to have
  $g_{s,\,\mathrm{min}}^{(2)}<1/2$, it is required
  $r<\log{(\sqrt{6}-1)}/2\approx 0.186$, which implies
  $g_d^{(2)}>31.7$ and
  $|\alpha_\mathrm{min}|>\sqrt{(2-\sqrt{6})/2}\approx 0.52>r$.}. This
is a fundamental result that we will find throughout the text in order
to find the conditions for and manipulate sub-Poissonian statistics
and antibunching in various systems under weak coherent driving.

An important fact for our classification of photon statistics is that,
since the sub-Poissonian behaviour is here due to an interference
effect, the set of parameters that suppresses the fluctuation at the
two-photon level does not suppress them at all $N$-photon levels,
which means that the multi-photon emission cannot be precluded
simultaneously at all orders. In other words, the condition in
Eq.~(\ref{eq:FriOct20154005CEST2017}) that minimizes $g_s^{(2)}$, also
minimizes the two-photon probability in the interference density
matrix~$\bra{2}\rho_s\ket{2}$, at low intensities. But this is not the
same condition that minimizes any other photon
probability~$\bra{n}\rho_s\ket{n}$. This incompatibility is revealed
by the \textit{$n$-norm}, as defined in
Ref.~\cite{arXiv_lopezcarreno16c}, which is the distance in the
correlation space between signal $s$ and a perfect single-photon
source:
\begin{equation}
\label{eq:crispsnorm}
\norm{(\g{k}_s )}_n = \sqrt[n]{\sum_{k=2}^{n+1} [\g{k}_s]^{n}}\,.
\end{equation}
In Figure~\ref{fig:1}(c) we show the~$5$-norm for the same range of
parameters of Panel~(b). The dashed black line indicates the minimum
values of~$\g{2}_s$, which lies in a high-fluctuation region when the
higher order correlation functions are taken into account.  Further
increasing~$n$ renders the correlation map completely red which means
that multiphoton emission is not suppressed even if we have $\g{2}_s$
close to zero. This is a feature typical of antibunching that arises
from a two-photon interference only, that suggests that their use as
single-photon sources may be an issue in the context of applications
for quantum technology where higher-photon correlations may jeopardize
two-photon suppression. This is related to the fact that this
antibunching stems from a Gaussian state, which is the most classical
of the quantum states.

The discussion presented above and, in particular, the decomposition
of the second-order correlation as in
Eq.~(\ref{eq:g2mixdecomposition}), are not limited to the particular
case of interfering pure states set as initial conditions. This can
also be applied to the dynamical case of a single system which
provides itself and directly a coherent component~$\alpha$ along with
another, and therefore quantum, type of component. Calling $s$ the
annihilation operator for a particular emitter which has such a
coherent---but not exclusively---component in its radiation, one can
thus express its emission as the interference (or superposition) of a
mean coherent field~$\mean{s}$ and its quantum fluctuations, with
operator~$d=s-\mean{s}$. That is, one can always write
\begin{equation}
  \label{eq:FriFeb23144144CET2018}
  s = \mean{s}+d\,.
\end{equation}
Following the terminology previously introduced in the
  literature for a similar purpose~\cite{carmichael85a}, we call this
interpretation of the emission a \emph{self-homodyne} effect. Since
$g_s^{(2)}$ is also given by Eq.~(\ref{eq:ThuFeb22121240CET2018}) with
the simplification brought by the fact that $\mean{d}=0$, by
replacing~$\alpha\rightarrow \mean{s}$ and~$d \rightarrow s-\mean{s}$,
we obtain the general expressions in terms of $\corr{s}{n}{m}$ for the
emission of a single-emitter~$s$, interfering its own components:
\begin{widetext}
      \begin{subequations}
	\label{eq:decompositiontermswhole}
	\begin{align}
          \mathcal{I}_0 &= \frac{\corr{s}{2}{2} - \av{\ud{s}s}^2 - 4
                          |\mean{s}|^4 + 6 |\mean{s}|^2
                          \pop{s}+2\Re[   {\mean{\ud{s}}}^2 \av{s^2} - 2
                          \mean{\ud{s}} \av{\ud{s} s^2} ]
                          }{\pop{s}^2}\, ,\\ 
          \mathcal{I}_1 &=4 \frac{\Re[ \mean{\ud{s}} \av{\ud{s} s^2} -
                          \mean{\ud{s}}^2 \av{s^2}] + 2
                          |\mean{s}|^2 \left(|\mean{s}|^2
                          -\mean{\ud{s}s}\right) }{\pop{s}^2}\, , \\
          \mathcal{I}_2 &=
                          2 \frac{ \Re[\mean{\ud{s}}^{2}  \av{s^2}] +  |\mean{s}|^2 \pop{s} - 2|\mean{s}|^4  }{\pop{s}^2}\, .
	\end{align}
      \end{subequations}
\end{widetext}
For completeness, in Appendix~\ref{ap:coh-sq} we present the possible
models (Hamiltonians and Liouvillians) that produce a coherent state
and a squeezed state in a cavity, and give the correspondence between
the dynamical parameters (such as the coherent driving and the
squeezing intensity) and the abstract quantities~$\alpha$ and~$r$.

In the following Sections we will use the self-homodyne decomposition
Eqs.~(\ref{eq:decompositiontermswhole}) and this understanding in
terms of interferences between coherent and quantum components, which
in our cases may or may not be of the squeezing type, to analyse some
statistical properties (anti- and super-bunching) of systems in their
low-driving regime, and contrast their statistics with conventional
blockade effects. We will focus on cases that are both fundamental and
tightly related to each other, namely, the 2LS (resonance
fluorescence) in Section~\ref{sec:FriFeb23095933CET2018}, the
anharmonic oscillator in Section~\ref{sec:jueene10122434CET2019}, the
Jaynes--Cummings Hamiltonian in
Section~\ref{sec:FriFeb23100104CET2018} and microcavity polaritons in
Section~\ref{sec:FriFeb23100219CET2018}.

\section{Resonance fluorescence in the Heitler regime}
\label{sec:FriFeb23095933CET2018}

\begin{figure}[th!]
\centering
\includegraphics[width=.9\linewidth]{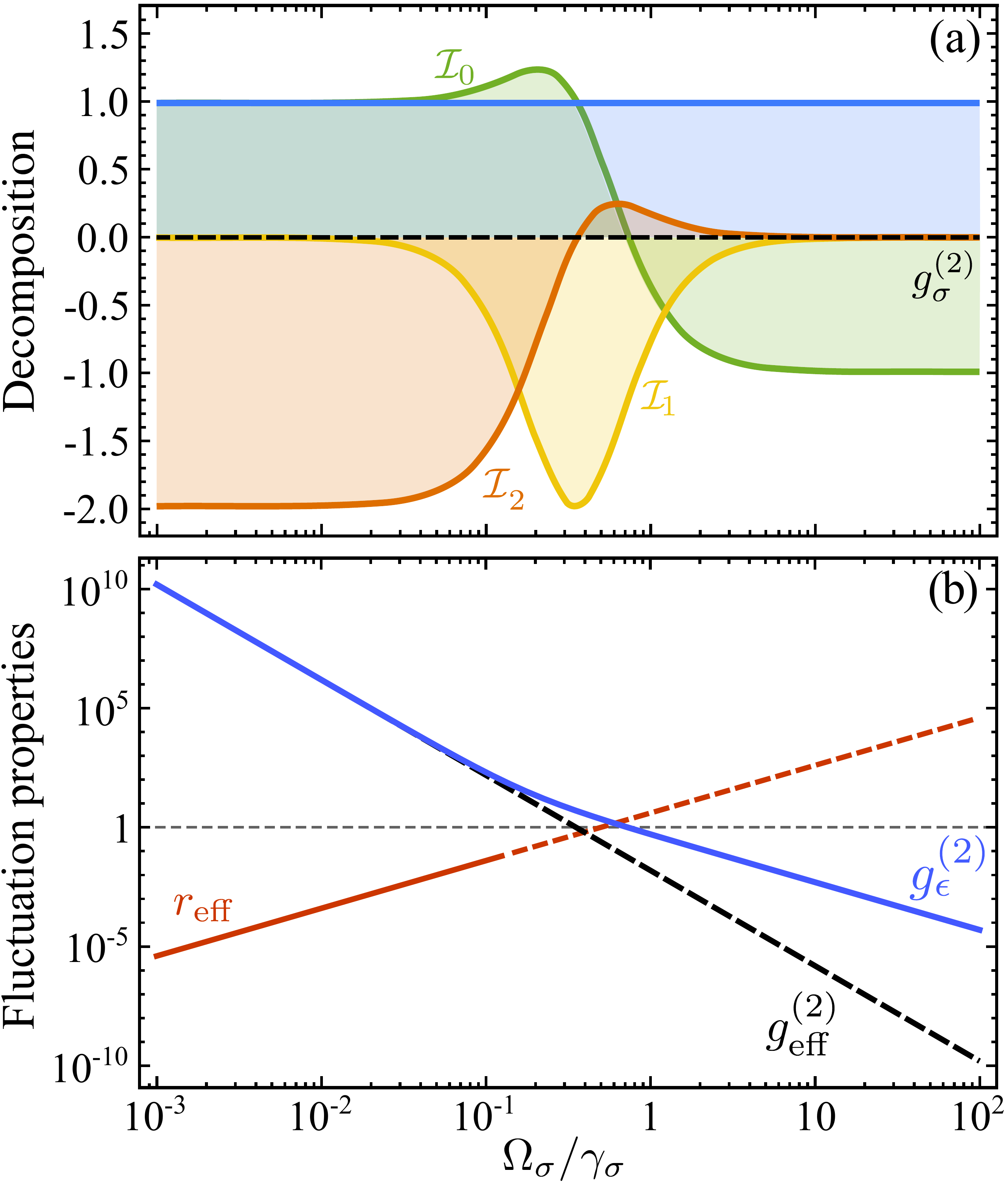}
\caption{Second-order coherence function $\g{2}_\sigma$ (dashed black)
  of resonance fluorescence, as a function of the driving-laser
  intensity~$\Omega_\sigma$ (at resonance). The Heitler regime,
  investigated in this work, is on the left
  ($\Omega_\sigma \ll\gamma_\sigma$). (a) Decomposition of
  $\g{2}_\sigma$, given by Eq.~(\ref{eq:decompositiontermswhole}) with
  $s\rightarrow \sigma$, which cancels out to give an exact zero. (b)
  Squeezing properties of the emitted light ($\Delta_\sigma = 0$),
  characterised by the effective squeezing parameter $r_\mathrm{eff}$
  (Eq.~\eqref{eq:2LSreff}) for the fluctuations only~$\epsilon$ (red
  line), and statistics of the fluctuations~$\g{2}_\epsilon$ (solid
  blue line), given by Eq.~\eqref{eq:FriFeb23174312CET2018}. The
  latter can be approximated by $\g{2}_\mathrm{eff}$
  (Eq.~\eqref{eq:2LSg2eff}) when the driving is low enough.  The red
  solid line indicates when the effective squeezing parameter fits
  properly the actual statistics and becomes dashed when the
  approximation fails. In both cases, the range of validity is
  $\Omega_{\sigma}\lesssim0.1\gamma_\sigma$.}
\label{fig:02}
\end{figure}

We first consider the excitation of a two-level system (2LS) driven by
a coherent source in the regime of low excitation---commonly referred
to as the \textit{Heitler regime}. Such a system is modeled by the
Hamiltonian ($\hbar=1$)
\begin{equation}
  \label{eq:2LShamiltonian}
  H_\mathrm{rf} = (\omega_\sigma-\omega_\mathrm{L}) \ud{\sigma} \sigma
  + \Omega_{\sigma} \left(\ud{\sigma} + \sigma \right)\,.
\end{equation}
This is the particular case of the general
Hamiltonian~(\ref{eq:Mon5Jun145532BST2017}) when only one mode is
considered and~$U\rightarrow\infty$. Here, the 2LS has a
frequency~$\omega_\sigma$ and is described with an annihilation
operator~$\sigma$ that follows the pseudospin algebra, whereas the
laser is treated classically, i.e., as a complex number, with
intensity~$\Omega_\sigma$ (taken real without loss of generality) and
frequency~$\omega_\mathrm{L}$. The dynamics only depends on the
frequency difference,
$\Delta_\sigma\equiv \omega_\sigma-\omega_\mathrm{L}$. The dissipative
character of the system is included in the dynamics with a master
equation
$\partial_t \rho = i \left[ \rho ,H_\mathrm{rf} \right] +
(\gamma_\sigma/2) \mathcal{L}_{\sigma} \rho$, where the Lindblad form
$\mathcal{L}_{\sigma} \rho = 2 \sigma \rho \ud{\sigma} - \ud{\sigma}
\sigma \rho - \rho \ud{\sigma} \sigma$ describes the decay of the 2LS
at a rate~$\gamma_\sigma$. The steady-state solution (computed as
indicated in Appendix~\ref{app:2}) can be fully written in terms of
two parameters: the 2LS population (or probability to be in the
excited state)~$\mean{n_\sigma}\equiv \pop{\sigma}$, and the coherence
or mean field~$\alpha \equiv \mean{\sigma}$~\cite{delvalle11a}:
\begin{equation}
\label{eq:2LSrhoSS}
\rho =
\begin{pmatrix}
1 - \mean{n_\sigma} & \alpha^\ast \\ \alpha & \mean{n_\sigma}
\end{pmatrix},
\end{equation}
where
\begin{subequations}
\label{eq:2LS_observables}
\begin{equation}
  \mean{n_\sigma} = \frac{4 \Omega_{\sigma}^2}{\gamma_\sigma^2 + 4
    \Delta_{\sigma}^2 + 8 \Omega_{\sigma}^2}\,, 
\end{equation}
\begin{equation}
\alpha = \frac{2 \Omega_{\sigma} (2 \Delta_{\sigma} + i
  \gamma_\sigma)}{\gamma_\sigma^2 + 4 \Delta_{\sigma}^2 + 8  
\Omega_{\sigma}^2}\,.
\end{equation}
\end{subequations}
As a consequence of the fermionic character of the 2LS, it can only
sustain one excitation at a time. Therefore, all the correlators
different from those in Eq.~(\ref{eq:2LS_observables}) vanish, and in
particular the $N$-photon correlations of the two-level system are
exactly zero, namely~$\g{N}_\sigma=0$ for~$N\geq 2$. We call this
perfect cancellation of correlations to all orders \emph{conventional
  blockade} or \emph{conventional antibunching} (CA), as it arises
from the natural Pauli blocking scenario. To investigate the
components of the correlations that ultimately provide the perfect
sub-Poissonian behaviour of the 2LS, we separate the mean field from
the fluctuations of the signal ($\sigma=\alpha+\epsilon$), in analogy
with Eq.~(\ref{eq:FriFeb23144144CET2018}).
Following~Eqs.~(\ref{eq:decompositiontermswhole}), $\g{2}_\sigma$ can
be decomposed as in Eq.~(\ref{eq:g2mixdecomposition}) with,
\begin{subequations}
  \label{eq:FriFeb23150914CET2018}
  \begin{align}
    \mathcal{I}_0 &= \frac{|\alpha|^2 \left( 6 \mean{n_\sigma}- 4
    |\alpha|^2 \right)}{n_\sigma^2} - 1\,,\\
    \mathcal{I}_1 &= -8 \frac{ |\alpha|^2 \left(
   \mean{n_\sigma}-   |\alpha|^2
    \right)}{n_\sigma^2}\,, \\ 
    \label{eq:2LSg2decompositiontermsI2}
    \mathcal{I}_2 &= 2 \frac{ |\alpha|^2 \left(
   \mean{n_\sigma}-  2 |\alpha|^2 \right)}{n_\sigma^2}\,.
  \end{align}
\end{subequations}
These are presented in Fig.~\ref{fig:02}(a) as a function of the
intensity of the driving laser. The decomposition shows that, although
the photon correlations of the 2LS are always perfectly
sub-Poissonian, or antibunched~\footnote{Equivalence of
  sub-Poissonianity with antibunching follows from
  $\lim_{\tau\rightarrow \infty}g_\sigma^{(2)}(\tau)=1$ for a
  continuously driven system, so that $g_\sigma^{(2)}=0$ also implies
  $g_\sigma^{(2)}<g_\sigma^{(2)}(\tau)$.}, the nature of their
cancellation varies depending on the driving
regime~\cite{lopezcarreno18b}. In the high-driving regime, the
coherent component is compensated by the sub-Poissonian statistics of
the quantum fluctuations ($\mathcal{I}_0<0$) since
$\lim_{\Omega_\sigma \rightarrow \infty} \alpha =0$ and fluctuations
become the total field, $\epsilon\rightarrow \sigma$. In contrast, in
the Heitler regime the coherent component is compensated by the
super-Poissonian but also squeezed
fluctuations~($\mathcal{I}_2<0$). The Heitler regime is, therefore, an
example of the type of self-homodyne interference that we discussed in
Sec.~\ref{sec:ThuFeb15174655CET2018}.

Let us then analyse more closely the fluctuations by looking into
their correlation functions:
\begin{equation}
  \label{eq:FriFeb23154045CET2018}
  \mean{\epsilon^{\dagger k}\epsilon^l}
  =(-1)^{k+l}\alpha^{\ast\,k-1}\alpha^{l-1}\big( |\alpha|^2
  (1-k-l+kl) +kl\,\mean{n_\epsilon}\big)\,,
\end{equation}
where~$\mean{n_\epsilon} = \mean{n_\sigma}-|\alpha|^2$ is the
contribution from the fluctuations to the total population of the
2LS. More details are given in
Appendix~\ref{sec:FriFeb23221121CET2018}. In particular,
the~$N$-photon correlations from the fluctuations alone is given by,
\begin{equation}
  \label{eq:FriFeb23155745CET2018}
  g^{(N)}_\epsilon =\frac{|\alpha|^{2(N-1)}
    \big(N^2\,\mean{n_\sigma}+(1-2N)|\alpha|^2\big)}{(\mean{n_\sigma}- |\alpha|^2)^N}\,,
\end{equation}
that in terms of the physical parameters reads
\begin{equation}
  \label{eq:FriFeb23174312CET2018}
  \g{N}_\epsilon = \frac{ \left(N-1\right)^2 \left(\gamma_\sigma^2 +4
        \Delta_{\sigma}^2 
    \right) + 8 N^2 \Omega_{\sigma}^2}{8^{N}\,
  \Omega_{\sigma}^{2N} \left(\gamma_\sigma^2 +4
    \Delta_{\sigma}^2 \right)^{N-1}}\,.
\end{equation}
In Fig.~\ref{fig:02}(b) we plot $\g{2}_\epsilon$ confirming that
fluctuations are sub-Poissonian or super-Poissonian depending on
whether the effective driving defined
as~$\Omega_\mathrm{eff}\equiv\Omega_\sigma/\sqrt{1+(2\Delta_\sigma/\gamma_\sigma)^2}$
is much larger or smaller than the system decay~$\gamma_\sigma$,
respectively (the figure is for the resonant case).

In the Heitler regime, we need to consider only the magnitudes up to
leading order in the effective normalised
driving~$p\equiv 2\Omega_\mathrm{eff}/\gamma_\sigma$. The main
contribution to the
intensity~$\mean{n_\sigma}=|\alpha|^2+\mean{n_\epsilon}$, in the
absense of pure dephasing, comes from the coherent part~$|\alpha|^2$
of the signal. Fluctuations only appear to the next order, having, up
to fourth order in $p$:
\begin{subequations}
  \begin{align}
    \mean{n_\sigma}&=p^2-2p^4 \,,\\
    |\alpha|^2&=p^2-4p^4\,,\\
    \mean{n_\epsilon}&=2p^4\,.
    \end{align}
\end{subequations}
The coherent contribution corresponds to the elastic (also known as
``Rayleigh'') scattering of the laser-photons by the two-level system,
while the fluctuations originate from the two-photon excitation and
re-emission~\cite{dalibard83a}. In the spectrum of emission, this
manifests as a superposition of a delta and a Lorentzian peaks with
exactly these weights, $|\alpha|^2$ and $\mean{n_\epsilon}$, both
centered at the laser frequency, with no width (for an ideal laser)
and $\gamma_\sigma$-width,
respectively~\cite{mollow69a,loudon_book00a,lopezcarreno18b}.
Fluctuations have no coherent intensity by construction,
$\mean{\epsilon}=0$. At the same time, their second momentum is not
zero but exactly the opposite of the coherent field one:
$\mean{\epsilon^2}=-\alpha^2$, thanks to the fact that
$\mean{\sigma^2}=\alpha^2+\mean{\epsilon^2}=0$. This means that both
contributions, coherent and incoherent, are of the same order in the
driving~$p$ when it comes to two-photon processes and can, therefore,
interfere and even cancel each other. This is precisely what happens
and is made explicit in the $g^{(2)}_\sigma$-decomposition above. The
strong two-photon interference ($\mathcal{I}_2$) can compensate the
Poissonian and super-Poissonian statistics of the coherent and
incoherent parts of the signal ($1+\mathcal{I}_0$). Since quadrature
squeezing is created by a displacement operator, or a Hamiltonian,
based on the operator~$\epsilon^2$, this situation corresponds to a
high degree of quadrature squeezing for the fluctuations. Further
analysis of the fluctuation quadratures (details of the calculation
are in Appendix~\ref{sec:FriFeb23221121CET2018}) shows that their
variances behave similarly to a \emph{squeezed thermal} state in the
Heitler regime, which allows us to derive an effective squeezing
parameter~$r_\mathrm{eff}=p^2$ to describe the state to lowest order
in the driving.  This is plotted in Fig.~\ref{fig:02}(b) as a function
of the driving, with the line becoming dashed when the interference
can no longer be described in terms of a squeezed thermal state.  Note
that the total signal has no squeezing at low driving, only
fluctuations do, because the coherent contribution is much larger.

\begin{figure}
  \centering \includegraphics[width=0.9\linewidth]{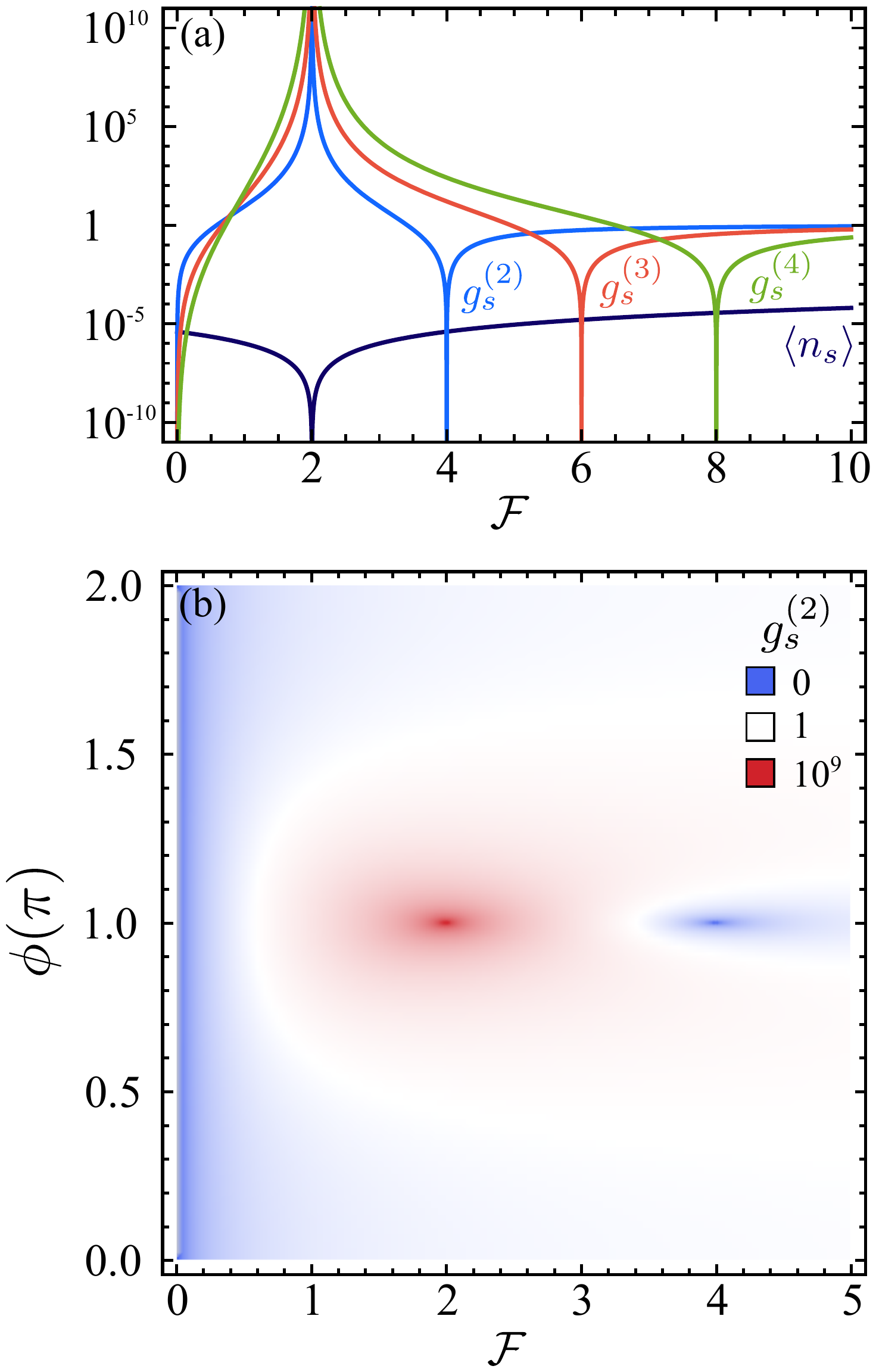}
  \caption{(Color online) Interference between the output of resonance
    fluorescence and an external laser of intensity proportional
    to~$\mathcal{F}$ and phase~$\phi$. All modes at resonance. (a)
    Intensity~$\mean{n_s}$ and $N$-th order coherence
    functions~$g_s^{(N)}$ of the resulting light as a function
    of~$\mathcal{F}$. (b) $g_s^{(2)}$ as a function of both
    $\mathcal{F}$ and~$\phi$ (in units of $\pi$) with the colour code
    in inset (logscale). Tuning the external laser allows to choose
    between various resonant conditions.}
  \label{fig:WedNov29150123CET2017}
\end{figure}

Resonance fluorescence by itself always provides antibunching, due to
the perfect cancellation of the various components. However, one can
disrupt this by manipulating the coherent fraction, simply by
interfering the signal~$\sigma$ in a beam splitter with an external
coherent state $\ket{\beta}$. This allows to change the photon
statistics of the total signal
$s= \mathrm{T} \sigma + i \mathrm{R} \beta $, where $T^2$ and $R^2$
are the transmittance and reflectance of the beam splitter. Actually,
since the decomposition affects correlators to all orders,
Eq.~(\ref{eq:FriFeb23155745CET2018}), one can target the $N$-photon
level instead of the 2-photon one. Namely, one can decide to set the
$N$-photon coherence function to zero. As a particular case, the
1-photon case cancels the signal altogether, which is obtained by
solving the condition
$\mean{n_s}=\mathrm{T}^2|\alpha + i \mathrm{R} \beta_1/\mathrm{T}|^2
=0$ (because $\mean{n_\epsilon} =0$ to second order in
$\Omega_\sigma$). We will show that the possibility to target one~$N$
in isolation of the others introduces a separate regime from
conventional blockade.  Given their relationship and in line with the
terminology found in the literature, we refer to this as
\emph{unconventional blockade} and \emph{unconventional antibunching}
(UA).

With this objective of tuning $N$-photon statistics and in order to
avoid referencing the specificities of the beam splitter which do not
change the normalized observables, let us
define~$\beta' \equiv \mathrm{R} \beta/\mathrm{T} \equiv
|\beta'|e^{i\phi}$ and parametrise its amplitude as a
fraction~$\mathcal{F}$ (always a positive number) of the laser field
exciting the 2LS:
\begin{equation}
  \label{eq:SatFeb24130524CET2018}
  |\beta'| = \frac{\Omega_\sigma}{\gamma_\sigma}\mathcal{F} \,.
\end{equation}
With this, the coherence function $g^{(N)}_s$ of the interfered field
in the Heitler regime is given by:
\begin{multline}
\label{eq:jueene17193740GMT2019}
g^{(N)}_s = \frac{\mathrm{T}^{2N}}{\av{n_s}^N}\frac{\mathcal{F}^{2 (N-1)}
  \Omega_{\sigma}^{2N}} {\gamma_{\sigma}^{2N}\left(\gamma_{\sigma}^2 +
  4 \Delta_\sigma^2\right) } \times \\ \big[\mathcal{F}^2
  \left(\gamma_{\sigma}^2 + 4 \Delta_{\sigma}^2 \right)+ 4 N
  \mathcal{F} \gamma_{\sigma} \left(\gamma_{\sigma} \cos \phi - 2
  \Delta_\sigma \sin \phi\right) +{}\\{}+ 4 N^2 \gamma_{\sigma}^2\big] \,.
\end{multline}
Since $g_s^{(1)}=1$, this expression also provides the
population~$\av{n_s}$ by considering the case~$N=1$. One can
appreciate the considerable enrichment brought by the interfering
laser by comparing $\av{n_s}$ (with the interfering laser,
$\mathcal{F}\neq0$) to Eq.~(\ref{eq:2LS_observables}(a)) (without,
$\mathcal{F}=0$) and even more so by comparing the $N$-photon
correlation function, which is identically zero without the
interfering laser, and that becomes
Eq.~(\ref{eq:jueene17193740GMT2019}) with the interfering laser.
Interestingly, there is now another condition that suppresses the
correlations and yields perfect antibunching at a given $N$-photon
order, in addition to the one obtained in the original system without
the interfering laser~(CA). The new conditions exist for any detuning
and are given by:
\begin{equation}
\label{eq:2LSconds}
\tan \phi_N = - \frac{2\Delta_\sigma}{\gamma_\sigma} \quad \quad
\mathrm{and} \quad\quad \mathcal{F}_N = -2 N \cos \phi_N\,.
\end{equation}
Focusing on the resonance case for simplicity, we have
$\mathcal{F}_N=2N$ and always the same phase, $\phi_N=\pi$, which
corresponds to the field $i\beta_N'=- N |\alpha|$. The total coherent
fraction changes phase for all $N$:
$\alpha + i \beta_N' = -(N-1)\alpha$.  The signal population
($\mean{n_s}=G^{(1)}_s$) vanishes due to a first order (or one-photon)
interference at the external laser parameter $\mathcal{F}_1 = 2 $,
which translates into $i\beta_1'=-i|\alpha|$. The external laser
completely compensates the coherent fraction of resonance
fluorescence, in this case $\alpha=i|\alpha|$ (with
$|\alpha|=2\Omega_\sigma/\gamma_\sigma$).  This situation corresponds
to a \emph{classical destructive interference}, which equally occurs
between two fully classical laser beams with the same intensity and
opposite phase.

In the case of highest interest, that of two-photon
correlations~$g^{(2)}_s$, we find a destructive two-photon
interference at the intensity $\mathcal{F}_2=4$, which corresponds to
an external laser $i\beta_2=- 2|\alpha|$, that fully inverts the sign
of the coherent fraction in the total signal:
$\alpha + i \beta_2' = -\alpha$. This coherent contribution leads to
perfect cancelation of the two-photon probability in a wavefunction
approach~\cite{visser95a} (see the details in
Appendix~\ref{sec:WedFeb28173330CET2018}).  Note that this does not,
however, satisfy all other $N$-photon interference conditions and
$g^{(N)}_s$ with $N>2$ do not vanish. This is a very different
situation as compared to the original resonance fluorescence where
$g^{(N)}_\sigma=0$ for all $N>1$. One, the conventional scenario,
arises from a an interference that takes place at all orders. The
other, the unconventional scenario, results from an interference that
is specific to a given number of photons.

\begin{table*}[ht]
  \centering
  \begin{ruledtabular}
    \begin{tabular}{c|c|c||c|c}
      $\g{N}$ & Squeezed Thermal & Heitler fluctuations &  Displaced Squeezed Thermal & Laser-corrected configuration \\  
      \hline 
      & & & & \\ [-2ex]
			$n_a$ & $\frac{32\Omega_\sigma^4}{ \Gamma_\sigma^4} + \frac{1792 \Omega_\sigma^8}{3 \Gamma_\sigma^8} + O\left(\Omega_\sigma^{12}\right)$ &
			$\frac{32\Omega_\sigma^4}{ \Gamma_\sigma^4} - \frac{512 \Omega_\sigma^6}{ \Gamma_\sigma^6} + O\left(\Omega_\sigma^{8}\right)$ 
& $\frac{4 \Omega_\sigma^2}{\Gamma_\sigma^2} + 
			\frac{32 \Omega_\sigma^4}{\Gamma_\sigma^4} + O\left(\Omega_\sigma^6\right)$ &
			$\frac{4 \Omega_\sigma^2}{\Gamma_\sigma^2} - 
			\frac{32 \Omega_\sigma^4}{\Gamma_\sigma^4}
			+ O\left(\Omega_\sigma^6\right)$ 
\\
      $\g{2}$ & $\frac{\Gamma_\sigma^4}{64 \Omega_\sigma^4} + \frac{11}{4} + O\left(\Omega_\sigma^4\right)$ &
                                                                                                             $\frac{\Gamma_\sigma^4}{64 \Omega_\sigma^4} + \frac{\Gamma_\sigma^2}{2 \Omega_\sigma^2} + O\left(\Omega_\sigma^0\right) $  
& $0 +\frac{32 \Omega_\sigma^2}{\Gamma_\sigma^2} + O\left(\Omega_\sigma^4\right)$ & $ 0 + \frac{128 \Omega_\sigma^2}{\Gamma_\sigma^2} + O\left(\Omega_\sigma^4\right) $ \\ [1.5ex]
      $\g{3}$ & $\frac{9 \Gamma_\sigma^4}{64 \Omega_\sigma^4} + \frac{51}{4} + O\left(\Omega_\sigma^4\right)$ & $ 4 - \frac{96 \Omega_\sigma^2}{\Gamma_\sigma^2} + O\left(\Omega_\sigma^4\right)$ & $16 +\frac{768 \Omega_\sigma^2}{\Gamma_\sigma^2} + O\left(\Omega_\sigma^{4}\right) $ & $\frac{\Gamma_\sigma^6}{128 \Omega_\sigma^6} + \frac{9 \Gamma_\sigma^4}{64 \Omega_\sigma^2} + O\left({1\over\Omega_\sigma^{2}}\right) $  \\ [1ex]
    \end{tabular}
  \end{ruledtabular}
  \caption{Two-level system. Comparison of first- (population),
    second- and third-order photon correlations, i) between a Squeezed
    Thermal state and the fluctuations in the Heitler regime and ii)
    between a Displaced Squeezed Thermal state and the fluctuations in
    the laser-corrected Heitler regime, to various orders in the
    driving~$\Omega_\sigma$.}
  \label{tab:fluctuations}
\end{table*}

\begin{table*}[ht]
	\centering
	\begin{ruledtabular}
		\begin{tabular}{c|c|c||c|c}
			$\g{N}$ & Displaced Squeezed Thermal & AO antibunching &  Displaced Squeezed Thermal & Laser-corrected configuration \\  
			\hline 
			& & & & \\ [-2ex]
			$n_a$ & $ 2.89 \, \Omega_b^2 + 4.63 \, \Omega_b^4 + O \left(\Omega_b^6 \right) $ &
			$ 2.89 \, \Omega_b^2 - 10.36 \, \Omega_b^4 + O \left(\Omega_b^6\right)$ 
			& $ 1.52 \, \Omega_b^2 + 4.63 \, \Omega_b^4 + O\left(\Omega_b^6\right)$ &
			$1.52 \, \Omega_b^2 - 3.25 \, \Omega_b^4
			+ O\left(\Omega_b^6\right)$ 
			\\
			$\g{2}$ & $ 0.38 + 5.18 \, \Omega_b^2 + O\left(\Omega_b^4\right)$ &
			$ 0.38 + 0.91 \, \Omega_b^2 + O\left(\Omega_b^4 \right) $  
			& $0 + 12.75 \, \Omega_b^2 + O\left(\Omega_b^4\right)$ & $ 0 + 47.84 \, \Omega_b^2 + O\left(\Omega_b^4\right) $ \\ [1.5ex]
			$\g{3}$ & $ 0.80 + 1.64 \, \Omega_b^2 + O\left(\Omega_b^4\right)$ & $ 0.06 + 0.37 \, \Omega_b^2 + O\left(\Omega_b^4\right)$ & $ 4 - 34.52 \, \Omega_b^2 + O\left(\Omega_b^{4}\right) $ & $ 0.71 + 0.78 \, \Omega_b^2 + O\left({\Omega_b^{4}}\right) $  \\ [1ex]
		\end{tabular}
	\end{ruledtabular}
	\caption{Anharmonic oscillator. Comparison of first-
          (population), second- and third-order photon correlations,
          i) between a Displaced Squeezed Thermal state and the
          anharmonic oscillator with $\Delta_b = \Delta_{-}$ (optimal
          antibunching) and ii) between a Displaced Squeezed Thermal
          state and the laser-corrected for the optimal $\g{2}$
          configuration ($\mathcal{F}_{2,2}$ and $\phi_{2,2}
          $). Selected parameters: $\gamma_b = U = 1$.}
	\label{tab:AOtable}
\end{table*}

All these interferences can be seen in
Fig.~\ref{fig:WedNov29150123CET2017}(a) where we plot them up to
$N=4$. When there is no interference with the external laser,
$\mathcal{F}=0$, antibunching is perfect to all orders recovering
resonance fluorescence. At the one-photon interference, the
denominator of $g^{(N)}_s$ becomes zero and the functions, therefore,
diverge. This produces a superbunching effect of a classical origin,
as previously discussed: a destructive interference effect that brings
the total intensity to zero. In this case, the external laser removes
completely by destructive interference the coherent fraction of the
total signal. Therefore, the statistics is that of the fluctuations
alone, what we previously called $g^{(N)}_\mathrm{\epsilon}$, given by
Eq.~\eqref{eq:FriFeb23155745CET2018}. We have already discussed how,
in the Heitler regime, fluctuations become super chaotic and
squeezed. We can see, on the left hand side of Fig.~\ref{fig:1}, that
in the limit of $\Omega_\sigma \rightarrow 0$, they actually diverge.
Such a superbunching is thus linked to noise.  The resulting state is
missing the one-photon component and, consequently, the next
(dominating) component is the two-photon one. Nevertheless, there is
not a suppression mechanism for components with higher number of
photons so that the relevance of such a configuration for multiphoton
(bundle) emission remains to be investigated, which is however better
left for a future work. We call this feature \emph{unconventional
  bunching} (UB) in contrast with bunching that results from a
$N$-photon de-excitation process that excludes explicitly the emission
of other photon-numbers.  This superbunching, as well as the
antibunching by destructive interferences, will be reappearing in the
next systems of study. The Heitler regime is, therefore, a simple but
rich system where all the squeezing-originated interferences already
occur although we need an external laser to have them
manifest. 

We now turn to the subtle point of which quantum state is realized by
the various scenarios. To lowest-order in the driving, the dynamical
state of the system can be described by a superposition of a coherent
and a squeezed state, insofar as only the lower-order correlation
functions (namely, population and~$\g{2}$) are considered.  This is
shown in Table~\ref{tab:fluctuations}, where we compare $g^{(N)}$ for
$1\le N\le 3$ (with $N=1$ corresponding to the population) for the
fluctuations in the Heitler regime vs the corresponding observables
for a squeezed thermal state, on the one hand, and the laser-corrected
configuration vs the displaced squeezed thermal state on the other
hand.  As explained in Appendix~\ref{sec:FriFeb23221121CET2018}, such
a comparison can be made by identifying the squeezing parameter and
thermal populations to various orders in a series expansion of the
quantum states with the corresponding observables from the dynamical
systems. One finds:
\begin{align}
  \mathrm{r}_\mathrm{eff} = \frac{4 \Omega_\sigma^2}{\Gamma_\sigma^2}, &&
                                                                          \mathrm{p}_\mathrm{eff} = \frac{16 \Omega_\sigma^4}{\Gamma_\sigma^4},
\end{align}
where we have defined
$\Gamma_{\sigma}^2 = \gamma_\sigma^2 + 4 \Delta_\sigma^2$.  By
definition, fluctuations have a vanishing mean, i.e.,
$\mean{\epsilon} = 0 $ so we must choose $\alpha = 0$.  On the other
hand, for the corrected emission, since one is blocking the two-photon
contribution (at first order, this gives $\g{2} = 0 $), the comparison
with a displaced thermal state is obtained by imposing the condition
for $\g{2}$ to vanish at first order ($r = |\alpha|^2$ and
$\theta = 2 \, \phi $). The results are compiled in the table up to
the order at which the results differ. Through the typical observables
that are the population and $g^{(2)}$, one can see how the system is
indeed well described to lowest order in the driving by a coherent
squeezed thermal state (displaced if there is a
laser-correction). However to next order, there is a departure,
showing that the Gaussian state representation is an approximation
valid up to second-order only. In fact, for three-photon correlations,
the disagreement occurs already at the lowest-order in the driving,
and is of a qualitative character, as is also shown in the
table. Therefore, such a description is handy but breaks down if a
high-enough number of photons or a too high-pumping is considered.

\section{Anharmonic blockade}
\label{sec:jueene10122434CET2019}

\begin{figure}[tbh!]
	\centering
	\includegraphics[width=.9\linewidth]{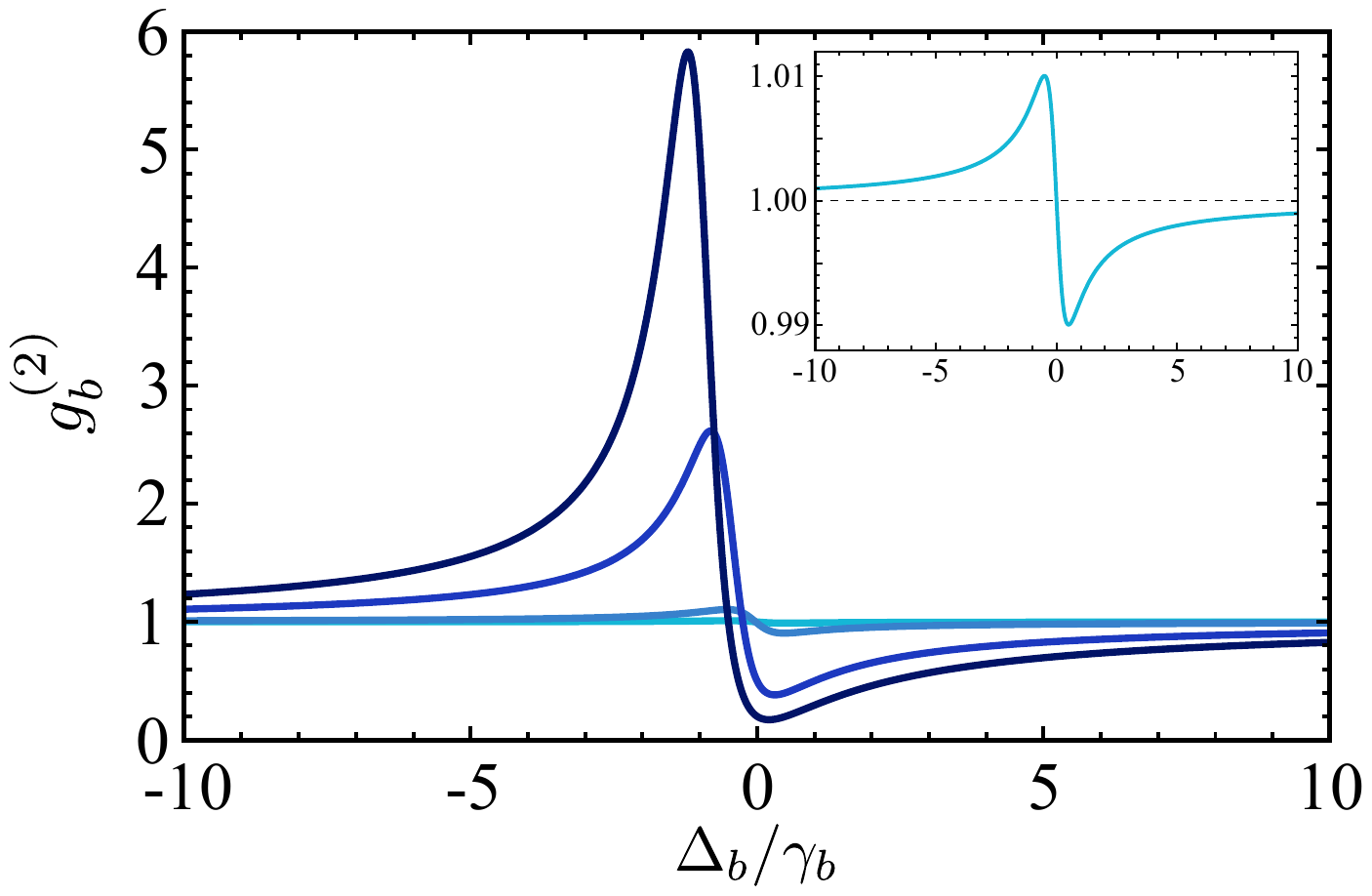}
	\caption{Second-order coherence function $\g{2}_b$ of an
          anharmonic oscillator, as a function of the
          detuning~$\Delta_b$ From light to dark blue, $U$ increases
          (the exact values are
          $U / \gamma_b = 0.01, \, 0.1, \, 1, \, 2 $, for fixed
          $\gamma_b = 1$). A zoom-in of the smallest case is shown in
          the inset.}
	\label{fig:ao1}
\end{figure}

To show that the effects of conventional (self-homodyne interference
at all~$N$) and unconventional (self-homodyne interference at a given
$N$ only) blockades take place in a general setting and are not
specific of strong quantum nonlinearities (such as a 2LS), we now
address the case of a single anharmonic oscillator, that describes an
interacting bosonic mode with a Kerr-type nonlinearity, which can be
very weak. With driving by a coherent source (a laser) at frequency
$\omega_{\mathrm{L}}$, its Hamiltonian reads
\begin{equation}
H_\mathrm{ao} = \Delta_b \, \ud{b} b + \frac{U}{2} \, \ud{b} \ud{b} b b +
\Omega_b (\ud{b} + b),
\end{equation}
where the cavity operators are represented by $\ud{b}$ and $b$,
$\Delta_b = \omega_b - \omega_{\mathrm{L}}$ is the detuning between
the cavity and the laser, $U$ denotes the particle interaction
strength (that provices the nonlinearity) and the driving amplitude is
given by $\Omega_b$.  This is the particular case of the general
Hamiltonian~(\ref{eq:Mon5Jun145532BST2017}) when only one mode is
considered and~$U$ remains finite and, generally, small. The level
structure of this system (at vanishing driving) is given by the simple
expression~$E^{(N)}=N\omega_b+N(N-1)U$. The condition for the laser
frequency to hit resonantly the $N$-photon level is
$\omega_\mathrm{L}=E^{(N)}/N$ (or $\Delta_b=-(N-1)U/2$).

We restrict our analysis of the dynamics
$\dot{\rho} = - i \, [H_\mathrm{ao}, \rho] + (\gamma_b/2)
\mathcal{L}_b \rho$, with $\gamma_b$ the decay rate of the mode, to
the case of vanishing pumping, i.e. $\Omega_b \ll \gamma_b$. Solving
the correlator equations in this limit gives the 
population
\begin{equation}
\label{eq:jueene17201516GMT2019}
\av{n_b} = \frac{\Omega_b^2}{\gamma_b^2 + 4 \Delta_b^2}\,,
\end{equation}
the 2nd-order Glauber correlator
\begin{equation}
\label{eq:jueene17201437GMT2019}
\g{2}_b = \frac{\av{\ud{b} \ud{b} b b}}{\av{\ud{b} b}^2} =
         \frac{\big(\gamma_a^2 + 4 \Delta_b^2 \big)}{\gamma_b^2 + (U + 2 \Delta_b)^2}\,,
\end{equation}
as well as the higher-order correlators
\begin{equation}
\g{N}_b = \frac{\big( \gamma_a^2 + 4 \Delta_b^2 \big)^{n-1}}{\prod_{k = 1}^{n-1} 
	\big[ \gamma_b^2 + \big(k U + 2 \Delta_b \big)^2 \big]}\,.
\end{equation}
This shows that, when scanning in frequency, $\g{2}_b$ has two
extrema, one minimum and one maximum, as can be seen in
Fig.~\ref{fig:ao1}, whose positions are given by
\begin{equation}
  \Delta_{\pm} = - \frac{1}{4} \Big(U \pm \sqrt{U^2 + 4 \gamma_b^2} \, \Big),
\end{equation}
with respective optimum antibunching~($-$) and bunching~$(+)$
\begin{equation}
\g{2}_b \big(\Delta_b = \Delta_{\pm}\big) =
      1 + \frac{U \Big( U \pm \sqrt{U^2 + 4 \gamma_b^2} \Big)}{2 \gamma_b^2}.
\end{equation}
Both of these features are linked to the level structure: the
antibunching condition is that of resonantly driving the first rung,
$E^{(1)}$ (note that $\Delta_-\sim 0$, especially when
$U\gg \gamma_b$), and the bunching condition, that of driving the
second rung, $E^{(2)}$ ($\Delta_+\sim -U/2$). In both cases, all other
rungs are off-resonance and will remain much less occupied. Therefore,
these effects are of a \emph{conventional} nature, as we have defined
it in the previous section: CA and \emph{conventional bunching} (CB),
respectively. The difference with resonance fluorescence is that here,
CA is not a perfect interference at all orders ($g^{(N)}=0$ for $N>1$)
but an approximated one. For instance, $g^{(2)}\approx(\gamma_b/U)^2$
(to leading order in $U/\gamma_b$), is only zero in the limit
$U\rightarrow \infty$, when the system converges to a 2LS. On the
other hand, there was not CB in resonance fluorescence due to the lack
of levels~$N>1$. Here, we see it appearing for the first time.

\begin{figure}[t!]
	\centering
	\includegraphics[width=.9\linewidth]{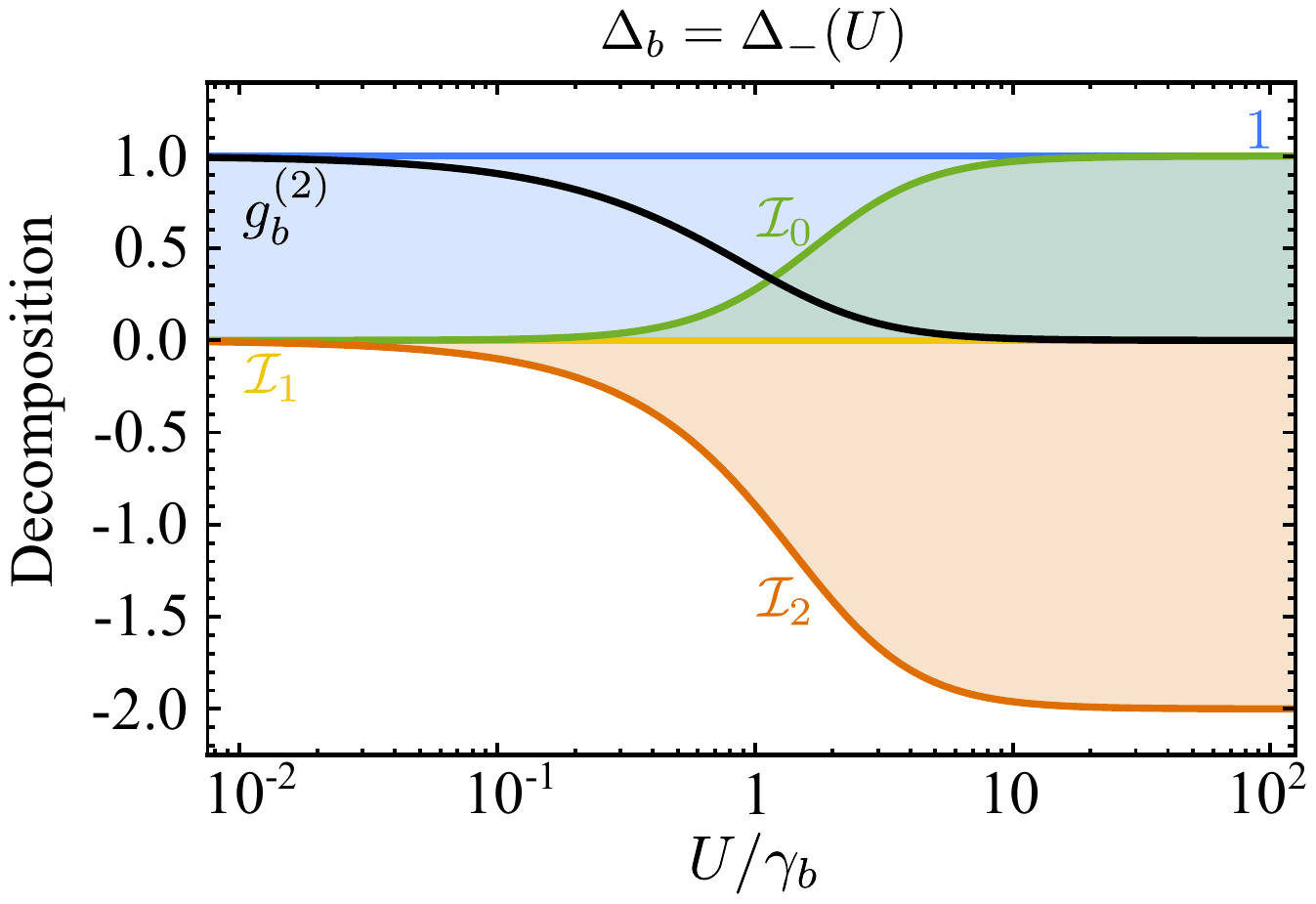}
	\caption{Second-order coherence function $\g{2}_b$ together
          with its decomposition, as a function of the non-linear
          interactionnonlinearity strength $U$ following the
          conventional antibunching frequency $\Delta_b = \Delta_{-}
          (U)$.}
	\label{fig:ao2}
\end{figure}

The decomposition~$\g{2}_b$ of according to
Eq.~(\ref{eq:g2mixdecomposition}) yields
\begin{subequations}
\begin{align}
\mathcal{I}_0 = & \, \frac{U^2}{\gamma_b^2 + (U + 2 \Delta_b)^2}   \,, \\
\mathcal{I}_1 = & \, 0  \,, \\
\mathcal{I}_2 = & \,-\frac{2 U \big(U + 2 \Delta_b \big)}{\gamma_b^2 + (U + 2 \Delta_b)^2}   \,.
\end{align}
\end{subequations}
$\mathcal{I}_0>0$ means that fluctuations are always
super-poissonian. $\mathcal{I}_1$ vanishes in the limit of low driving
(as in the case of the Heitler regime), which means that there are no
anomalous correlations to leading order in~$\Omega_b$.  The remaining
term $\mathcal{I}_2$ can take positive (for $\Delta_b > - U/2$) and
negative (for $\Delta_b < - U/2$) values, resulting in
super-Poissonian statistics or on the contrary favouring
antibunching. The various terms and the total $\g{2}_b$ are shown in
Fig.~\ref{fig:ao2} as a function of $U$, for the case
$\Delta_b=\Delta_{-}(U)$ that maximizes antibunching, showing the
evolution from Poissonian fluctuations in the linear regime of a
driven harmonic mode to antibunching as the two-level limit is
recovered with $\mathcal{I}_0 \rightarrow 1$ and
$\mathcal{I}_2 \rightarrow -2$ (cf.~Fig.~\ref{fig:02}(a)).

\begin{figure*}[th!]
	\centering
	\includegraphics[width=\linewidth]{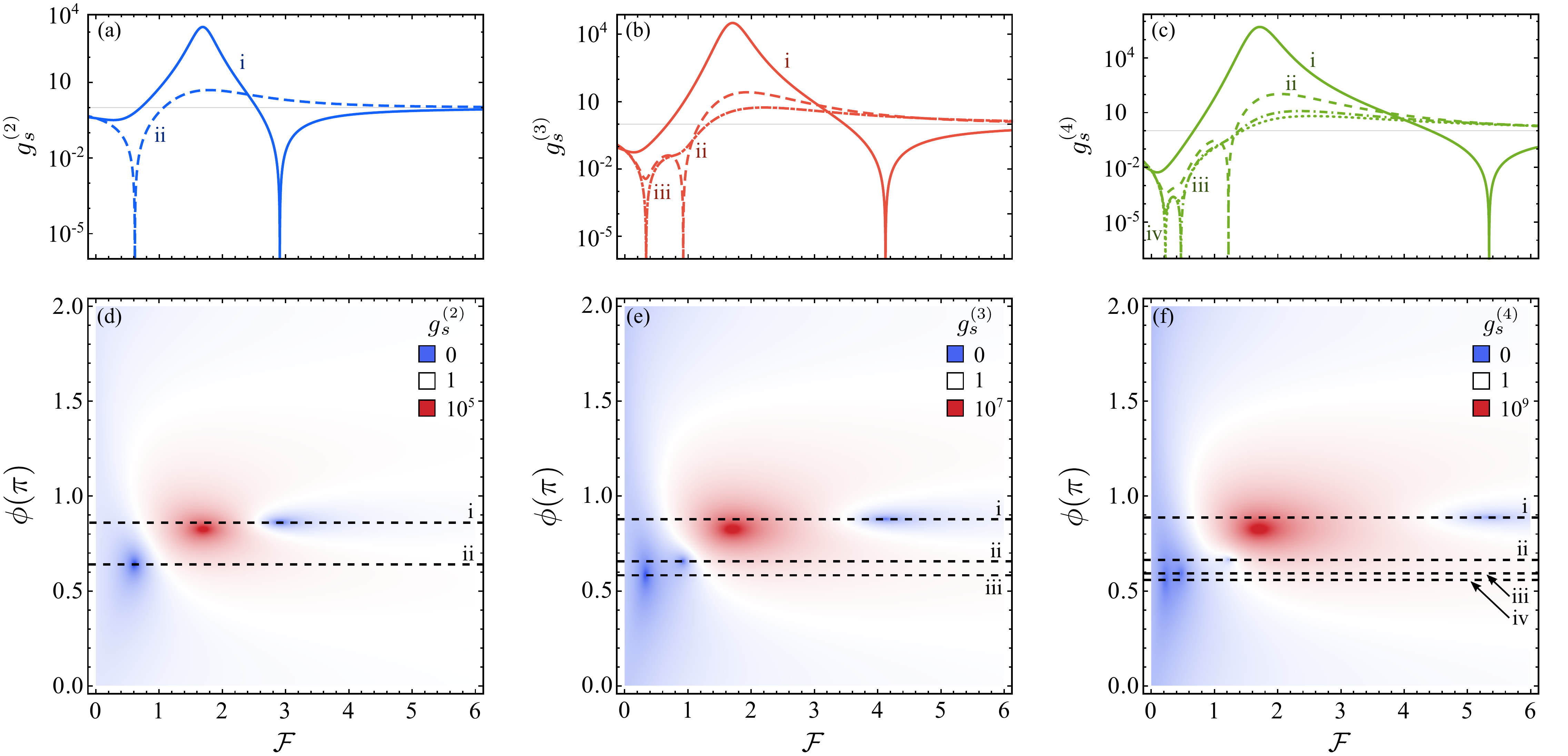}
	\caption{(Color online) Interference between the output of a
          driven anharmonic system and an external laser of intensity
          proportional to~$\mathcal{F}$ and phase~$\phi$. Various
          columns shown increasing orders of photon correlations, with
          the upper row showing a phase-cut along the lines that
          intercept one of the resonances in the full landscape shown
          in the bottow row (the superbunching case is not shown).
          Parameters: $U = \gamma_b$, $\Omega_b = 0.001 \, \gamma_b$,
          $\mathrm{T} = 0.5$ and $\gamma_b = 1$.}
	\label{fig:ao3}
\end{figure*}

As previously, the statistics can be modified by adjusting the
coherent component of the original signal $b$ with an external laser
$\beta = |\beta|e^{i \phi}$.  The resulting signal is then described
by the operator $s = \mathrm{T} b + i \, \mathrm{R} \beta$, with
coherent contribution now given by
$\av{s} = \mathrm{T} \av{b} + i \, \mathrm{R} \beta $. To simplify
further the calculations, we choose
$\beta = \frac{\mathrm{T}}{\mathrm{R}} \beta'$, where $\beta'$ is also
written in terms of the driving amplitude and an adimensional
amplitude $\mathcal{F}$:
\begin{equation}
\beta' = \frac{\Omega_b}{\gamma_b} \mathcal{F}\,.
\end{equation}
Additionally, we shift the phase $\phi \rightarrow \phi + \pi$ so in
the limit of high $U$, all the results are consistent with the previous
case.  Then, the total population becomes:
\begin{widetext}
\begin{equation}
\label{eq:jueene17201652GMT2019}
\av{n_s} = \frac{\mathrm{T}^2 \Omega_b^2}
{\gamma_b^2 \big(\gamma_b^2 + 4 \Delta_b^2\big)} 
 \big[\mathcal{F}^2\big(\gamma_b^2 + 4 \Delta_b^2\big) 
+ 4 \gamma_b \mathcal{F}  \big( \gamma_b \cos \phi - 2 \Delta_b \sin \phi \big) 
 + 4 \gamma_b^2 \big]\, ,
\end{equation}
%
and the 2-photon correlations become
\begin{equation}
\label{eq:jueene17201723GMT2019}
\begin{split}
\g{2}_s = & \tilde{\Gamma}_b^2 
\Big\lbrace \tilde{\Gamma}_b^2 \big[\gamma_b^2 + (U + \Delta_b)^2\big] \mathcal{F}^4 
+ 8 \gamma_b \big[\gamma_b^2 + (U + \Delta_b)^2\big] 
( \gamma_b \cos \phi - 2 \Delta_b  \sin \phi)  \mathcal{F}^3 +
 \big[ 16 \gamma_b^2 \big(\gamma_b^2 + (U + 2 \Delta_b)^2\big) +  \\
 &  8 \gamma_b^2 \cos 2 \phi \, \big( \gamma_b^2 - 2 \Delta_b (U + 2 \Delta_b) \big) -
 8 \gamma_b^3 \sin 2 \phi (U + 4 \Delta_b)\big] \mathcal{F}^2 + 
32 \gamma_b^3 \big[\gamma_b \cos \phi - (U + 2 \Delta_b) \sin \phi \big] \mathcal{F} +
 16 \gamma_b^2 \Big\rbrace \biggm/ \\
 & \Big\lbrace \big[ \gamma_b^2 + (U + 2 \Delta_b)^2 \big]
 \big[ \mathcal{F}^2\big(\gamma_b^2 + 4 \Delta_b^2\big) 
 + 4 \gamma_b \mathcal{F}  \big( \gamma_b \cos \phi - 2 \Delta_b \sin \phi \big) 
 + 4 \gamma_b^2\big]^2  \Big\rbrace,
\end{split}
\end{equation}
\end{widetext}
where we have used $\tilde{\Gamma}_b^2 = \gamma_b^2 + 4
\Delta_b^2$. Here as well, we can compare the enrichment brought by
the interfering laser by comparing
Eqs.~(\ref{eq:jueene17201516GMT2019})
and~(\ref{eq:jueene17201652GMT2019}) for populations and
Eqs.~(\ref{eq:jueene17201437GMT2019}) and
(\ref{eq:jueene17201723GMT2019}) for second-order correlations, with
and without the interference, respectively. In this case, higher-order
correlators could also be given in closed-form but is too awkward to
be written here. The cases $g_s^{(k)}$ for $2\le k\le 4$ are shown in
Fig.~\ref{fig:ao3} as a function of the parameters of the interfering
laser. By comparing this to Fig.~\ref{fig:WedNov29150123CET2017} for
the 2LS, one can see that the anharmonic system is significantly more
complex, with resonances for the correlations that occur for specific
conditions of the phase for each~$\mathcal{F}$ that leads to
unconventional forms of antibunching or superbunching, rather than to
be simply out-of-phase previously. This makes salient the punctual
character of the unconventional mechanism: each strong correlation at
any given order must be realized in a very particular way: the one
that matches the corresponding interference.

The maximum bunching (UB) accessible with the interfering laser is
reached when the coherent-fraction population goes to zero (1-photon
suppression) for which the conditions on the phase and amplitude read
\begin{equation}
\label{}
\tan \phi_1 = - \frac{2\Delta_b}{\gamma_b} \quad \quad
\mathrm{and} \quad\quad \mathcal{F}_1 = -2  \cos \phi_1\,.
\end{equation}
Those conditions are exactly the same as Eq.~\eqref{eq:2LSconds} for
$N = 1$. Analogous conditions for the multi-photon cases can be found
solving $\g{N}_s=0$. For the case $N=2$, we find four different
roots:
\begin{subequations}
\begin{align}
\mathcal{F}_{2,1/2} = & \frac{2i e^{i \phi} \gamma_b}{(U + 2 \Delta_b) + i \gamma_b}
\bigg\lbrace 1 \pm \sqrt{\frac{U}{(U + 2 \Delta_b) + i \gamma_b}} \bigg\rbrace \,, \\
\mathcal{F}_{2,3/4} = & \frac{2i e^{-i \phi} \gamma_b}{(U + 2 \Delta_b) - i \gamma_b}
\bigg\lbrace 1 \pm \sqrt{\frac{U}{(U + 2 \Delta_b) - i \gamma_b}} \bigg\rbrace^{-1} \,.
\end{align}
\end{subequations}
Since these should be, by definition, real, this imposes another
constrain on~$\phi$. Although the expression for real-valued~$\phi$ to
make~$\mathcal{F}$ real cannot be given in closed form, they are
readily found numerically. It is possible to get four real solutions,
that are however degenerate. There are only two different conditions
for $\phi$ since the real part is the same for each pair of roots,
i.e., $\Re(\mathcal{F}_{2,1}) = \Re(\mathcal{F}_{2,4})$ and
$\Re(\mathcal{F}_{2,2}) = \Re(\mathcal{F}_{2,3}))$.  This yields two
physical solutions. For instance, for $U = \gamma_b$ and
$\Delta_b = \Delta_{-}$ (the case shown in Fig.~\ref{fig:ao3}),
$\g{2}_s$ vanishes at $\mathcal{F}_{2,1} \approx 0.615$ and
$\mathcal{\phi}_{2,1} \approx 0.659 \, \pi$ for one solution and at
$\mathcal{F}_{2,2} \approx 2.907 $ and
$\mathcal{\phi}_{2,2} \approx 0.860 \, \pi$ for the other one. Similar
resonances in higher-order correlations could be found following the
same procedure.

Regarding the quantum state realized in the system, similar
conclusions can be drawn for the anharmonic oscillator than for the
two-level system (previous Section, cf.~Tables~\ref{tab:fluctuations}
and~\ref{tab:AOtable}). Specifically for this case, the system can be
described by a displaced squeezed thermal state, properly
parameterized, but to lowest-order in the driving and for the
population and the two-photon correlation only. Departures arise to
next-order in the pumping or to any-order for three-photon
correlations and higher. The main differences is that the anharmonic
oscillator case has to be worked out numerically, so the prefactors
are given by the solutions that optimize the antibunching, for the
system parameters indicated in the caption. The same result otherwise
holds that the Gaussian-state description is a low-driving
approximation valid for the population and two-photon statistics. The
same holds for the systems studied in the following Sections, although
this point will not be stressed anymore.

\section{Jaynes--Cummings blockade}
\label{sec:FriFeb23100104CET2018}

Now that we have considered the two-level system on the one hand
(Section~\ref{sec:FriFeb23095933CET2018}) and the bosonic mode on the
other hand (Section~\ref{sec:jueene10122434CET2019}), we turn to the
richer and intricate Physics that arises from their coupling. We will
show how the themes of the previous Sections allow us to unify in a
fairly concise picture the great variety of phenomena observed and/or
reported in isolation.

We thus consider the case where a cavity mode, with bosonic
annihilation operator~$a$ and frequency~$\omega_a$, is coupled to a
2LS, with operator~$\sigma$ and frequency~$\omega_\sigma$, with a
strength given by~$g$. Such a system is described by the
Jaynes--Cummings Hamiltonian~\cite{jaynes63a,shore93a},
\begin{multline}
  \label{eq:Thu31May103357CEST2018}
  H_\mathrm{jc} = \Delta_\sigma \ud{\sigma} \sigma + \Delta_a \ud{a} a +
  g \left(\ud{a} \sigma + \ud{\sigma} a \right) +{}\\
  {}+\Omega_a \left(e^{i \phi} \ud{a} + e^{- i \phi} a\right) +
  \Omega_\sigma \left(\ud{\sigma} + \sigma\right),
\end{multline}
where we also include both a cavity and a 2LS driving term by a laser
of frequency~$\omega_\mathrm{L}$, with respective
intensities~$\Omega_a$ and~$\Omega_\sigma$ and relative phase~$\phi$.
We assume $g$ and $\Omega_\sigma$ to be real numbers, without loss of
generality since the magnitudes of interest ($G_a^{(N)}$) are
independent of their phases. The relative phase~$\phi$ between dot and
cavity drivings is, on the other hand, important.  We also limit
ourselves in this text to the case where the frequencies of the dot
and cavity drivings~$\omega_\mathrm{L}$ are identical and the analysis
could be pushed further to the case where this limitation is
lifted. The dissipation is taken into account through the master
equation~$\partial_t \rho = i \left[ \rho ,H_\mathrm{jc} \right] +
(\gamma_\sigma/2) \mathcal{L}_{\sigma}
\rho+(\gamma_a/2)\mathcal{L}_a\rho$ where~$\gamma_a$ is the decay rate
of the cavity. We solve the steady state in the low-driving
regime,~i.e., when $\Omega_a\ll \gamma_a\,,\gamma_\sigma$, as
indicated in Appendix~\ref{app:2} and find the populations:
\begin{widetext}
  \begin{equation}
    \label{eq:Tue29May182623CEST2018}
    \av{n_{\substack{a\\\sigma}}} =  4 \, \frac{4 g^2 \Omega_{\substack{a\\\sigma}}^2
	+ \tilde{\Gamma}_{\substack{\sigma\\a}}^2 \Omega_{\substack{a\\\sigma}}^2  - 4 g \Omega_a \Omega_\sigma
	\left(\pm 2 \Delta_{\substack{a\\\sigma}}\cos \phi + \gamma_{\substack{\sigma\\a}} \sin \phi \right)}
{16 g^4 + 8 g^2 \left(\gamma_a \gamma_\sigma - 4 \Delta_a \Delta_\sigma\right) +
	\tilde{\Gamma}_a^2 \tilde{\Gamma}_\sigma^2	} \, ,
  \end{equation}
  with matching upper/lower indices (including~$\pm$) and with
  $\tilde{\Gamma}^2_i = \gamma_i^2 + 4 \Delta_i^2$ (for
  $i = a, \sigma$).
  Similarly, we find the two-photon coherence function from the
  cavity:
\begin{equation}
\label{eq:Tue29May184632CEST2018}
\begin{split}
\g{2}_a = & \Big\lbrace \Big[ 16 g^4 + 8 g^2 \left(\gamma_a \gamma_\sigma -4 \Delta_a
\Delta_\sigma \right) + \tilde{\Gamma}_a^2 \tilde{\Gamma}^2_\sigma \Big]
\Big[16 g^4 \left(1 + \chi^2 \right) + 8 g^2 
\big(2 \chi^2 \tilde{\Gamma}_{11}^2  + 4 \Delta_\sigma \tilde{\Delta}_{11}- \gamma_\sigma
\tilde{\gamma}_{11}\big) + \tilde{\Gamma}^2_{\sigma} \tilde{\Gamma}_{11}^2 -\\
& 16 g \chi \big( \Delta_\sigma \tilde{\Gamma}_{11}^2 + 4 g^2 \tilde{\Delta}_{11}
[1+\chi^2] \big) \cos \phi + 8 g^2 \chi^2
\big(4 g^2 - \gamma_\sigma \tilde{\gamma}_{11} + 4 \Delta_\sigma 
\tilde{\Delta}_{11}\big) \cos 2 \phi \, - \\
& 8 g \chi \big(\gamma_\sigma \tilde{\Gamma}_{11}^2 + 4 g^2 \tilde{\gamma}_{11} [\chi^2-1] \big)
\sin \phi + 16 g^2 \chi^2  \big(\gamma_a \Delta_\sigma + \gamma_\sigma \tilde{\Delta}_{12}\big)
\sin 2 \phi  \Big] \Big\rbrace \Big/ \\
& \Big\lbrace \Big[16 g^4 + 8 g^2 \Big(\gamma_a \tilde{\gamma}_{11}- 4
\Delta_a \tilde{\Delta}_{11}\Big) + \tilde{\Gamma}_a^2 \tilde{\Gamma}_{11}^2\Big] \Big[4 g^2 \chi^2 + \tilde{\Gamma}_{\sigma}^2 - 4 g \chi \big(2 \Delta_\sigma \cos \phi +
\gamma_\sigma \sin \phi\big)\Big]^2 \Big\rbrace ,
\end{split}
\end{equation}
\end{widetext}
where $\tilde{\Delta}_{ij} \equiv i \Delta_a + j \Delta_\sigma$,
$\tilde{\gamma}_{ij} = i \gamma_a + j \gamma_\sigma$,
$\tilde{\Gamma}_{ij}^2 \equiv \tilde{\gamma}_{ij}^2 + 4
\tilde{\Delta}^2_{ij}$ and $\chi=\Omega_\sigma/\Omega_a$ is the ratio
of excitation. The range of $\chi$ extends from 0 to $\infty$ so that
it is convenient to use the derived
quantity~$\tilde{\chi}=\frac{2}{\pi} \atan(\chi)$ which varies between
0 and 1.  Equation~(\ref{eq:Tue29May184632CEST2018}) is admittedly not
enlightening per se but it contains all the physics of conventional
and unconventional photon statistics that arises from self-homodyning,
including bunching and antibunching, for all the regimes of
operations. It is quite remarkable that so much physics of
dressed-state blockades and interferences can be packed-up so
concisely.

\begin{figure}
  \centering
  \includegraphics[width=0.8\linewidth]{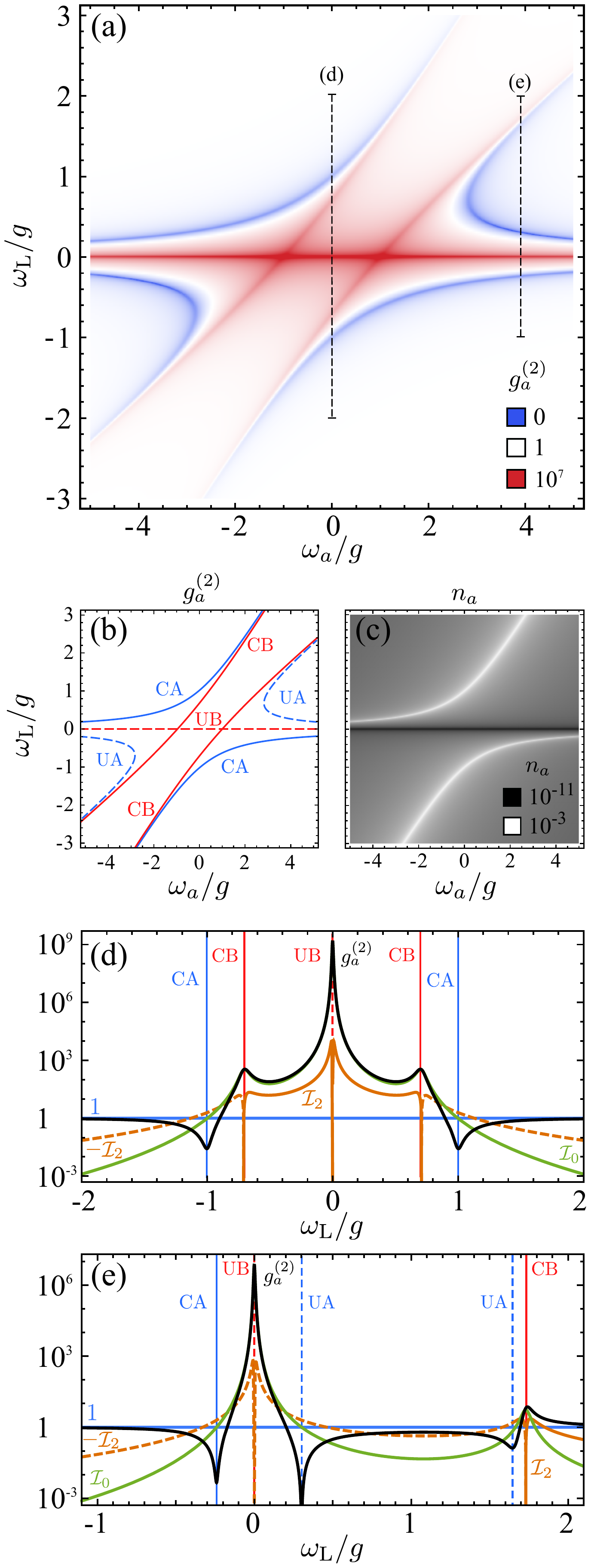}
  \caption{(Color online) Jaynes--Cummings model. (a) Photon
    statistics $g^{(2)}_a$ (log scale).  (b) Structure in terms of
    conventional C (solid) and unconventional U (dashed) features for
    B, bunching (blue) and A, antibunching (red).  CA (CB) is given by
    the resonant condition with the first (second) JC rung,
    cf.~Eqs.~(\ref{eq:eigenstates}).  UA is given by the interference
    condition Eq.~(\ref{eq:UAcondition}) and UB by
    $\omega_\mathrm{L}=0$. (c) Population~$n_a$, showing that i) only
    polariton emit strongly and ii) how the depletion (not an exact
    zero) at~$\omega_\mathrm{L}=0$ accounts for UB. Other features are
    not visible in the population only.  (d-e) $g^{(2)}_a$ (black
    line) for the two cuts shown in dashed lines in (a) with the
    cavity frequency either (d) resonant with the 2LS or (e) chosen to
    optimise UA ($\omega_a = 29.5\gamma_\sigma$).  The decomposition
    in~$\mathcal{I}_j$ components (from Appendix~\ref{app:4}) is also
    shown, with the sign of $\mathcal{I}_2$ changed when it is
    negative and plotted dashed.  Parameters: $\omega_\sigma=0$,
    $\gamma_a= 0.1 \, g$ and $\gamma_\sigma=0.01 \, g$.}
  \label{fig:JCantibunchingplat}
\end{figure}

We plot a particular case of this formula as a function of~$\omega_a$
and~$\omega_\mathrm{L}$ in Fig.~\ref{fig:JCantibunchingplat}(a),
namely, only driving the cavity ($\Omega_\sigma=\chi=0$).  Its reduced
expression, Eq.~(\ref{eq:JCg2}), is given in
Appendix~\ref{sec:Wed30May110352CEST2018}.  The general case is
available through an applet~\cite{wolfram_casalengua18a} and we will
shortly discuss other cases as well. The structure that is thus
revealed can be decomposed in two classes, as shown in panel~(b): the
conventional statistics that originates from the nonlinear properties
of the quantum levels, in solid lines, and the unconventional
statistics that originates from interferences, in dashed lines. Both
can give rise to bunching (in red) and antibunching (in blue). We now
discuss them in details.

\subsection{Conventional statistics}

Conventional features arise from the laser entering in resonance with
a dressed state of the dissipative JC
ladder~\cite{delvalle09a,laussy12e}, which energy is the real part of
\begin{multline}
  \label{eq:eigenstates}
  E^{(N)}_{\pm}=N\omega_{a}+\frac{\omega_\sigma-\omega_a}{2}-i
  \frac{(2N-1)\gamma_a+\gamma_\sigma}{4}\\\pm\sqrt{(\sqrt{N}g)^2+\left(
      \frac{\omega_a-\omega_\sigma}{2}-i\frac{\gamma_a-\gamma_\sigma}{4}\right)^2
  }\,.
\end{multline}
The first rung $E^{(1)}_{\pm}$ yields the CA lines in
Fig.~\ref{fig:JCantibunchingplat}(b). This corresponds to an increase
in the cavity population, as shown in
Fig.~\ref{fig:JCantibunchingplat}(c) as two white lines, corresponding
to the familiar lower and upper branches of strong coupling.  The
system effectively gets excited, but through its first rung only. The
second rung blocks further excitation according to the conventional
antibunching (CA), or photon-blockade, scenario, so that with the
increase of population goes a decrease of two-photon excitation,
leading to antibunching. This is in complete analogy with the CA that
appears in the Heitler regime of resonance fluorescence. This is not
an exact zero in $g^{(2)}_a$ in the low driving regime (the imaginary
part of the root does not vanish) because the conditions for perfect
interference are no longer met having a strongly coupled cavity with a
decay rate. We recently showed in Ref.~\cite{lopezcarreno18b}
that even in the vanishing coupling regime, $g\rightarrow 0$, when the
cavity acts as a mere detector of the 2LS emission, perfect
antibunching is spoiled due to the finite decay rate ($\gamma_a$
representing the precision in frequency detection). This is due to the
fact that the cavity is effectively filtering out some of the
incoherent fraction of the emission while the coherent fraction is
still fully collected. The interference condition in the $g^{(2)}_a$
decomposition, $1+\mathcal{I}_1=-\mathcal{I}_2=2$, is no longer
satisfied (see Fig.~2 of Ref.~\cite{lopezcarreno18b}).

On the other hand, driving resonantly the second rung,
$E^{(2)}_{\pm}$, leads to conventional bunching (CB), shown as red
solid lines in Fig.~\ref{fig:JCantibunchingplat}(b).  These quantum
features are well known and also found with incoherent driving of the
system in the spectrum of emission~\cite{laussy12e}, they are not
conditional to the coherence of the driving. This also corresponds to
an increase in the cavity population although this is not showing in
Fig.~\ref{fig:JCantibunchingplat}(c), where only first order effects
appear.

\subsection{Unconventional statistics}

We now turn to the other features in $g^{(2)}_a$ that do not
correspond to a resonant condition with a dressed state: these are,
first, a superbunching line at $\omega_\mathrm{L}=0$ (dashed red in
Fig.~\ref{fig:JCantibunchingplat}(b)) and second, two symmetric
antibunched lines (dashed blue).  All correspond to a self-homodyne
interference that the coherent field driving the cavity can produce on
its own, without the need of a second external laser.  In this case,
this also involves several modes (degrees of freedom) and more
parameters than in resonance fluorescence, so the phenomenology is
richer, but can be tracked down to the same physics.  We call them
again unconventional antibunching (UA) and unconventional bunching
(UB) in full analogy with the Heitler regime of resonance fluorescence
and in agreement with the literature that refers to particular cases
of this phenomenology as ``unconventional blockade''~\cite{bamba11a}
(the term of ``tunnelling'' has also been employed but the underlying
physical picture might be misleading~\cite{faraon08a}).

We first address antibunching (UA).  This is found by minimizing
$\g{2}_a$ in regions where there is no CA, which yields (for the
particular case $\chi = 0$):
\begin{equation}
  \label{eq:UAcondition}
  \Delta_a = -\Delta_\sigma \left(1 + \frac{4 g^2}{\gamma_\sigma^2 + 4 \Delta_\sigma^2}\right) \,,
\end{equation}
which is the analytical expression for the the dashed blue lines in
Fig.~\ref{fig:JCantibunchingplat}(b) (we remind that
$\Delta_i\equiv \omega_i-\omega_\mathrm{L}$ for~$i=a$, $\sigma$).  The
most general case when both the emitter and cavity are excited is
given in Appendix \ref{app:perfectantibunching}. In the minimization
process, we also find the condition for CA, due to the first-rung
resonance, but this can be disconnected from UA beyond the fact that
CA is already identified, because of UA also admits an exact zero,
which is found by either solving $\g{2}_a=0$ or setting to zero the
two-photon probability in the wavefunction
approximation~\cite{visser95a} (see
Appendix~\ref{sec:WedFeb28173330CET2018}). This gives the conditions
on the detunings as function of the system
parameters~\footnote{Eq.~(\ref{eq:FriMar2105208CET2018}) generalizes
  the condition given in Ref.~\cite{carmichael85a} for the resonant
  case and coincides with it with the
  notation~$\gamma_a /2 \rightarrow \kappa$
  and~$\mathscr{C} \rightarrow \mathscr{C}/2$).}:
\begin{subequations}
  \label{eq:FriMar2105208CET2018}
  \begin{align}
    \label{eq:FriMar2105208CET2018a}
    \Delta_\sigma^2&=\frac{\gamma_\sigma^2}{4}\left(\frac{4g^2}{\gamma_\sigma(\gamma_{\sigma }+\gamma_a)}-1\right)\,,\\
    \label{eq:FriMar2105208CET2018b}
    \Delta_{a} &= - \bigg (2+ \frac{\gamma_a}{\gamma_\sigma} \bigg)
                 \Delta_\sigma\,.
  \end{align}
\end{subequations}
These conditions are met in Fig.~\ref{fig:JCantibunchingplat}(a) at
the lowest point where the blue UA line intersects the (e) cut (and on
the symmetric point $\omega_a < 0$). When the laser is at resonance
with the 2LS ($\Delta_\sigma=0$) and cavity losses are large
($\gamma_a \gg \gamma_\sigma$), this occurs when the cooperativity
$\mathscr{C}\equiv \frac{4 g^2}{\gamma_a \gamma_\sigma}=1$.  This type
of UA interference is second-order, so it is not apparent in the
cavity population at low driving,
Fig.~\ref{fig:JCantibunchingplat}(b).  One has to turn to two-photon
correlations instead. Note also that UA requires a cavity-emitter
detuning that is of the order of~$g$.

\begin{figure*}[t]
	\centering
	\includegraphics[width=\linewidth]{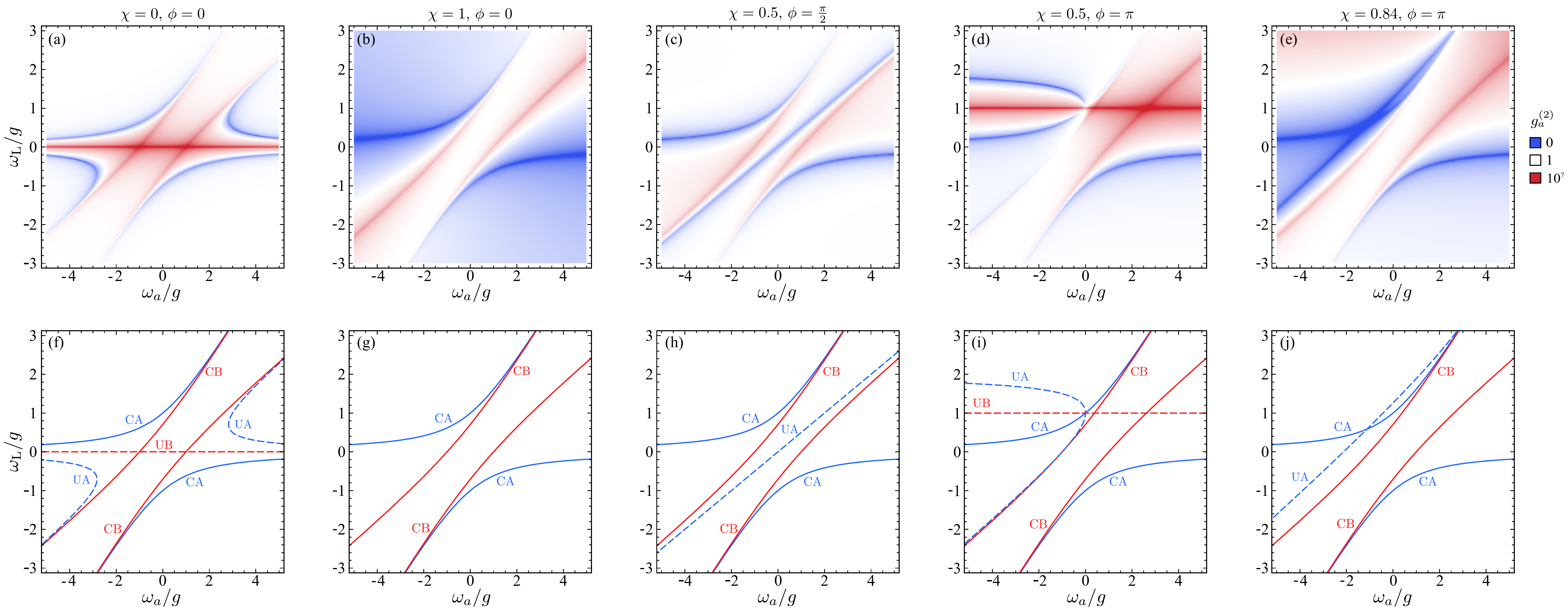}
	\caption{(Color online)~Effect of the type of driving on the
          two-photon statistics, for the Jaynes--Cumming model as
          measured in its cavity emission. The upper row shows
          Eq.~(\ref{eq:Tue29May184632CEST2018}) for the parameters
          indicated in each panel and the lower row identifies the
          various features through the structure of conventional
          (Eq.~(\ref{eq:eigenstates}c), solid) and unconventional
          (Eq.~(\ref{eq:UAcondition}), dashed) lines. Chosen parameters:
      $\gamma_a = 0.1 \, g$, $\gamma_\sigma = 0.01 \, g$.}
	\label{fig:Wed30May111214CEST2018}
\end{figure*}

Since this is an interference effect, we perform the same
decomposition of $\g{2}_a$ in terms of coherent and incoherent
fractions, as in previous Sections, given by
Eq.~(\ref{eq:g2mixdecomposition}), and show the terms that are not
zero in Fig.~\ref{fig:JCantibunchingplat}(d-e).  The full expressions
are in Appendix~\ref{app:4}. The term $\mathcal{I}_1$ is exactly zero
to lowest order in the driving and only the
fluctuation-statistics~$\mathcal{I}_0$ and the two-photon
interference~$\mathcal{I}_2$ play a role, like in the Heitler regime
of resonance fluorescence. We observe that, in this decomposition,
there is no difference between the CA and UA, since both occur
approximately when the statistics of the laser and fluctuations,
$1+\mathcal{I}_0=2$, are compensated by their two-photon interference,
$\mathcal{I}_2=-2$, again as in the Heitler regime. The fundamental
differences between these two types of antibunching will be discussed
later on. Before that, we discuss the last feature: the unconventional
bunching at~$\omega_\mathrm{L}=0$.

The reason for the super-bunching peak labelled as UB in
Fig.~\ref{fig:JCantibunchingplat}(b) is also the same as in resonance
fluorescence: the cancellation of the coherent part, in this case, of
the cavity emission, and the consequent dominance of the fluctuations
only, which are super-Poissonian in this region. Therefore, contrary
to the CB, this superbunched statistics is not directly linked to an
enhanced $N$-photon (for any~$N$) emission and it does not appear one
could harvest or Purcell-enhance it, for instance, by coupling the
system to an auxiliary resonant cavity. Since it is pretty much wildly
fluctuating noise, the actual prospects of multi-photon physics in
this context remains to be investigated. In any case, the conditions
that yield the super-Poissonian correlations can thus be obtained by
minimising the cavity population $\mean{n_a}$ or, from the
wavefunction approximation detailed in
Appendix~(\ref{sec:WedFeb28173330CET2018}), by minimising the
probability to have one photon, given by
Eq.~(\ref{eq:WedFeb28145309CET2018}), which coincide with the coherent
fraction to lowest order in $\Omega_a$. One cannot achieve an exact
zero in this case but the cavity population is clearly undergoing a
destructive interference, as shown by the black horizontal line in
Fig.~\ref{fig:JCantibunchingplat}(c). The resulting condition links
the laser frequency with the 2LS one:
\begin{equation}
  \label{eq:FriFeb23173509CET2018}
\Delta_\sigma = \chi g \cos \phi	 \,,
\end{equation}
which reduces to simply $\Delta_\sigma=0$ (laser in resonance with the
2LS) if i) the dot and cavity are driven with a $\pi/2$-phase
difference or ii) the laser drives the cavity only ($\chi = 0)$.

So far, we have focused on the particular case of
Eq.~(\ref{eq:Tue29May184632CEST2018}) where~$\Omega_\sigma=0$ (i.e.,
Eq.~(\ref{eq:JCg2})). This is the case dominantly studied in the
literature and the one assumed to best reflect the experimental
situation. It is also for our purpose a good choice to clarify the
phenomenology that is taking place and how various types of statistics
cohabit. It must be emphasised, however, that while the physics is
essentially the same in the more general configuration, the results
are, even qualitatively, significantly different in configurations
where the two types of pumping are present. This is shown in
Fig.~\ref{fig:Wed30May111214CEST2018}. While conventional features are
stable, being pinned to the level structure, the unconventional ones that
are due to interferences are very sensible to the excitation
conditions and get displaced or, in the case of QD excitation only,
even completely suppressed. If one is to regard conventional features
as more desirable for applications, this figure is therefore again an
exhortation at focusing on the QD excitation configuration.

\begin{figure}[th]
  \centering
  \includegraphics[width=.9\linewidth]{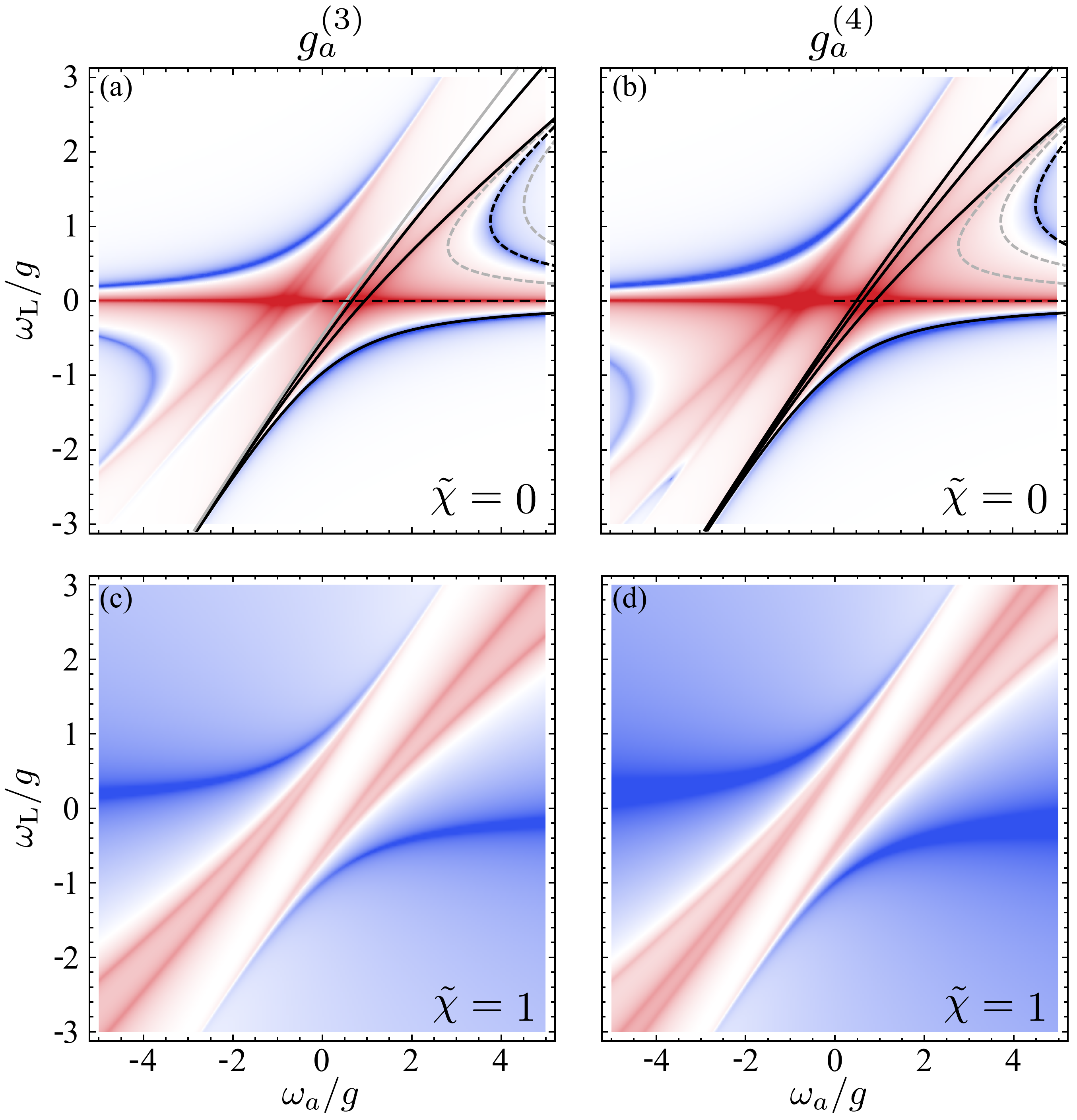}
  \caption{(Color online). Higher-order photon statistics, at
    (left column) three- and (right) four-photon level. Top row is for
    cavity excitation and bottom row for 2LS excitation. In the top
    row, we have superimposed the right-half of the conventional
    (solid) and unconventional (dashed) features, putting them in grey
    when not present for a given order of the correlations. The
    conventional features grow in numbers and stay pinned at the same
    positions while the unconventional ones remain in the same number
    but at different positions.  Parameters:
    $\gamma_a = 0.1 \, g$, $\gamma_\sigma = 0.001 \, g$.}
  \label{fig:Tue29May164357CEST2018}
\end{figure}

While we have focused on the two-photon statistics, both the
conventional and unconventional effects occur at the~$N$-photon level,
in which case they manifest through higher-order coherence
functions~$g^{(N)}$, and their~$N$th-order behaviour is one of the key
differences between conventional and unconventional statistics.
Regarding conventional features, resonances happen at the $N$-photon
level when $N$ photons of the laser have the same energy than one of
the dressed states (and only one, thanks to the JC nonlinearities):
$\omega_\mathrm{L}=\mathrm{Re}\{E^{(N)}_{\pm}\}/N$. The blockade that
is realised is a real blockade in the sense that all the correlation
functions are then depleted simultaneously. In
Fig.~\ref{fig:Tue29May164357CEST2018}, the counterpart of
Fig.~\ref{fig:JCantibunchingplat}(a) is shown for~$\g{3}_a$
and~$\g{4}_a$ and shows how more conventional features appear with
increasing~$N$ but otherwise stay pinned to the same conditions, while
the number of unconventional features stays the same, but their
positions drift with~$N$, so that one cannot simultaneously realise
$g^{(N)}_a<1$ for all~$N$. This is an important difference between a
convex mixture of Gaussian states, which is a semi-classical state,
and a state beyond this class, which is genuinely quantum, as
previously mentioned. The latter requires the ability to imprint
strong correlations at several and possibly all photon-orders. This
suggests that CA could be more suited than UA for quantum
applications. Note how with the 2LS direct excitation, shown in the
second row of Fig.~\ref{fig:Tue29May164357CEST2018}, one only finds
conventional statistics, with magnified features such as broader
antibunching in the photon-like branch and narrower one in the
exciton-like branch.  The $N$-photon resonances are neatly separated
for large-enough detuning, which is the underlying principle to
harness rich $N$-photon resources~\cite{sanchezmunoz14a}.

\begin{figure}
  \centering
  \includegraphics[width=0.7\linewidth]{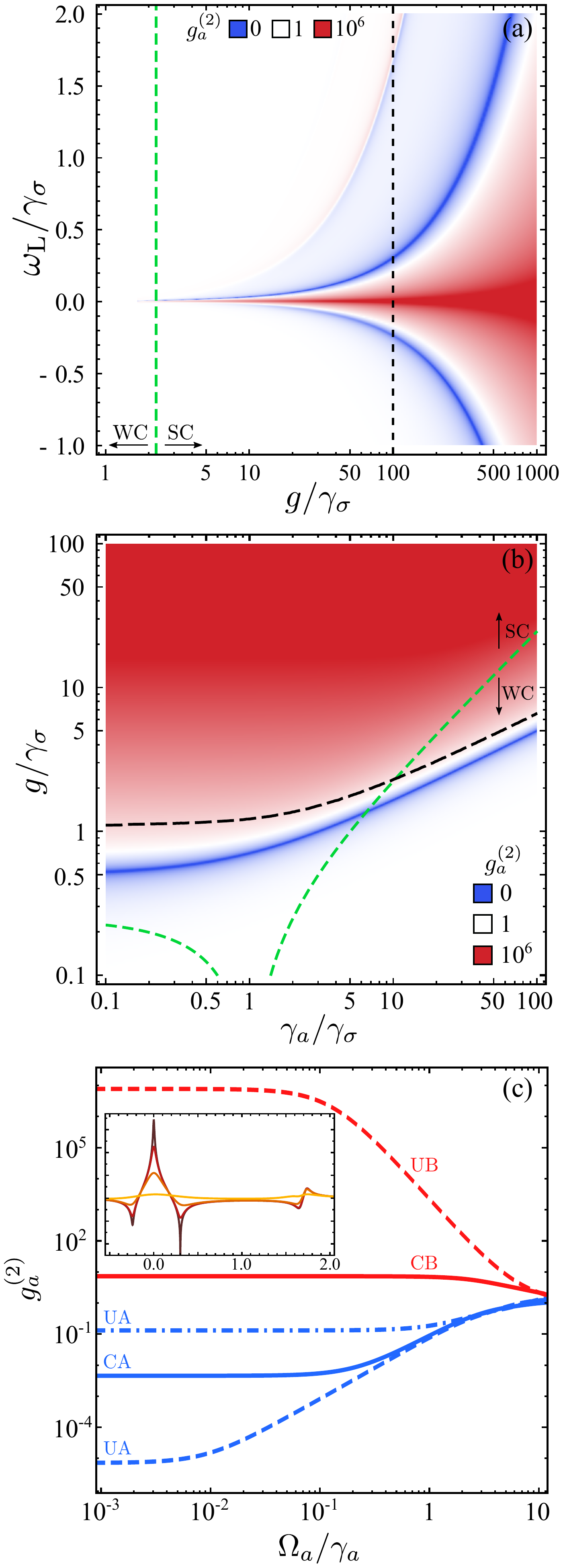}
  \caption{(Color online) Transition to weak coupling and
    non-vanishing pumping. (a) Evolution of $g^{(2)}_a$ along the cut
    (e) in Fig.~\ref{fig:JCantibunchingplat} as a function of the
    coupling strength~$g$, showing how the structure collapses on the
    bare modes.  CA (lowest blue line) is the first feature to vanish
    with decreasing~$g$, as the underlying dressed states disappear.
    (b) Behaviour of the ``center'' point
    at~$\omega_\mathrm{L}=\omega_a=0$, that typically features the UB
    peak, although not exclusively as it can also exhibit
    antibunching, even in strong-coupling (Black dashed line indicates
    when the behaviour of $\g{2}_a$ changes from local minimum to
    maximum).  (c) Effect of increasing pumping, computing the
    features shown in Fig.~\ref{fig:JCantibunchingplat}(e) (vanishing
    pumping) as a function of (finite) driving $\Omega_a$.  For high
    pumping, $\Omega_a \gtrsim \gamma_a$, every feature (both bunching
    and antibunching) is spoiled and eventually disappears
    ($\g{2}_a \rightarrow 1$). The inset figure compares the same cut
    of $\g{2}_a$ for the cases of vanishing driving
    ($\Omega_a \rightarrow 0$) and finite driving
    ($\Omega_a / \gamma_a = 0.5, 2 \ \text{and} \ 10$).  Lighter
    colors correpond to higher driving amplitudes.  }
  \label{fig:JCsuper}
\end{figure}

We now turn to another noteworthy regime, out of the many
configurations of interest that are covered by
Eq.~(\ref{eq:Tue29May184632CEST2018}), namely, the transition from
weak to strong coupling. The so-called strong-coupling, when
$g>|\gamma_a-\gamma_\sigma|/4$, is one of the coveted attributes of
light-matter interactions, leading to the emergence of dressed states
and to a new realm of physics. It is also, however, an ill-defined
concept in presence of detuning~\cite{laussy12e} and one would still
find the dressed-state structure of
Fig.~\ref{fig:Tue29May164357CEST2018} in the largely detuned regime
when driving the 2LS, even up to large
photon-order~\cite{kavokin_book17a}.  The restructuration of the
statistics when crossing over to the weak-coupling regime is explored
in Fig.~\ref{fig:JCsuper}(a), where we track the impact on $g^{(2)}_a$
of changing the coupling~$g$, on the cut in
Fig.~\ref{fig:JCantibunchingplat}(e) that intersects from top to
bottom CB, UA (twice), UB and CA. One can see how the features
converge as the coupling is reduced, with the conventional ones
disappearing first, which is expected from the disappearance of the
underlying dressed states, that are responsible for the conventional
effects. The unconventional antibunching, on the other hand, is more
robust and can be tracked well into weak coupling where all effects
ultimately vanish at the same time as they merge. Conventional
antibunching is the most robust feature, as can be seen by tracking
for instance the UB peak at the point where it is the most isolated
from the other features, namely, at resonance
where~$\omega_\mathrm{L}=\omega_a=\omega_\sigma=0$. Spanning over the
two main parameters that control strong coupling, the coupling
strength~$g$ (in units of~$\gamma_\sigma$) and the rates of
dissipation rates~$\gamma_a/\gamma_\sigma$, one sees that the strong
bunching is not always sustained but can be instead overtaken by
conventional antibunching, which is the well-defined blue line in the
figure (given by Eq.~(\ref{eq:UAcondition})). The region where the UB
peak is well-defined can be identified by inspecting the second
derivative of $\g{2}_a$ as a function of the laser frequency,
$\partial^2_{\omega_\mathrm{L}}\g{2}_a$ at $\omega_{\mathrm{L}} = 0$
and is shown in Fig.~\ref{fig:JCsuper}(b) as a dashed black line.  The
white line that separates the antibunching region from the bunching
one corresponds to the critical coupling strength~$g_P$ between the
cavity and the 2LS that leads to $\g{2}_a=1$ (its expression is given
in the Appendix, Eq.~(\ref{eq:FriMar2104433CET2018})). The strong-weak
coupling frontier $g/\gamma_\sigma<|\gamma_a/\gamma_\sigma-1|/4$ is
indicated with a dotted green line as a reference, illustrating again
the lack of close connection between strong-coupling and the
photon-statistics features.

We conclude our discussion of the Jaynes-Cummings system with the
second main difference between conventional and unconventional
statistics, namely their resilience to higher driving. All our results
are exact in the limit of vanishing driving, that is to say, in the
approximation of neglecting $\Omega$ terms of higher-orders than the
smallest contributing one. For non-vanishing driving, numerically
exact results can be obtained instead (and can be made to agree with
arbitrary precision to the analytical expressions, as long as the
driving is taken low enough, what we have consistently checked). A
characteristic of the unconventional features is that, being due to an
interference effect for a given photon number only, it is fragile to
driving, unlike the conventional features which display more
robustness. This is illustrated in Fig.~\ref{fig:JCsuper}(c) for the
case of cavity driving~$\Omega_a$, where we compare the analytical
result from Eq.~(\ref{eq:Tue29May184632CEST2018}) or, in this case,
Eq.~(\ref{eq:JCg2}), in black, with the numerical solution
for~$\Omega_a=0.25\gamma_a$, so still fairly small. One can see how
the conventional features are qualitatively preserved and
quantitatively similar to the analytical result, while the
unconventional antibunching has been completely washed out.

One could consider still other aspects of the Physics embedded in
Eq.~(\ref{eq:Tue29May184632CEST2018}). We invite the inquisitive
and/or interested readers to explore them through the
applet~\cite{wolfram_casalengua18a} which is helpful to get a sense of
the complexity of the problem. Instead of discussing these further, we
now turn to another platform of interest that bears many similarities
with the Jaynes--Cummings results.

\section{Microcavity-polariton blockade}
\label{sec:FriFeb23100219CET2018}

Microcavity polaritons~\cite{kavokin_book17a} arise from the
strong coupling between a planar cavity photon and a quantum well
exciton, both of which are bosonic fields with annihilation
operators~$a$ and~$b$, respectively. These fields are coupled with
strength~$g$ and have frequencies~$\omega_a$ and~$\omega_b$. Moreover,
the excitons, being electron-hole pairs, have Coulomb interactions
that we parametrise as~$U/2$.  Thus, the Hamiltonian describing the
polariton system is given by
\begin{multline}
  \label{eq:FriMar2182347CET2018}
  H_\mathrm{pol} = \Delta_a \ud{a} a + \Delta_{b} \ud{b} b + 
  g \left(\ud{a} b + \ud{b} a \right)
   +{}\\
  {}  \Omega_a \left(e^{i \phi}\ud{a} + e^{-i \phi} a\right) + \Omega_b (\ud{b} + b)
  + \frac{U}{2} \ud{b} \ud{b} b b\,,
\end{multline}
where~$\Delta_{a,b} = \omega_{a,b} - \omega_{\mathrm{L}}$ are the
frequencies of cavity/exciton referred to $\omega_\mathrm{L}$, which
is the frequency of the laser that drives the photonic/excitonic field
with amplitudes~$\Omega_{a,b}$.  The phase difference between them is
represented by $\phi = \phi_a - \phi_b$ and since the absolute phase
can be chosen freely, $\phi_{a,b}$ are fixed to be $\phi$ and $0$.
The dissipative dynamics of the polaritons is given by a master
equation
$\partial_t \rho = i \left[ \rho ,H_\mathrm{pol} \right] +
(\gamma_b/2) \mathcal{L}_{b} \rho+ (\gamma_a/2) \mathcal{L}_{a} \rho$,
where~$\gamma_a$ and~$\gamma_b$ are the decay rates of the photon and
the exciton, respectively.  As compared to the Jaynes--Cummings
Hamiltonian~(\ref{eq:Thu31May103357CEST2018}), the polariton
Hamiltonian substitutes the 2LS by a weakly interacting Bosonic mode,
$b\rightarrow\sigma$ with nonlinearities~$\ud{b}\ud{b}bb$ thus
slightly displacing the state with two excitations while the 2LS
forbids it entirely. In the case where~$U\rightarrow\infty$, the
Jaynes--Cummings limit is recovered, but in most experimental cases,
$U/\gamma_a$ is very small.  In all cases, in the low driving regime
($\Omega_a \rightarrow 0$) the steady-state populations of the photon
and the exciton are given by the same expressions as in the
Jaynes--Cummings model, Eq.~(\ref{eq:Tue29May182623CEST2018})
with~$\sigma\rightarrow b$, since the 2LS converges to a bosonic field
in the linear regime. The differences arise in the two-particle
magnitudes (cf.~Eq.~(\ref{eq:Tue29May184632CEST2018})):
%
%
\begin{widetext}
  \begin{subequations}
\label{eq:Thu31May102314CEST2018}  	
\begin{equation}
\begin{split}
\g{2}_a = & \Big\lbrace \Big[16 g^4 + 8 g^2 \big(\gamma_a \gamma_b - 4 \Delta_a
\Delta_b \big) + \tilde{\Gamma}_a^2 \tilde{\Gamma}_b^2 \Big]
\Big[\tilde{\Gamma}_b^2 \tilde{\Gamma}_{11}^2 \big(\gamma_b^2 + \tilde{U}_{12}^2\big)
+ 8 g^2 \Big(U^2 [4 \Delta_b \tilde{\Delta}_{11}- \gamma_b \tilde{\gamma}_{11}]+ 2 \tilde{\Gamma}_{11}^2 [\gamma_b^2 + \tilde{U}_{12}^2] \chi^2 +\\
& 8 U \Delta_b^2 \tilde{\Delta}_{11} -2 U \gamma_b^2 \tilde{\Delta}_{13} - 4 U \gamma_a \gamma_b
\Delta_b \Big) + 16 g^4 \Big(U^2 + [\tilde{\gamma}_{11}^2 + (U + 2 \tilde{\Delta}_{11})^2] \chi^4\Big) - 16 g \chi \cos \phi \, \Big( \Delta_b \tilde{\Gamma}_{11}^2 [\gamma_b^2 +4 
\tilde{U}_{12}^2] \\ 
& + 2 g^2 [U(2 \tilde{\Delta}_{11} \tilde{U}_{12}- \gamma_b \tilde{\gamma}_{11})
+ (2 U^2 \tilde{\Delta}_{11}+ 2 \Delta_b \tilde{\Gamma}_{11}^2+ U \lbrace \gamma_a
\tilde{\gamma}_{11} + 4 \tilde{\Delta}_{11} \tilde{\Delta}_{12} \rbrace ) \chi^2 ]\Big) +
8 g^2 \chi^2 \cos 2 \phi \, 
\Big(4 g^2 U [U + 2 \tilde{\Delta}_{11}] -  \\
& U^2 [\gamma_b \tilde{\gamma}_{11}- 4 \Delta_b \tilde{\Delta}_{11}] - [\gamma_b^2 - 4 
\Delta_b^2] \tilde{\Gamma}_{11}^2 + 2 U [\gamma_a^2 \Delta_b + \tilde{\Delta}_{12}
(4 \Delta_b \tilde{\Delta}_{11} - \gamma_b^2)] \Big) -8 g \chi \sin \phi \, 
\Big( \gamma_b \tilde{\Gamma}_{11}^2 [\gamma_b^2 + \tilde{U}_{12}^2] + \\
& 4 g^2 [\gamma_b \tilde{\Gamma}_{11}^2 \chi^2 + U (\chi^2-1) (U \tilde{\gamma}_{11}  + 2 
\gamma_b \Delta_a +2 \tilde{\gamma}_{12} \Delta_b)] \Big) + 8 g^2 \chi^2 \sin 2 \phi \,
\Big( -4 g^2 U \tilde{\gamma}_{11} + 4 \gamma_b \Delta_b \tilde{\Gamma}_{11}^2 + 2 U^2
[\gamma_a \Delta_b + \gamma_b \tilde{\Delta}_{12}] + \\
& U [\gamma_a^2 \gamma_b + 4 \gamma_b \tilde{\Delta}_{12}^2 + \gamma_a \tilde{\Gamma}_b^2] \Big) \Big]
\Big\rbrace \Big/ \\
&  \Big\lbrace \Big( \tilde{\Gamma}_a^2 \tilde{\Gamma}_{11}^2 \big[\gamma_b^2 + \tilde{U}_{12}^2 \big] + 16 g^4 \big[\tilde{\gamma}_{11}^2 + \big(U + 2 
\tilde{\Delta}_{11}\big)^2\big] + 8 g^2 \big[U^2 \big(\gamma_a \tilde{\gamma}_{11} -
4 \Delta_a \tilde{\Delta}_{11}\big) + \tilde{\Gamma}_{11}^2 \big(\gamma_a \gamma_b -
4 \Delta_a \Delta_b\big) - \\
& 2U \big(\gamma_a^2 \tilde{\Delta}_{1 \bar{1}} - 2 \gamma_a \gamma_b \Delta_b
+ 4 \Delta_a \tilde{\Delta}_{11} \tilde{\Delta}_{12}\big) \big] \Big)
\Big(4 g^2 \chi^2 + \tilde{\Gamma}_b^2  - 4 g \chi
\big[2 \Delta_b \cos \phi + \gamma_b \sin \phi \big]\Big)^2 \Big\rbrace
\, , 
\end{split}
\end{equation}
\begin{equation}
\begin{split}
\g{2}_b = & \big\lbrace \tilde{\Gamma}_{11}^2  \big[16 g^4 + 8 g^2 \big( \gamma_a 
\gamma_b - 4 \Delta_a \Delta_b \big) + \tilde{\Gamma}_a^2 \tilde{\Gamma}_b^2\big] \big\rbrace  
\Big/ \\
& \big\lbrace \tilde{\Gamma}_a^2 \tilde{\Gamma}_{11}^2 \big[\gamma_b^2 +\tilde{U}_{12}^2 \big] + 16 g^4 \big[\tilde{\gamma}_{11}^2 + \big(U + 2 
\tilde{\Delta}_{11}\big)^2\big] + 8 g^2 \big[U^2 \big(\gamma_a \tilde{\gamma}_{11} -
4 \Delta_a \tilde{\Delta}_{11}\big) + \tilde{\Gamma}_{11}^2 \big(\gamma_a \gamma_b -
4 \Delta_a \Delta_b\big) - \\
& 2U \big(\gamma_a^2 \tilde{\Delta}_{1 \bar{1}} - 2 \gamma_a \gamma_b \Delta_b
+ 4 \Delta_a \tilde{\Delta}_{11} \tilde{\Delta}_{12}\big) \big] \big\rbrace
\, ,
\end{split}
\end{equation}
  \end{subequations}
where we have used the short-hand
notation~$\gamma_{+}=\gamma_a+\gamma_b$, $\Delta_{\pm}=\Delta_a \pm
\Delta_b$ and~$\Gamma_{c}^2 = \gamma_c^2 + 4\Delta_c^2$ for
$c=a,b,+$ as well as $\tilde{\Delta}_{ij} \equiv i \Delta_a + j
\Delta_b$, $\tilde{\gamma}_{ij} = i \gamma_a + j \gamma_\sigma$,
$\tilde{\Gamma}_{ij}^2 \equiv \tilde{\gamma}_{ij}^2 + 4 \tilde{\Delta}^2_{ij}$,
$\tilde{U}_{ij} = i U + j \Delta_b$
and $\bar{j}$ denotes negative integer values~($\bar{j} = -j$).
\end{widetext}
Note that a major conceptual difference with the Jaynes--Cummings
model is that it now becomes relevant to consider the emitter (in this
case, excitonic) two-photon coherence function, $g^{(2)}_b$, while in
the Jaynes--Cummings case one has the trivial
result~$g^{(2)}_\sigma\equiv0$. The exciton statistics enjoys
noteworthy characteristics, as we shall shortly see.

\begin{figure}
  \centering
\includegraphics[width=0.8\linewidth]{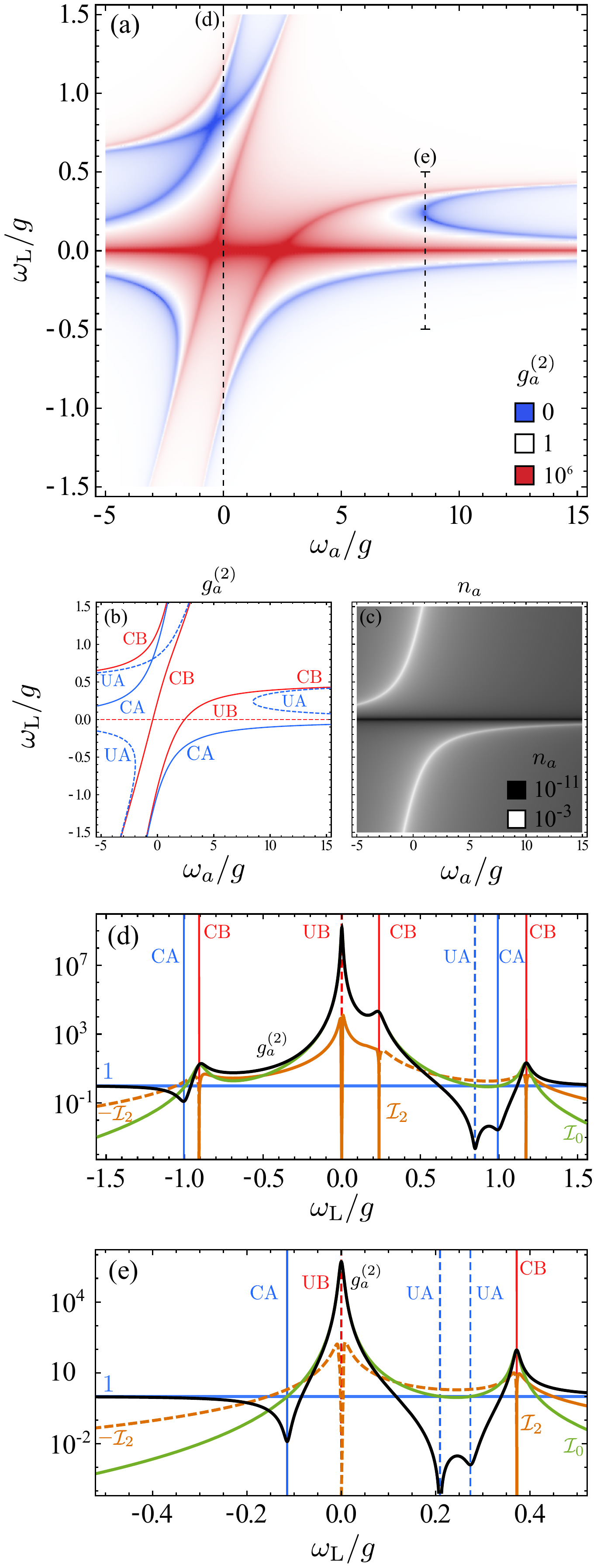}
\caption{(Color online) Polaritonic counterpart of
  Fig.~\ref{fig:JCantibunchingplat}. (a)~$g^{(2)}_a$ and (b) its
  structure in terms of conventional and unconventional features. CA
  (CB) is given by the resonant condition with the first (second)
  polariton manifold, cf.~Eqs.~\eqref{eq:eigenstates} (for $N = 1$) and
  \eqref{eq:pb2ndman}.  UA is given by the interference
  condition Eq.~\eqref{eq:UAcurve} and UB by $\omega_\mathrm{L}=0$. (c)
  Population~$n_a$ and (d-e) $g^{(2)}_a$ (black line) for the two cuts
  shown in dashed lines in (a) with the cavity frequency either (d)
  resonant with the 2LS or (e) chosen to optimise UA
  ($\omega_a = 8.63 \, g$).  The decomposition in~$\mathcal{I}_j$ components
  is also shown, with the same conventions as in
  Fig.~\ref{fig:JCantibunchingplat}.  Parameters:
  $\gamma_a= 0.1 \, g$ , $\gamma_b = 0.01 \,g$ and $U= 10 \gamma_a$.}
\label{fig:08}
\end{figure}

We repeat in Fig.~\ref{fig:08} the same plots for the polariton system
as for the Jaynes--Cummings case
(Fig.~\ref{fig:JCantibunchingplat}). The
applet~\cite{wolfram_casalengua18a} also covers this more general
case. The cavity population is exactly the same, as already mentioned,
and all other panels bear clear analogies. The two-photon coherence
function converges to the Jaynes--Cummings one in the infinite
interation limit ($\lim_{U\rightarrow\infty}g^{(2)}_a$) but is
distinctly distorted for high-energy laser driving in the positive
photon-exciton detuning region, and features an additional UA and CB
couple of lines in the negative detuning region. The decomposition of
$\g{2}_a$ as in Eq.~(\ref{eq:g2mixdecomposition}) can be made (the
expressions are however bulky and relegated to Appendix~\ref{app:4})
and are shown in Fig.~\ref{fig:08}(d) and~(e).

\subsection{Conventional statistics}

Like in the Jaynes-Cummings model, one can identify the conventional
antibunching (CA) and bunching (CB) by mapping the observed features
to an underlying blockade mechanism, namely, the positions at which
$N$-photon excitation occurs, which is when the laser is resonant with
one of the states in the $N$-photon rung. The first rung that provides
CA is given by the same Eq.~(\ref{eq:eigenstates}), with $N=1$, since
this corresponds to the linear regime where both systems
converge. One finds, therefore, that the two CA blue lines in
Fig.~\ref{fig:08}(a), marked in solid blue in~(b),
are the same as in the Jaynes--Cummings model. They coincide as well
with the white regions in Fig.~\ref{fig:08}(c) where
the cavity emission is enhanced. This is the standard one-photon
resonance, with a blockade of photons into higher rungs due to the
non-linearity now introduced by the interactions (instead of the 2LS).

Higher rungs are different from the Jaynes--Cummings model, but their
effects otherwise follow from the same principle and they are
similarly obtained by diagonalizing the effective Hamiltonian in the
corresponding $N$-excitation Hilbert subspace, that is, in the basis
$\{\ket{N,0},\, \ket{N-1,1},\,\hdots \ket{0,N} \}$ (where each state
is characterised by the photon and exciton number). At the two-photon
level, one is interested in the second rung, which contains three
eigenstates. The expressions for the general eigenenergies are rather
large but we can provide here the first order in the interactions~$U$
in the strong coupling limit~($g\gg \gamma_a$, $\gamma_b$):
\begin{subequations}
\label{eq:pb2ndman} 
  \begin{align}
  E_0^{(2)}&=\omega_a + \omega_b +\frac{g^2}{2 R^2}U\,,\\
  E^{(2)}_{\pm}&=\omega_a + \omega_b \pm 2R +\frac{2g^2+(\omega_a - \omega_b)[(\omega_a - \omega_b)\mp 2R]}{8 R^2}U\, ,
  \end{align}
   \label{eq:eigenstatesPol}
\end{subequations}
with~$R=\sqrt{g^2+(\omega_a - \omega_b)^2/4}$ the normal mode
splitting typical of strong coupling. In this limit,
$E^{(1)}_{\pm}=(\omega_a + \omega_b)/2 \pm R $. The CB lines are
positioned, therefore, according to the conditions for two-photon
excitation by the laser: $\omega_\mathrm{L}=\Re{E_-^{(2)}}/2$,
$\Re{E_0^{(2)}}/2$, $\Re{E_+^{(2)}}/2$, in increasing order, as they
appear in Fig.~\ref{fig:08}(a), marked with solid red lines in
Fig.~\ref{fig:08}(b). The upper CB line, corresponding to $E_+^{(2)}$,
is the faintest one in the cavity emission due to the fact that it has
the most excitonic component. It is monotonically blueshifted with
increasing~$U$ and becomes linear in the density plot as
$E_+^{(2)}\rightarrow U$. The other two levels converge to those in
the Jaynes--Cummings model in such case:
$E_-^{(2)} \rightarrow -\sqrt{2}g$ and
$E_0^{(2)}\rightarrow \sqrt{2}g$.

\subsection{Unconventional statistics}

We now shift to the unconventional features in polariton
blockade. Superbunching, or UB, is found by minimization of
$\mean{n_a}$ and, therefore, also occurs for the same condition as the
Jaynes--Cummings model~$\Delta_b = \chi g \cos \phi$.  The maximum
superbunching is obtained at one of the crossings of UB and CB.
 
Now turning to the more interesting Unconventional Antibunching (UA)
features, they are found, in the polariton case as well, from the
minimization of $\g{2}_a$. Since the equations are quite bulky, only
the case of cavity excitation ($\Omega_b = 0$) is included here (the
full expression can be consulted at
App. \ref{app:perfectantibunching}). The UA curve is given by the
solution of
\begin{equation}
\label{eq:UAcurve}
\Delta_a = - \Delta_b - \frac{4 g^2 \Delta_b}{\gamma_b^2 + 4 \Delta_b^2} +
\frac{2 g^2 (U + 2 \Delta_b)}{\gamma_b^2 + (U+2 \Delta_b)^2},
\end{equation}
and the conditions for perfect antibunching come from solving the equation
\begin{equation}
  \gamma_b \left[1 + 4 g^2 \left(
- \frac{4 g^2 \Delta_b}{\gamma_b^2 + 4 \Delta_b^2} +
\frac{2 g^2 (U + 2 \Delta_b)}{\gamma_b^2 + (U+2 \Delta_b)^2}\right)\right] = - \gamma_a
\end{equation}
and subsequently imposing that every parameter must be real (or the
more restrictive case: real and positive) that lead to additional
restrictions.

\begin{figure}[th]
  \centering
  \includegraphics[width=\linewidth]{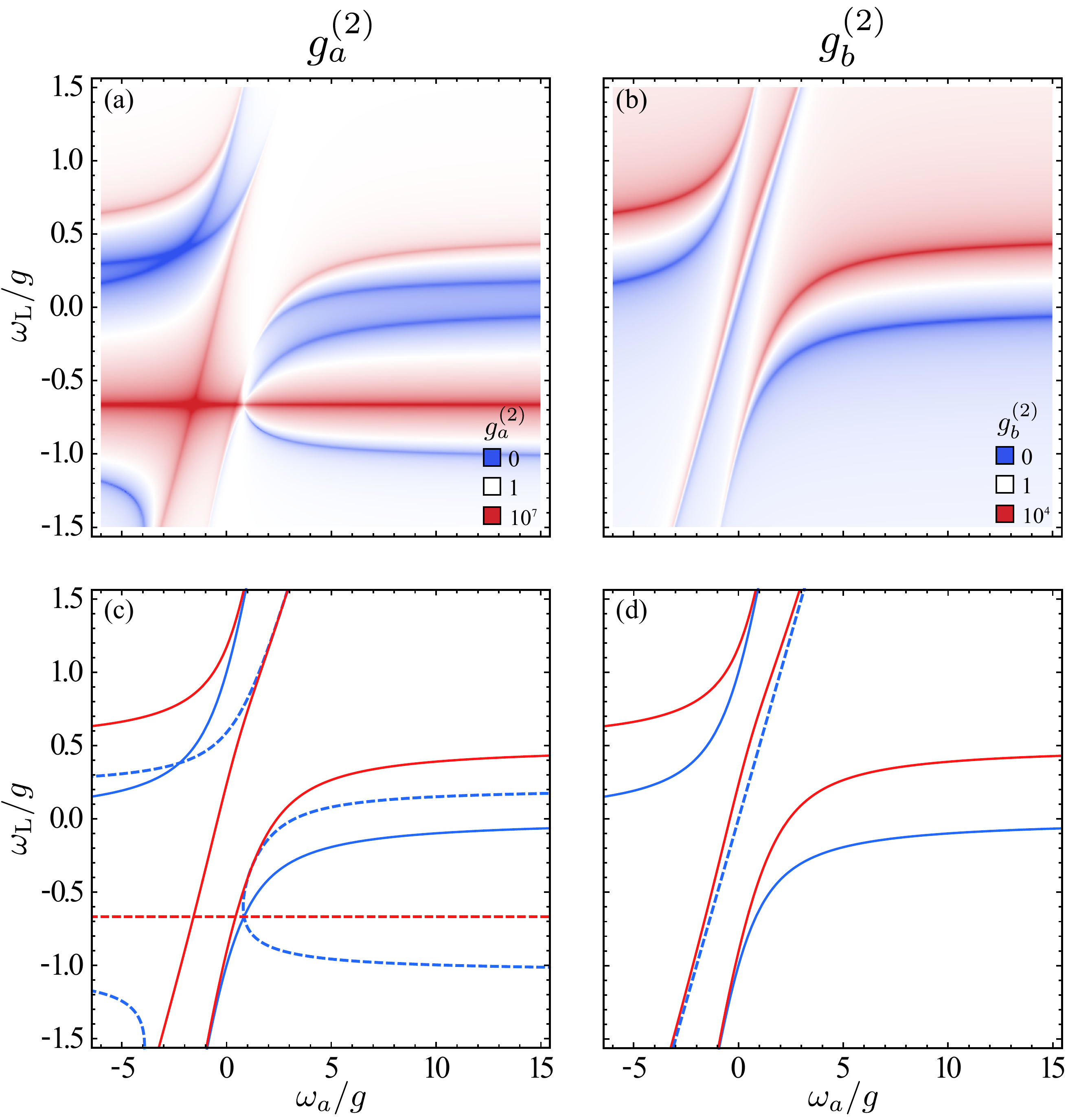}
  \caption{(Color online) Effects on the polaritonic photon
    statistics. Same as Fig.~\ref{fig:08} but for (a) nonzero
    exciton-driving and~(b) observed through the direct excitonic
    emission~$g^{(2)}_b$. In the latter case, all the unconventional
    features have disappeared. The bottom row identifies the features
    through the structure of conventional and unconventional
    lines. Parameters: $\gamma_a = 0.1 \, g$, $\gamma_b = 0.01 \, g$,
    $\chi = 0$ and $U = 10 \, \gamma_a$.}
  \label{fig:Fri8Jun183859CEST2018}
\end{figure}

\begin{figure}[t!]
  \centering \includegraphics[width=.85\linewidth]{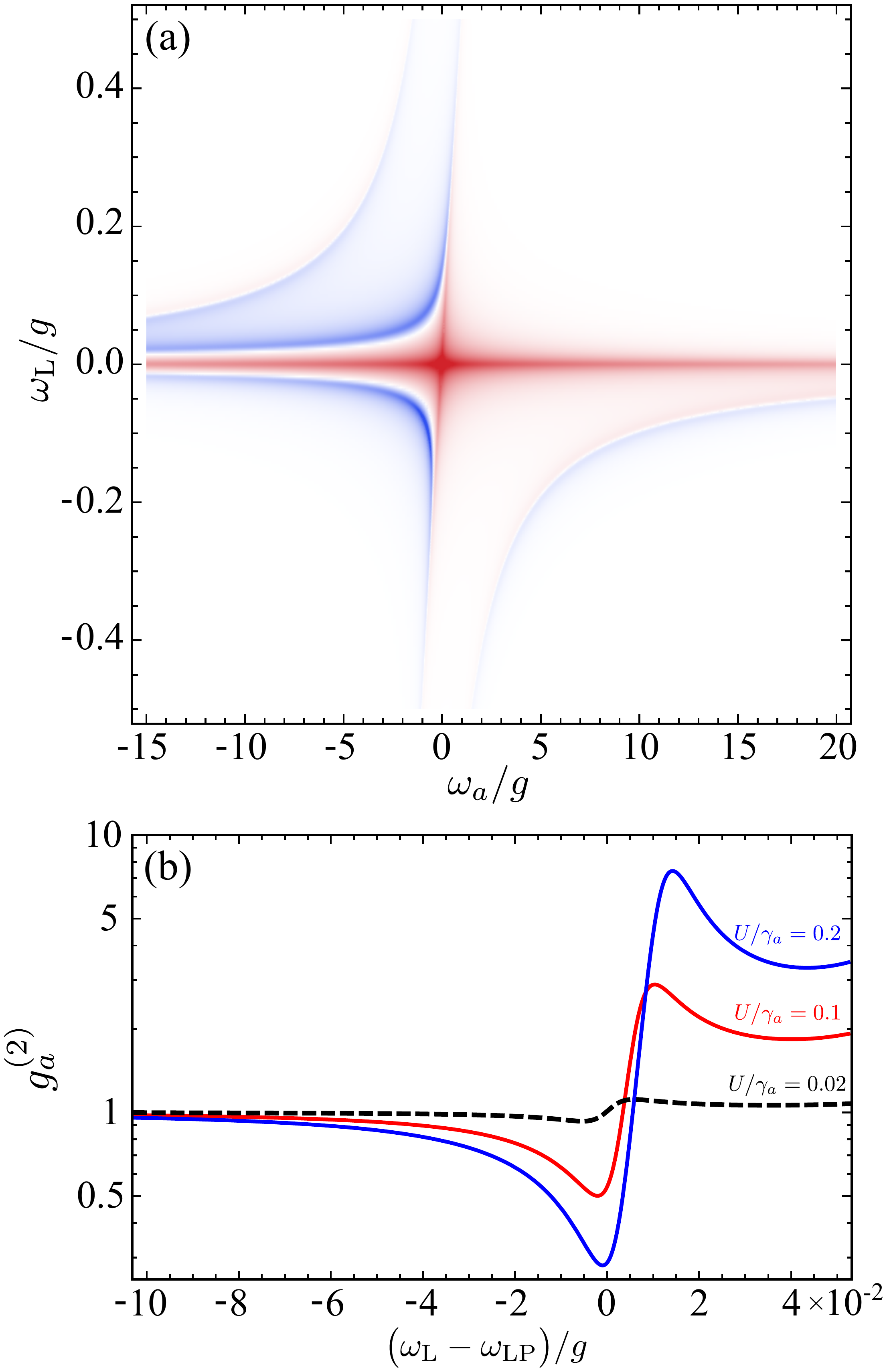}
  \caption{Polariton blockade. (a) Same as Fig.~\ref{fig:08} but for
    an experimentally realistic weak value of $U/\gamma_a=0.1$,
    showing how the UB and CA lines merge into the characteristic
    ``polariton blockade'' dispersive shape shown in Panel~(b) (fixed
    $\omega_a = 1$, the rest of parameters have got the same value as
    the indicated figure).}
  \label{fig:Fri8Jun183957CEST2018}
\end{figure}

As already noted, the polariton case adds a third CA line as compared
to its Jaynes--Cummings counterpart. The correspondence between both
cases is still clear, but this is largely thanks to the large
interaction strength chosen in Fig.~\ref{fig:08}, namely,
$U/\gamma_a=10$. This choice will allow us to survey quickly the
polaritonic phenomenology based on the more thoroughly discussed
Jaynes--Cummings one. Figure~\ref{fig:Fri8Jun183859CEST2018}, for
instance, shows the polaritonic counterpart of
Fig.~\ref{fig:Wed30May111214CEST2018} on its left panel but for one
case of mixed-pumping only, highlighting the considerable reshaping of
the structure and the importance of controlling, or at least knowing,
the ratio of exciton and photon driving. The right panel of
Fig.~\ref{fig:Fri8Jun183859CEST2018} provides $g^{(2)}_b$, which, if
compared to Fig.~\ref{fig:08}, shows that the main result is to remove
all the unconventional features and retain only the conventional
ones. The peaks are also less in the excitonic emission, producing a
smoother background. Another dramatic feature of the excitonic
correlations, which is apparent from
Eq.~(\ref{eq:Thu31May102314CEST2018}), is that it is independent from
the ratio~$\chi$ of driving, i.e., the same result is obtained if
driving the cavity alone, the exciton alone, or a mixture of both, in
stark contrast of the cavity correlations (cf. Fig.~\ref{fig:08}(a)
and \ref{fig:Fri8Jun183859CEST2018}(a) where the only difference is
that half the excitation drives the 2LS in the second case rather than
going fully to the cavity in the first case). This could be of
tremendous value for spectroscopic characterization of such systems
since it is typically difficult to know the exact type of pumping,
while experimental evidence shows that both fields are indeed being
driven under coherent excitation~\cite{dominici14a}. If measuring the
excitonic correlations, there is no dependence from the particular
type of coupling of the laser to the system. On the other hand,
excitonic emission is much less straighftorward of access.  Note,
finally, that one could similarly consider the lower and upper
polariton statistics, but they are even less featureless, with
correlations of the signal that merely follow the polariton branches
(their expression is given in the appendix for completeness).

Finally, in Fig.~\ref{fig:Fri8Jun183957CEST2018}, we focus on the
effect of the interaction strength and how to optimise the observation
of antibunching. We have already emphasized that for clarity of the
connection between the Jaynes--Cummings and the polariton case, we
have considered a value of~$U/\gamma_a$ substantially in excess even
of the few cases themselves largely in excess of the bulk of the
literature~\cite{sun17b,rosenberg18a,togan18a}. While it is not excluded that
such a regime will be available in the near future, it is naturally
more relevant to turn to the most common experimental configuration
where $U/\gamma_a\ll 1$. We show such a case in
Fig.~\ref{fig:Fri8Jun183957CEST2018}(a), where $U/\gamma_a=0.1$. We
see how, as a result, the CB and CA lines of the positively-detuned
case collapse one onto the other. The UA line previously in between
has, in the process, disappeared. The CA and CB however do not cancel
each other but merge into a characteristic dispersive-like shape,
shown in panel~(b), the observation of which, predicted over a decade
ago~\cite{verger06a}, has been a long-awaited result for polaritons
and has indeed been just recently reported from two independent
groups~\cite{arXiv_delteil18a,arXiv_munozmatutano17a}.  While this
shape has been regarded as an intrinsic and fundamental profile of
polariton blockade, our wider picture shows how it arises instead from
different features brought to close proximity by the weak
interactions. The difficulty in reporting polariton blockade lies in
the weak value of antibunching, which is largely due to the fact that
no optimisation over the full structure of photon correlations, that
was unknown till this work, has been made for the driving
configuration that yields the best antibunching for given system
parameters.

\section{Summary, Discussion and Conclusion}
\label{sec:jueene24231247GMT2019}

We have connected an hitherto disparate and voluminous phenomenology
of photon statistics in the light emitted by a variety of optical
systems, into a unified picture that identifies two classes of
conventional and unconventional features, covering both the cases of
antibunching and bunching, which leads us to a classification of CA,
UA, CB and UB.  One class (conventional), linked to real states
repulsion, occurs at all orders and for all photon numbers while the
other (unconventional) occurs for a given photon-number with no a
priori underlying level structure.

To lowest order in the driving, the dynamical response can be
described by an interferences between a squeezed component and a
coherent component, and thus, in this picture, one can understand the
photon statistics emitted by many optical systems as simply arising
from the particular way each implementation finds to produce some
squeezing on the one hand and some coherent field on the other hand,
and interfere them during its emission. {In agreement with the
  previous literature, we call} this phenomenon
``self-homodyning''. With this understanding, one can bring
considerable tailoring of photon correlations by modifying the
relative importance of coherence vs squeezing, which is conveniently
achieved by superimposing a fraction of the driving laser to the
output of the system {(``homodyning'')}.  This Gaussian-state
approximation holds to lower orders in the driving for the first
correlation functions only. For instance, for the case of the
laser-corrected two-level system (Table~\ref{tab:fluctuations}), the
squeezed-coherent Gaussian description holds up to the second-order in
the driving for the population and to the first-order in the driving
for~$\g{2}$. Deviations occur for these observables to higher-orders
in the driving while higher-order correlation functions already differ
to the lowest order in the driving. Such deviations seem to arise from
the non-Gaussian nature of the quantum fluctuations in these highly
non-linear systems. This remains to be studied in detail.

Such a general picture can explain under a unified mechanism a wealth
of observations that could otherwise appear to be peculiarities that
are specific to a particular configuration. To take one recent example
from a group that has been leading in the development and applications
of the type of homodyning and self-homodyning that we have studied
above, in Ref.~\cite{arXiv_trivedi19a}, Trivedi \emph{et al.} study
the generalization of the Jaynes--Cummings system to~$N$ emitters: the
so-called Tavis--Cummings Hamiltonian.  Here, it is found that driving
resonantly the eigenstates~\cite{laussy11a} produces conventional
antibunching, flanked by unconventional antibunching for laser
frequencies detuned from the one- and two-photon resonances. This is
the counterpart of the situation of
Fig.~\ref{fig:JCantibunchingplat}(d) (resonance) and~(e) (detuning),
both also shown in Panel~(a), where increasing~$N$ has the effect of
bosonizing the interacting (matter-like) part of the system or
decreasing the effective nonlinearity, similarly to decreasing~$g$
for~$N=1$.  Interestingly, it is reported that while for the case of
resonance, antibunching is spoiled with an increasing number of
emitters~$N$, in presence of a detuning, one of the antibunching peaks
is, on the opposite, enhanced with increasing~$N$.  This apparently
puzzling behaviour is easily understood once the conventional and
unconventional nature of the respective antibunching lines are
recognized. In the resonant case, antibunching is always conventional,
and as such it is spoiled by the bosonization of the system due to it
increasing number of emitters~\cite{dicke54a}, or by reducing the
coupling. Since both weaken the nonlinearity in the level structure,
this destroys the conventional blockade that is based on it.  With
detuning, on the other hand, one finds not only conventional but also
unconventional antibunching,
cf.~Fig.~\ref{fig:JCantibunchingplat}(b). Their CA is also spoiled
with increasing~$N$, as reported, but their UA, however, increases,
which can be expected since it is due to an self-homodyning
interference between the coherent and incoherent parts of the emission
at the two-photon level, as explained above, and this does not suffer
from a reduced nonlinearity (or increasing~$N$). It can in fact be
also optimized (i.e., reduced) like all types of UA and as a result,
should even reach $g^{(2)}=0$ to lowest order for a proper choice of
the detuning, that will depend on~$N$ in a way that remains to be
computed. Since we have shown, however, that the interference nature
of UA makes it sensitive to dephasing, and that
detuning~\cite{lopezcarreno19a} results in fast oscillations in
autocorrelation times, with a narrowing plateau of antibunching, one
can also expect this antibunching to be particularly fragile and
difficult to resolve when including a realistic model for its
detection.  This is consistent with the finding of the Authors that
inhomogeneous broadening quickly spoils UA.  Finally, they also find
in both detuned and resonant cases the unconventional bunching, as the
large bunching central peak that is a typical feature of the general
mechanism (cf.~Fig.~\ref{fig:JCantibunchingplat}). This is therefore
the super-chaotic noise due to self-homodyning stripping down the
emission to its mere fluctuations. As such, the interpretation in
terms of two-photon bound states that is offered in
Ref.~\cite{arXiv_trivedi19a} and in other
works~\cite{faraon08a,dory17a} should be further analyzed and
quantified.  We suspect the emission in UB to be less efficient for
multiphoton Physics as compared to leapfrog
emission~\cite{sanchezmunoz14a}, due to the lack of a suppression
mechanism for higher photon-number processes, and despite the large
values of the correlation functions that they produce.

As a conclusion, our picture brings considerable simplification in the
interpretation and identification of the various phenomena observed in
a plethora of systems, in particular with respect to connecting them
between each other throughout platforms. It clarifies the value but
also the limitation of a description in terms of Gaussian-states.  It
also opens a new route to control and fine-tune such photon
correlations and make a more informed and better use of them for
quantum applications.

\bibliographystyle{naturemag}
\bibliography{Sci,Books,arXiv,pb} 


\onecolumngrid
\appendix
\setcounter{equation}{0}
\setcounter{figure}{0}

\section*{Appendices}

\section{Interference between a coherent and squeezed state.}
\label{app:1}

In a beam splitter (with a certain transmittance $\mathrm{T}^2$ and
reflectance $\mathrm{R}^2$), the relation between the input light
fields ($a$, $d$) and the output fields ($o$, $s$) is a simple unitary
transformation:
\begin{equation}
\label{eq:BStransformations}
o=\mathrm{T} d +i\mathrm{R} a \, , \quad s = i\mathrm{R} d+\mathrm{T} a \, , \quad \text{and} \quad d = \mathrm{T}  o - i \, \mathrm{R}  s  \, , \quad
a =  -i \, \mathrm{R} o + \mathrm{T} s \, ,
\end{equation}
where the real coefficients $\mathrm{T}$ and $\mathrm{R}$ must fulfil
$1 \geq \mathrm{T}, \mathrm{R} \geq 0$ and $\mathrm{T}^2 +
\mathrm{R}^2 = 1$. In our case of interest, the input 0 is a squeezed
state $\ket{\xi} = \mathcal{S}_d \left(\xi\right) \ket{0}$ and input
1, a coherent state $\ket{\alpha} = \mathcal{D}_a \left(\alpha\right)
\ket{0}$. The squeezing operator is $\mathcal{S}_d \left(\xi\right) =
\exp \left[\frac{1}{2} \left(\xi^* d^2 - \xi \ud{d}^2\right)\right]$
with squeezing parameter $\xi = r e^{i \theta}$ and the displacement
operator is $\mathcal{D}_a \left( \alpha \right) = \exp \left( \alpha
\ud{a} - \alpha^* a \right)$ with coherent parameter $\alpha =
|\alpha| e^{i \phi}$. The total input state can be written as:
\begin{equation}
\ket{\psi_{\mathrm{in}}} = \mathcal{D}_a \left(\alpha\right) \mathcal{S}_d \left(\xi\right)
\ket{0}_{da},
\end{equation}
where the state subscript indicates the input/output subspaces where
operators are acting upon. These operators are written in the input
basis. Now, applying transformations of
Eq.~\eqref{eq:BStransformations} and rearranging terms, we obtain, in
first place, for the displacement operator: $\mathcal{D}_a
\left(\alpha\right) = \exp \left( \alpha \ud{o} - \alpha^* o + \alpha
\ud{s} - \alpha^* s \right)$, where $\alpha_o = i \mathrm{R} \alpha$
and $\alpha_s = \mathrm{T} \alpha$.  Exponentials of operators can be
factorized since both output are independent from each other and
commute. This leads to the simple expression: $\mathcal{D}_a
\left(\alpha \right) = \mathcal{D}_o \left(\alpha_o \right)
\mathcal{D}_s \left(\alpha_s\right) = \mathcal{D}_s
\left(\alpha_s\right) \mathcal{D}_o \left(\alpha_o \right)$, where
each displacement operator $\mathcal{D}_j \left(\alpha_j\right)$
($j=o,s$) only acts over the assigned output. Then, the squeezing
operator in the output basis reads:
\begin{equation}
\mathcal{S}_d \left(\xi \right) = \exp \left[ \frac{1}{2}
  \left(\xi_o^* o^2 - \xi_o \ud{o} \right) + \frac{1}{2}
  \left(\xi_s^* s^2 - \xi_s \ud{s} \right) + \right.
  \left. \left(\xi_{os}^* o s - \xi_{os} \ud{o} \ud{s} \right)
  \right] =\exp(S_o+S_s+S_{os})\,,
\end{equation}
where $\xi_o = \mathrm{T}^2 \xi$, $\xi_s = - \mathrm{R}^2 \xi$ and
$\xi_{os} = i \mathrm{R} \mathrm{T} \, \xi$. This exponential can
be simply split into two different contributions only if
$\lbrack S_o + S_s , S_{os} \rbrack =0$, which is fulfilled only
in the particular case $\mathrm{T} = \mathrm{R}$ (symmetrical BS).
Although this constriction could suppose a huge difficulty to overcome,
first correction term grows proportional to $r^2 \mathrm{T} \mathrm{R} \left(\mathrm{T}^2-\mathrm{R}^2 \right)$. Thus, for either low squeezing signal ($r \ll 1$) or almost symmetrical BS ($\mathrm{T}-\mathrm{R} \approx 0$), the output signal 
could be described as follows. The commutator $\lbrack S_o , S_s \rbrack$ vanishes for any possible values, so the exponential simplifies into
$\mathcal{S}_d \left(\xi \right) = \mathcal{S}_o \left(\xi_o \right)
\mathcal{S}_s \left(\xi_s \right) \mathcal{S}_{os} \left(\xi_{os}
\right)$. With the previous results, the output state can be written
as:
\begin{equation}
\ket{\psi_{\mathrm{out}}} = \mathcal{D}_o \left(\alpha_o\right)
\mathcal{S}_o \left(\xi_o\right) \mathcal{D}_s \left(\alpha_s\right)
\mathcal{S}_s \left(\xi_s\right) \mathcal{S}_{os}
\left(\xi_{os}\right) \ket{0}_{os} \\ = \mathcal{D}_o
\left(\alpha_o\right) \mathcal{S}_o \left(\xi_o\right) \mathcal{D}_s
\left(\alpha_s\right) \mathcal{S}_s \left(\xi_s\right)
\ket{\xi_{os}}_{os},
\end{equation}
where $\ket{\xi_{os}}$ represents a two-mode squeezed state. In the
Fock basis, this state can be written as (from
Ref.~\cite{guerry_book05a}, Chapter 7):
\begin{equation}
\ket{\xi_{os}} = \frac{1}{\cosh r_{os}} \sum_{n=0}^{\infty} \left(\tanh r_{os}\right)^n
\ket{n, n}_{os},
\end{equation}
where $r_{os} = |\xi_{os}| = \mathrm{R} \mathrm{T} \, r$. The
corresponding density matrix for this pure state reads
$\rho_{\mathrm{out}} = \ket{\psi_{\mathrm{out}}}
\bra{\psi_{\mathrm{out}}}$. Tracing out output $o$, we obtain the
density matrix for output $s$ only (our signal of interest): $\rho_s =
\Tr_o\{ \rho_{\mathrm{out}}\}$. For the next step, we will use the properties of the
trace to move clockwise the operators acting over the output $o$
subspace and the identities $\ud{\mathcal{D}_o} \left(\alpha_o\right)
\mathcal{D}_o \left(\alpha_o\right) = \ud{\mathcal{S}_o}
\left(\xi_o\right) \mathcal{S}_o \left(\xi_o\right) =
\hat{\mathds{1}}_o $, where $\hat{\mathds{1}}_o $ is the identity
operator. Furthermore, any operator that only acts on the $s$-subspace
can be taken out of the trace. With this,
\begin{equation}
\rho_s= \mathcal{D}_s \left(\alpha_s\right) \mathcal{S}_s
\left(\xi_s\right) \big(\Tr_o \lbrace \ket{\xi_{os}} \bra{\xi_{os}}
\rbrace \big) \ud{\mathcal{S}_s} \left(\xi_s\right) \ud{\mathcal{D}_s}
\left(\alpha_s\right) \, .
\end{equation}
Computing the partial trace:
\begin{equation}
\begin{split}
	\Tr_o \lbrace \ket{\xi_{os}} \bra{\xi_{os}} \rbrace = &
        \sum_{p=0}^{\infty} \bra{p}_o \left(\frac{1}{\cosh^2 r_{os}}
        \sum_{n,m=0}^{\infty} \left(\tanh r_{os}\right)^{n+m}
        \ket{n}_o \ket{n}_s \bra{m}_s \bra{m}_o\right) \ket{p}_o \\ =
        & \, \frac{1}{\cosh^2 r_{os}} \sum_{n}^{\infty} \left(\tanh
        r_{os}\right)^{2n} \ket{n}_s \bra{n}_s ,
\end{split}
\end{equation}
from the second to the third equality we have used $\braket{m}{p}_o =
\delta_{m,p}$ and $\braket{p}{n}_o = \delta_{p,n}$. The resulting
density matrix has the form of a thermal state $\rho_{\mathrm{th}}$
with mean population $\mathrm{p_{th}} \equiv \pop{s}_{\mathrm{th}} =
\sinh^2 r_{os}$. To sum up, the output field detected at a single arm
of the system corresponds to a \textit{displaced squeezed thermal}
state. So, to sum up,
\begin{equation}
\rho_s = \mathcal{D}_s \left(\alpha_s\right) \mathcal{S}_s
\left(\xi_s\right) \rho_{\mathrm{th}} \ud{\mathcal{S}_s}
\left(\xi_s\right) \ud{\mathcal{D}_s} \left(\alpha_s\right),
\end{equation}
with parameters $\alpha_s = \mathrm{T}|\alpha| e^{i \phi}$, $\xi_s =
r_s e^{i \theta_s} = \mathrm{R}^2 e^{i \left( \theta + \pi \right)}$
and $ \mathrm{p_{th}} = \sinh[2](\mathrm{R}\mathrm{T} \,
r)$. Even though $\mathrm{T}$ and $\mathrm{R}$ appear as
parameters, last equation only works for $\mathrm{R} \approx
\mathrm{T}$. We restrict ourselves to the case of \textit{50:50} beam splitter
($\mathrm{T}^2 = \mathrm{R}^2  = 1/2$).
Thermal population can be expressed in terms of
squeezed population of the input signal $\av{n_d} = \sinh^2 r$:
\begin{equation}
\mathrm{p_\mathrm{th}} + \frac{1}{2} = \frac{1}{2} 
\sqrt{1+ \av{n_d}}.
\end{equation}
From $\rho_s$ we can compute the observables for the mixed signal:
\begin{subequations}
\begin{equation}
\av{n_s} = \frac{|\alpha|^2}{2} + \frac{\av{n_d}}{2} \, , \quad |\av{s^2}| = \left(\mathrm{p_{th}} + \frac{1}{2}\right) \sinh(r)\,,
\end{equation}
\begin{equation}
\label{eq:DST_g2}
\g{2}_s =  1 + \av{n_s}^{-2} \sinh^2 r
\left[\cosh 2r + 2 |\alpha|^2 \left(1- \cos(\theta-2\phi) \coth r\right) \right]  \, ,
\end{equation}
\begin{multline}
\g{3}_s =  1 + \av{n_s}^{-3} \sinh^2 r \, \Big\lbrace \sinh^2 2r + 5 \sinh^2 r \cosh 2r
+ 6 |\alpha|^4 \left(1- \cos(\theta-2\phi) \coth r \right) + \\
  \: 
3 |\alpha|^2 \left[3 \coth^2 r - 1 + 6 \left(1- \cos(\theta-2\phi) \coth r\right) \right]
\Big\rbrace  \,,
\end{multline}
\end{subequations}

\section{Decomposition of $g_s^{(3)}$ in powers $\alpha$}
\label{app:g3decomp}

We provide the third-order coherence function of signal $s$, in terms
of two interfering fields, $s=\alpha + d$. As in the case for
$\g{2}_s$, the highest order contributions in powers of $\alpha$ can
be gathered into the coherent term (given by 1), yielding:
\begin{equation}
\g{3}_s = 1 + \sum_{m=0}^{4} \mathcal{J}_m \, ,
\end{equation}
with
\begin{align}
\mathcal{J}_0 &=\frac{\corr{d}{3}{3} - \pop{d}^3 }{\av{n_s}^3} \, ,\\
\mathcal{J}_1 &= 6 \frac{ \Re[ \alpha^* \left( \corr{d}{2}{3} - \coh{d} \pop{d} \right)] }{\av{n_s}^3} \, ,\\
\mathcal{J}_2 &= 3\Big[ 2\Re[{\alpha^*}^2 \av{\ud{d} d^3} ] +
|\alpha|^2 \left(3\corr{d}{2}{2} +\pop{d}^2 \right)
 - 4 \Re[ \alpha^* \coh{d}]^2 \Big]/ \av{n_s}^3 \,,\\ \mathcal{J}_3 &= 2 \Big[ \Re[ {\alpha^*}^3 \av{d^3} ] +  |\alpha|^2 \Re[
\alpha^* \left( 9 \av{\ud{d} d^2} - \coh{d} \pop{d} \right)]
 - 4 \Re[ \alpha^* \coh{d}]^3 \Big]/ \av{n_s}^3 \, ,\\
\mathcal{J}_4 &= 6|\alpha|^2 \frac{ \Re[ {\alpha^*}^2 \av{d^2} ] + |\alpha|^2 \pop{d} - 2 \Re[\alpha^*  \coh{d}]^2}{\av{n_s}^3}\,.
\end{align}

\section{Coherent and squeezed steady states}
\label{ap:coh-sq}

The coherent and the squeezed state described in
Sec.~(\ref{sec:ThuFeb15174655CET2018}) can be obtained as the
steady-state solutions of a driven cavity, described through the
master equation (hereafter $ \hbar = 1 $ is assumed)
\begin{equation}
  \label{eq:FriOct20195045CEST2017}
  \partial_t \rho = i[\rho,H_c]+\frac{\gamma_c}{2}\mathcal{L}_c
  \rho\,.
\end{equation}
Here~$\mathcal{L}_c=(2c\rho \ud{c} - \ud{c}c\rho - \rho\ud{c}c)$
and~$c=a,d$ are the annihilation operator of the field generating the
coherent and squeezed state. The Hamiltonians are set as
$H_a=\Delta_a\ud{a}a + \Omega_a (\ud{a}+a)$ for the coherent state and
$H_d=\Delta_d \ud{d}d + \lambda_d(d^{\dagger 2}+d^2)$ for the squeezed
state. In the former case, $\Delta_a$ is the detuning between the
cavity and the laser that excites the cavity with intensity~$\Omega_a$
and in the latter case, it is the detuning between the cavity and the
driving mode that squeezes the cavity with intensity~$\lambda_d$. The
relation between the dynamical quantities and the displacement and the
squeezing parameters is the following: $\alpha=\mean{a}$,
$\sinh^2(r)=\mean{\ud{d}d}$ and
$\tan\theta= i (\mean{d^{\dagger 2}}-\mean{d^2})/(\mean{d^{\dagger
    2}}+\mean{d^2})$. These two systems can be solved exactly, so the
steady-state solutions for the parameters defined above are:
\begin{equation}
    \label{eq:FriOct20213722CEST2017a}
    |\alpha|= \frac{2\Omega_a}{\sqrt{\gamma_a+4\Delta_a^2}}\quad\text{and}\quad
    \tan\phi = -\frac{\gamma_a}{2\Delta_a}\,,
\end{equation}
for the coherent state and
\begin{equation}
    \label{eq:FriOct20213722CEST2017b}
    \tanh(2r)=\frac{4\lambda_d}{\sqrt{\gamma_d^2+4 \Delta_d^2}}\quad\text{and}\quad\tan\theta=\frac{\gamma_d}{2\Delta_d}\,,
\end{equation}
for the squeezed state. For the latter case, there is also a thermal contribution given by
$\mathrm{p_{th}} = \sinh[2](r)$, so the resulting state is actually
a \textit{thermal squeezed} state.

\section{Steady states of light matter coupling at vanishing laser driving}
\label{app:2}

In this work we need to solve the steady state dynamics of
light-matter interaction in the low coherent driving regime. The light
field is a cavity mode~$a$ and the matter field can be either a
2LS~$\sigma$ or another bosonic mode~$b$. We solve the dynamics in
terms of a general mean value, a product of any system operator, which
in its most general normally ordered form
reads~$C_{\{m,n,\mu,\nu\}}=\mean{\sigma^{\dagger m}\sigma^n a^{\dagger
    \mu} a^\nu}$ (with~$m$, $n \in\{0,1\}$ and $\mu$, $\nu\in
\mathbb{N}$) if the matter field is a 2LS or
~$C_{\{m,n,\mu,\nu\}}=\mean{b^{\dagger m}b^n a^{\dagger \mu} a^\nu}$
(with~$m$, $n$, $\mu$, $\nu\in \mathbb{N}$) if the matter field is
bosonic. This general element follows the master equation described in
the main text, which can be expressed in a matricial form:
\begin{equation}
\label{eq:TueMay5174356GMT2009}
\partial_t C_{\{m,n,\mu,\nu\}}=\sum_{{m',n',\mu',\nu'}}\mathcal{M}_{\substack{m,n,\mu,\nu\\m',n',\mu',\nu'}}C_{\{m',n',\mu',\nu'\}}\,.
\end{equation}
The regression matrix
elements~$\mathcal{M}_{\substack{m,n,\mu,\nu\\m',n',\mu',\nu'}}$, in
the case of a coupled 2LS, are given by:
\begin{subequations}
	\label{eq:TueDec23114907CET2008}
	\begin{align}
	&\mathcal{M}_{\substack{m,n,\mu,\nu\\m,n,\mu,\nu}}=-\frac{\gamma_a}2(\mu+\nu)-\frac{\gamma_\sigma}2(m+n)
	+ i (\mu - \nu ) \delta_{a} + i (m-n) \Delta_\sigma 
	\end{align}
	\begin{align}
	&\mathcal{M}_{\substack{m,n,\mu,\nu\\1-m,n,\mu,\nu}}=i\Omega_\sigma[m+2n(1-m)]\,,\quad
	&&\mathcal{M}_{\substack{m,n,\mu,\nu\\m,1-n,\mu,\nu}}=-i\Omega_\sigma[n+2m(1-n)]\,\\
	&\mathcal{M}_{\substack{m,n,\mu,\nu\\m,n,\mu-1,\nu}}=i \Omega_a\mu\,,\quad
	&&\mathcal{M}_{\substack{m,n,\mu,\nu\\m,n,\mu,\nu-1}}=- i\Omega_a\nu\,,\\
	&\mathcal{M}_{\substack{m,n,\mu,\nu\\m,1-n,\mu,\nu-1}}=-i g(1-n)\nu\,,\quad 
	&&\mathcal{M}_{\substack{m,n,\mu,\nu\\1-m,n,\mu-1,\nu}}=i g(1-m)\mu\,, \\
	&\mathcal{M}_{\substack{m,n,\mu,\nu\\m,1-n,\mu,\nu+1}}=-i g n \,,\quad
	&&\mathcal{M}_{\substack{m,n,\mu,\nu\\1-m,n,\mu+1,\nu}}=i g m \,, \\
	&\mathcal{M}_{\substack{m,n,\mu,\nu\\1-m,n,\mu,\nu+1}}=2 i n g(1-m) \,,\quad
	&&\mathcal{M}_{\substack{m,n,\mu,\nu\\m,1-n,\mu + 1,\nu}}= -2 i n g(1-m)\,,
	\end{align}
\end{subequations}
and zero everywhere else. In the main text, we discuss first the case
of resonance fluorescence, which corresponds to having only the 2LS
operator~$\sigma$ and no cavity mode~$a$ (taking $g$, $\Omega_a=0$
here). Second, we solve the Jaynes--Cummings model with both cavity and dot driving
with a phase difference between the sources,
which corresponds to setting $\Omega_\sigma / \Omega_a = \chi e^{-i \phi} $ here.

Similarly, for the polariton model where the matter field is an
exciton (boson), we have:
\begin{subequations}
	\begin{align}
	&\mathcal{M}_{\substack{m,n,\mu,\nu\\m,n,\mu,\nu}}=-\frac{\gamma_a}2(\mu+\nu)-\frac{\gamma_b}2(m+n)
	+ i (\mu - \nu ) \Delta_{a} + i (m-n) \Delta_b + i \frac{U}{2} \left[m (m-1) - n(n-1) \right],
	\end{align}
	\begin{align}
	&\mathcal{M}_{\substack{m,n,\mu,\nu\\m,n,\mu-1,\nu}}=i\Omega_a \mu ,
	&&\mathcal{M}_{\substack{m,n,\mu,\nu\\m,n,\mu,\nu-1}}=-i\Omega_a \nu \,\\
	&\mathcal{M}_{\substack{m,n,\mu,\nu\\m+1,n,\mu-1,\nu}}= i g \mu\,,\quad
	&&\mathcal{M}_{\substack{m,n,\mu,\nu\\m,n+1,\mu,\nu-1}}=-i g \nu\,,\\
	&\mathcal{M}_{\substack{m,n,\mu,\nu\\m-1,n,\mu+1,\nu}}= i g m\,,\quad
    &&\mathcal{M}_{\substack{m,n,\mu,\nu\\m,n-1,\mu,\nu+1}}=-i g n\,,\\
	&\mathcal{M}_{\substack{m,n,\mu,\nu\\m+1,n+1,\mu,\nu}}=i U (m-n) ,\quad
	\end{align}
\end{subequations}
and, again, the remaining matrix elements are zero.

These equations can be solved numerically, by choosing a high enough
truncation in the number of excitations, in order to obtain the steady
state ($\partial_t C_{\{m,n,\mu,\nu\}}=0$) for any given
pump. However, we are interested in an analytical solution when
applying vanishing driving limit ($\Omega_\sigma \rightarrow 0$ or
$\Omega_a\rightarrow 0$). In this case, it is enough to solve
recursively sets of truncated equations. That is, we start with the
lowest order correlators, with only one operator, that we write in a
vectorial form for convenience (using the JC model as an example):
$\mathbf{v}_1=(
\mean{a}~\mean{a^\dagger}~\mean{\sigma}~\mean{\sigma^\dagger
})^\mathrm{T}$. Its equation, $\partial_t \mathbf{v}_1= M_1
\mathbf{v}_1+A_1+~\mathrm{h.~o.~t.}$, provides the steady state
value~$\mathbf{v}_1= -M_1^{-1} A_1+~\mathrm{h.~o.~t.}$, to lowest
order in $\Omega_a$ (with h. o. t. meaning \emph{higher order
  terms}). We proceed in the same way with the two-operator
correlators $\mathbf{v}_2=( \mean{a^2}~\mean{a^{\dagger
    2}}~\mean{a^\dagger
  a}~\mean{\sigma^\dagger\sigma}~\mean{\sigma^\dagger
  a}\,\hdots)^\mathrm{T}$, only, in this case, we also need to include
the steady state value for the one-operator correlators as part of the
independent term in the equation: $\partial_t \mathbf{v}_2= M_2
\mathbf{v}_2+A_2+X_{21} \mathbf{v}_1+~\mathrm{h.~o.~t.}$. The steady
state reads~$\mathbf{v}_2= -M_2^{-1} (A_2+X_{21}
\mathbf{v}_1)+~\mathrm{h.~o.~t.}$ with an straightforward
generalization $\mathbf{v}_N= -M_N^{-1} (A_N+\sum_{j=1}^{N-1}X_{N j}
\mathbf{v}_j)+~\mathrm{h.~o.~t.}$.

We, in particular, aim at obtaining photon correlators of the
type~$\mean{a^{\dagger N} a^N }$ that follow~$\mean{a^{\dagger N} a^N
}\sim (\Omega_a)^{2N} $, to lowest order in the driving
$\Omega_a$. The normalized correlation
functions~$g_a^{(N)}=\mean{a^{\dagger N} a^N }/\mean{a^\dagger a}^N $
are independent of $\Omega_a$ to lowest order, their computation
requiring to solve the $2N$ sets of recurrent equations and taking the
limit~$\lim_{\Omega_a\rightarrow 0}g_a^{(N)}$.

\section{Homodyne interference with resonance fluorescence: correlations and squeezing from the fluctuations and the total signal}
\label{sec:FriFeb23221121CET2018}

The correlations from the fluctuations of resonance fluorescence, with
operator $\epsilon = \sigma -\alpha$, can be accessed using the
technique of homodyne detection explained in
Sec.~(\ref{sec:ThuFeb15174655CET2018}). In this case, we feed one of
the beam splitter arms with resonance fluorescence ($d\rightarrow
\sigma$) and the other with a coherent field ($a \rightarrow \beta$).
The correlators of the output of the arms as defined in
Appendix~\ref{app:1}, $s= i\mathrm{R} \beta + \mathrm{T} \sigma$, are
given by:
\begin{equation}
\label{eq:2LScorrBS}
\corr{s}{n}{m} = \sum_{p=0}^n \sum_{q=0}^m \binom{n}{p} \binom{m}{q}
(-i \mathrm{R}\beta^*)^{n-p} (i \mathrm{R}\beta)^{m-q}
\corr{\sigma}{p}{q} \, .
\end{equation}
Since $\corr{\sigma}{p}{q} =0$ for $p,q >1$, this simplifies to
\begin{equation}
\corr{s}{n}{m} = (-i \mathrm{R}\beta^*)^{n}
(i \mathrm{R}\beta)^{m} - i \mathrm{R} \mathrm{T}  \ (-i \mathrm{R}\beta^*)^{n-1}
(i \mathrm{R}\beta)^{m-1} \left(m \beta \coh{\sigma} - n \beta^*
  \coh{\sigma}^*  \right)  + 
n m \  (-i \mathrm{R}\beta^*)^{n-1}
(i \mathrm{R}\beta)^{m-1} \mathrm{T}^2 \av{n_{\sigma}}\,.
\end{equation}
For instance, the coherent fraction and total population of the output
field are:
\begin{subequations}
  \label{eq:FriFeb23210546CET2018}
  \begin{align}
    \label{eq:FriFeb23210546CET2018a}
    \coh{s} &= i \mathrm{R} \beta + \mathrm{T} \coh{\sigma}\,,\\
    \label{eq:FriFeb23210546CET2018b}
    \pop{s} &= \mathrm{R}|\beta|^2 + \mathrm{T}^2 \av{n_\sigma} +2 \mathrm{R} \mathrm{T}
    \Re [ i  \beta^* \coh{\sigma}]\,.
  \end{align}
\end{subequations}
Therefore, we can chose the coherent field to compensate exactly the
coherent component of the 2LS $\alpha=\mean{\sigma}$,~i.e.,
setting~$\beta= i \frac{\mathrm{T}}{\mathrm{R}}\mean{\sigma}$ so that
we hace only the transmitted fluctuations~$s=T\epsilon$. In such a
case the correlators simplify even further,
\begin{equation}
\label{eq:2LSincohCorrs}
\corr{\epsilon}{n}{m}=\corr{s}{n}{m}/ \mathrm{T}^{n+m} = (-1)^{m+n}
\alpha^{m-1} \alpha^{* (n-1)} \left[ nm \ \av{n_\sigma} -
  \left(n+m-1 \right) \big |\alpha \big |^2\right]\,.
\end{equation}
With this general expression, we obtain the population and coherence
functions, Eq.~(\ref{eq:FriFeb23155745CET2018}) of the main text.

We want to recover and analyse light properties from the original 2LS
(before the beam splitter).  In order to do this, the factor
$\mathrm{T}^2$ on the population should be eliminated (making
$\mathrm{T}^2 \av{n_\sigma} \rightarrow \av{n_\sigma}$,
$\mathrm{T}\coh{\sigma} \rightarrow \coh{\sigma}$).  Note that this
change is not inconsistent, given that the beam splitter divides by
two the incoming signal and thus merely attenuating it by a
factor~$\mathrm{T}^2$. Moreover, since the photon correlations are
normalized objects, a global attenuation in the \textit{unnormalized}
correlators do not change them. However, suppressing the coherent
contribution of the emission is not the only possibility. We can also
tune the coherent contribution by choosing $\beta' = e^{i \phi}
|\beta'|$, where the amplitude is parametrize as $|\beta'|=
\frac{\mathrm{R}}{\mathrm{T}} |\beta|$.  Thus, we are broadening the
range of possible output configurations~\cite{lopezcarreno18b}. So, 
for the most general case, the
$N$-particle correlators have the following form:

\begin{equation}
\label{}
\g{N}_s = 
\frac{|\beta'|^{2(n-1)} \left( |\beta'|^2 + n^2 \av{n_\sigma}
+2 n \Im \lbrace \av{\sigma} |\beta'| e^{-i \phi} \rbrace	\right)}
{\left(|\beta'|^2 +  \av{n_\sigma}+2 
	\Im \lbrace \av{\sigma} |\beta'| e^{-i \phi} \rbrace	\right)^n},
\end{equation}
After this general expression, the amplitude $|\beta'|$ is usually expressed
in a more suitable way, referencing it to the driving intensity of the laser: $|\beta'| = \frac{\Omega_{\sigma}}{\gamma_{\sigma}} \mathcal{F}$.

Other two important quantities are the mean and variance of the quadratures,
$\av{X_{s,\chi}} = \frac{1}{2} (e^{i\chi} \av{\ud{s}} + \mathrm{c.c.})$ and
their dispersion $\av{\Delta X_{s,\chi}^2} = \av{X_{s,\chi}^2 - \av{X_{s,\chi}}^2}$,
respectively:
\begin{equation}
\av{X_{s,\chi}} = \frac{1}{2} \left(e^{i \phi} \av{s}^* + \mathrm{c.c} \,\right),
\end{equation}
The mean value only depends on
the total coherent contribution $\av{s}=   \mathrm{T} (i \beta' + \alpha)$. 

The maximum and minimum of the (normal-ordered) quadrature variance for a single-mode
can be inferred independently of specific nature of the field:
\begin{equation}
\label{eq:sQuadDisp}
\av{:\Delta X_{s}^2:}_{\mathrm{max}/\mathrm{min}} =
\av{\Delta X_{s}^2}_{\mathrm{max}/\mathrm{min}} - \frac{1}{4} =
\frac{1}{2} \left[\pm |\av{s^2}-\av{s}^2| + \pop{s} - |\av{s}|^2\right],
\end{equation}
where the sign correspond to maximum and minimum, respectively.
While the variance is strictly a positive-valued quantity, its normal-ordered 
counterpart is not. This latter indicates the deviation of the variance 
from the vacuum value (which is $\frac{1}{4}$) so values below 0 will reveal some degree of 
quadrature squeezing. Likewise the angle of squeezing is generically given by:
\begin{equation}
\theta = \mathrm{arg} \left[\av{s^2} - \av{s}^2\right].
\end{equation}
After substituting the correlators \eqref{eq:2LSincohCorrs} on \eqref{eq:sQuadDisp}
and, then, using the steady-state solution given in
Eq.~(\ref{eq:2LSrhoSS}):
\begin{align}
\av{:\Delta X_{s}^2:}_\mathrm{min} = -\frac{2 \mathrm{T}^2 \Omega_{\sigma}^2 \left(\gamma_{\sigma}^2 + 4 \Delta_\sigma^2 - 8 \Omega_\sigma^2 \right)}{\left(\gamma_{\sigma}^2 + 4 \Delta_\sigma^2 + 8 \Omega_\sigma^2\right)^2}, & \ \
\av{:\Delta X_{s}^2:}_\mathrm{max} = \frac{2 \mathrm{T}^2 \Omega_{\sigma}^2}{\gamma_{\sigma}^2 + 4 \Delta_\sigma^2 + 8 \Omega_\sigma^2}, 
\end{align}
and the angle of squeezing will be:
\begin{equation}
\theta = \mathrm{arg} \left[\left(\gamma_{\sigma}-2 i \Delta_\sigma \right)^2\right]
\end{equation}
It is not surprising that factor $\mathcal{F}$ does not appear since all the 
squeezing properties exclusively come from the fluctuations. The strength of this effect
is reduced by the factor $\mathrm{T}^2$ as the input signal $\sigma$ ($\alpha + \epsilon$) is divided
by the beam splitter, which can naturally absorbed into $\Omega_\sigma^2$.
Now we are interested in the low driving regime ($\Omega_{\sigma} \rightarrow 0$), so
the previous expressions at the lowest order in $\Omega_\sigma$ simplify
\begin{equation}
\label{eq:2LSlowpumpDisp}
\av{:\Delta X^2_s:}_{\mathrm{max}/\mathrm{min}} \approx \pm 
\frac{2  \Omega_{\sigma}^2}{ \gamma_{\sigma}^2 + 4 \Delta^2_\sigma},	
\end{equation}
In this limit, both dispersion are symmetrical (but with opposite signs).
We can regard these expression as a limit of low squeezing (and coherent
intensity $|\alpha|^2$) from a displaced squeezed thermal state 
(although the coherent part does not contribute to the variance).
Such states has got the following dispersion when $r \rightarrow 0$:
\begin{equation}
\label{eq:DSTlowsqueezedDispersion}
\av{:\Delta X^2:}_{\mathrm{max}/\mathrm{min}}^{\mathrm{DST}} \approx
\frac{1}{4}
\left[\left(1 \pm 2 r\right) \left(1 + 2 \av{n_\mathrm{th}}\right)-1\right]
\approx \pm \, \frac{r}{2},
\end{equation}
where the superscript on the average
reminds that the observable corresponds to an \textit{displaced squeezed thermal} state.
We have approximated $1 + 2 \av{n_\mathrm{th}}$ as $1$ since
the thermal population grows like $\Omega_{\sigma}^4$ (which comes from the first order
of the incoherent population). Comparing \eqref{eq:2LSlowpumpDisp}
with \eqref{eq:DSTlowsqueezedDispersion}, we found that the incoherent
population in the Heitler regime behaves like a squeezed thermal state
with squeezing parameter
\begin{equation}
\label{eq:2LSreff}
r_\mathrm{eff} = 
\frac{4 \Omega_{\sigma}^2}{ \gamma_{\sigma}^2 + 4 \Delta_{\sigma}^2},
\end{equation}
and the \emph{effective} thermal population $\mathrm{p}_\mathrm{th}$:
\begin{equation}
\label{eq:2LSntheff}
\mathrm{p}_\mathrm{th} \approx \frac{16	\Omega_\sigma^4}{\left( \gamma_{\sigma}^2 +
4 \Delta_{\sigma}^2\right)^2 },
\end{equation}
From these two parameters an effective $\g{2}$, namely $\g{2}_\mathrm{eff}$,
can be obtained for the fluctuations.
Supposing that, in the low excitation regime, fluctuations would behave similar to an squeezed thermal state, then $\g{2}_\epsilon$ should have the same form.
Fixing $|\alpha| = 0$ in Eq.~\eqref{eq:DST_g2} and taking the limit $r^2 \rightarrow
0$ and $\mathrm{p}_\mathrm{th} \rightarrow 0$ (both go to 0 with the same power dependence),
we get
\begin{equation}
\g{2}_\mathrm{eff} \approx \frac{r_\mathrm{eff}^2}{\left(r_\mathrm{eff}^2 +\mathrm{p}_\mathrm{th}\right)^2},
\end{equation}
which, after substituting Eqs.~\eqref{eq:2LSreff}-\eqref{eq:2LSntheff}, reads
\begin{equation}
\label{eq:2LSg2eff}
\g{2}_\mathrm{eff} \approx \frac{\left(\gamma_{\sigma}^2 + 4 \Delta_{\sigma}^2\right)^2}
{\Omega_{\sigma}^4}.
\end{equation}

\section{Wavefunction approximation method at vanishing pumping
  regime} 
\label{app:3}
In the context of this paper, the wavefunction
approximations~\cite{visser95a} consist of assuming that the state of
the system composed by two fields, with annihilation operators~$\xi$
and~$c$ following either fermionic or bosonic algebra, can be
approximated by a pure state, which in the Fock state basis reads,
\begin{equation}
  \label{eq:WedFeb28112316CET2018}
\ket{\psi} = \sum_{n,m} \mathcal{C}_{nm} \ket{n}_c\ket{m}_\xi \equiv
\sum_{n,m} 
\mathcal{C}_{nm} \ket{n\,,m} \,,
\end{equation}
where~$\mathcal{C}_{nm}$ are the probability amplitude of having~$n$
photons in the field described with operator~$\xi$ and~$m$ photons in
the field described with operator~$c$; and the summation is done until
the allowed number of photons: one for a fermionic field and~$N$ for a
bosonic one. Given that the dynamics of the system is given by the
master equation
\begin{equation}
\label{eq:WedFeb28114318CET2018}
\partial_t \rho = i[\rho,H] + \sum_k (\tilde \Gamma_k/2)
\mathcal{L}_{j_k}\rho\,,
\end{equation}
where~$H$ is the Hamiltonian of the system and we have assumed that
the dissipation is given by ``jump operators''~$j_k$ at
rates~$\tilde\Gamma_k$, the dynamics of the wavefunction is given by
Schr\"odinger
equation
\begin{equation}
  \label{eq:WedFeb28113458CET2018}
  \partial_t \ket{\psi} = - i H_\mathrm{eff}\ket{\psi}
\end{equation}
where~$H_\mathrm{eff}$ is a non-hermitian Hamitonian constructed as~$
H_\mathrm{eff}=H-i \sum_k \tilde\Gamma_k\, \ud{j_k} j_k$, and the
coefficients evolve as,
\begin{equation}
\label{eq:coeffeqs}
\partial_t \, \mathcal{C}_{nm} = -i  \sum_{p,q} \bra{n\,,m}H_\mathrm{eff}
\ket{p\,,q} \mathcal{C}_{pq}\,. 
\end{equation}
In the following sections we make explicit both the effective
Hamiltonians and the differential equations for the coefficients for
all the systems considered in the main text. 

\subsection{Two-level system in the Heitler regime}

The Hamiltonian describing the excitation of a sensor (a cavity) by
the emission of a 2LS, which in turn is driven in the Heitler regime
by a laser, is given by the Hamiltonian \eqref{eq:Thu31May103357CEST2018}.
To complete the analogy of beam splitter setting and be consistent with the main text,
both driving and coupling for the sensor has to be defined in terms of coherent source
amplitude $|\beta|$ and BS parameters $\mathrm{T}$ and $\mathrm{R}$: $\Omega_a  \rightarrow
i \mathrm{R}|\beta|$, $g \rightarrow \mathrm{T} g$.
%
%
The system and driving source are not necessarily at resonance
so we define the detuning as $\Delta_{\sigma} = \omega_\sigma - \omega_{\mathrm{L}}$.
The effective Hamiltonian that describes
the dynamics in the wavefunction approximation is
\begin{equation}
  \label{eq:MonMar5160912CET2018}
  H_\mathrm{eff} = H -\frac{i}{2}\left ( \gamma_\sigma
    \ud{\sigma}\sigma + \Gamma \ud{a}a \right)\,,
\end{equation}
where~$H$ is the Hamiltonian in
Eq.~\eqref{eq:Thu31May103357CEST2018}. Replacing the effective
Hamiltonian in Eq.~(\ref{eq:MonMar5160912CET2018}) on the expression
in Eq.~(\ref{eq:WedFeb28113458CET2018}), we obtain the differential
equations for the coefficients of interest:
\begin{subequations}
  \label{eq:MonMar5161258CET2018}
  \begin{align}
    i \partial_t \mathcal{C}_{01} &= \Omega_\sigma + \mathrm{T}\,g\,
                                    \mathcal{C}_{10} - i
                                    \mathrm{R}|\beta|e^{-i\phi}
                                    \mathcal{C}_{11}+\left(\Delta_{\sigma}-
                                    i\frac{\gamma_\sigma}{2} \right)
                                    \mathcal{C}_{01}\,,\\
    i \partial_t \mathcal{C}_{10} &= i \mathrm{R}
                                    |\beta|e^{i\phi} + \Omega_\sigma
                                    \mathcal{C}_{11}+ 
                                    \mathrm{T}\,g\, \mathcal{C}_{01} - i  \sqrt{2}
                                    \mathrm{R}|\beta|e^{-i\phi}
                                    \mathcal{C}_{20}-i\frac{\Gamma}{2}
                                    \mathcal{C}_{10}\,,\\
    i \partial_t \mathcal{C}_{11} &= \Omega_\sigma \mathcal{C}_{10} + i
                                    \mathrm{R}|\beta| e^{i\phi}
                                    \mathcal{C}_{01}+\sqrt{2}
                                    \mathrm{T}\,g\,\mathcal{C}_{20} + 
                                    \left(\Delta_\sigma-i\frac{\gamma_\sigma+\Gamma}{2}
                                    \right) \mathcal{C}_{11}\,,\\
    i \partial_t \mathcal{C}_{20} & = \sqrt{2}\mathrm{T}\,g\,\mathcal{C}_{11} +
                                    i \sqrt{2} \mathrm{R}|\beta|e^{i
                                    \phi}\mathcal{C}_{10} - i\Gamma
                                    \mathcal{C}_{20}\,,  
  \end{align}
\end{subequations}
where we have assumed that the driving to the 2LS is low
enough so that the states with three or more excitation can be safely
neglected. Assuming that the coherent field that drives the sensor can
be written as a fraction of the field that drives the two-level
system ($|\beta| = g \frac{\mathrm{T} \Omega_{\sigma}}{\mathrm{R} \gamma_\sigma} \mathcal{F} $), 
very similar as Eq.~(\ref{eq:SatFeb24130524CET2018}), and to leading
order in the coupling and the driving intensity of the two-level
system, the solution to Eq.~(\ref{eq:MonMar5161258CET2018}) is
\begin{subequations}
  \label{eq:MonMar5165614CET2018}
  \begin{align}
    \mathcal{C}_{01} &= -\frac{2i\Omega_\sigma}{\gamma_\sigma + 2 i \Delta_\sigma}\,,\\
    \mathcal{C}_{10} &= -\frac{2g\Omega_\sigma \mathrm{T}}{\Gamma}
                       \left(\frac{2}{\gamma_\sigma + 2 i \Delta_\sigma}-\frac{\mathcal{F}e^{i\phi}}{\gamma_\sigma}\right)\,,\\
    \mathcal{C}_{11} &= -\frac{4 i g \Omega_\sigma^2 \mathrm{T}
                       }{\Gamma \gamma_\sigma \left(\gamma_\sigma + 2 i \Delta_{\sigma}\right) \left(\gamma_\sigma + \Gamma 
                       + 2 i \Delta_{\sigma}\right) }
                   [-2\gamma_\sigma
                       +\mathcal{F}(\gamma_\sigma +\Gamma+ 2 i \Delta_\sigma)e^{i\phi}]\,,\\
    \mathcal{C}_{20}&= \frac{2\sqrt{2} g^2\Omega_\sigma^2 \mathrm{T}^2}{\gamma_\sigma^2
                      \Gamma^2 \left(\gamma_\sigma + 2 i \Delta_{\sigma}\right) \left(\gamma_\sigma + \Gamma 
                      + 2 i \Delta_{\sigma}\right)} \times \\
                  & \ \ [4 \gamma _{\sigma }^2+\mathcal{F}^2 e^{2 i \phi } \left(\gamma _{\sigma }+2 i \Delta _{\sigma }\right) \left(\gamma _{\sigma }+\Gamma +2 i \Delta _{\sigma }\right)-4 \mathcal{F} e^{i \phi } \gamma _{\sigma } \left(\gamma _{\sigma }+\Gamma +2 i \Delta _{\sigma }\right)]\,.
  \end{align}
\end{subequations}
The population of both the 2LS and the cavity, and
the~$\g{2}_a$ can be obtained from the coefficients
in Eq.~(\ref{eq:MonMar5165614CET2018}). However, to recover
some information from the unfiltered signal ($\Gamma \rightarrow \infty$),
this limit has to be performed carefully and the previous expressions
need some manipulation first. A new set of coefficients is defined as:
$\mathcal{C}'_{ij} = \left(\frac{\Gamma}{2\mathrm{T} g} \right)^i \mathcal{C}_{ij}$ so
any explicit dependence of sensor parameters disappears, resulting in a non-vanishing
(finite) solution after the proper limit are taken.
After the substitution and taking the limit, the new coefficients are
\begin{subequations}
	\begin{align}
	\mathcal{C}'_{01} &=  -\frac{2i\Omega_\sigma}{\gamma_\sigma + 2 i \Delta_\sigma} \,,\\
	\mathcal{C}'_{10} &= \Omega_{\sigma} \left(\frac{e^{i \phi} \mathcal{F}}{\gamma_\sigma} - \frac{2}{\gamma_\sigma + 2 i \Delta_\sigma}\right) \,,\\
	\mathcal{C}'_{11} &=- \frac{2 i e^{i \phi} \mathcal{F} \Omega_\sigma^2}{\gamma_\sigma
	                    \left(\gamma_\sigma + 2 i \Delta_{\sigma}\right)}\,,\\
	\mathcal{C}'_{20}&= \frac{e^{i \phi} \mathcal{F} \Omega_\sigma^2}{\gamma_\sigma^2
		\left(\gamma_\sigma + 2 i \Delta_{\sigma}\right)} 
	\left[e^{i \phi} \mathcal{F} \left(\gamma_\sigma+ 2 i \Delta_\sigma \right)
	-4 \gamma_\sigma\right] \,.
	\end{align}
\end{subequations}
Now, these solutions provide useful information about the equivalent filtered signal:
$\av{n_a} \approx |\mathcal{C}'_{10}|^2$,
$\mathrm{P}_{20} = |\mathcal{C}'_{20}|^2$ (probability of two-photon state)
and~$\g{2}_a \approx 2|\mathcal{C}'_{20}|^2/ |\mathcal{C}'_{10}|^4$. The cancellation of the coefficient~$\mathcal{C}'_{20}$,and therefore of~$\g{2}_a$, yields the condition on the attenuation
factor
\begin{equation}
  \label{eq:TueMar6111957CET2018}
  \mathcal{F} = \frac{4 e^{-i \phi}\gamma_\sigma}{\gamma_\sigma + 2 i \Delta_{\sigma}}\,,
\end{equation}
which can only be satisfied---$\mathcal{F}$~is an attenuation factor,
and thus a real number---when the relative phase between the driving field
and the 2LS coherent contribution is either 0 or $\pi$ (opposite phase),
in agreement with Fig.~\ref{fig:WedNov29150123CET2017}(a-d) in the main text. 
Note, as well, that the cancellation of the coefficient~$\mathcal{C}_{10}$, and
therefore of the population of the cavity, is obtained
when~$\mathcal{F} =  \frac{2 e^{-i \phi}\gamma_\sigma}{\gamma_\sigma + 2 i \Delta_{\sigma}}$ which is a real number for the same
phases for which Eq.~(\ref{eq:TueMar6111957CET2018}) is a real
number.

\subsection{Jaynes--Cummings blockade}
\label{sec:WedFeb28173330CET2018}

%

The Hamiltonian describing the Jaynes--Cummings model in given in
Eq.~(\ref{eq:Thu31May103357CEST2018}), and the dynamics is complemented
with a master equation that takes into account the decay of the
2LS with rate~$\gamma_\sigma$ and of the cavity with
rate~$\gamma_a$. As such, the effective Hamiltonian that described the
dynamics in the wavefunction approximation is 
\begin{equation}
  \label{eq:WedFeb28144028CET2018}
  H_\mathrm{eff}  = 
  \left(\Delta_a - i \frac{\gamma_a}{2} \right) \ud{a} a \ +
  \left(\Delta_\sigma - i 
    \frac{\gamma_\sigma}{2} \right) \ud{\sigma} \sigma + 
  g (\ud{a} \sigma  + \ud{\sigma} a ) + \Omega_a (e^{i \phi}\ud{a} + e^{-i \phi} a) + \Omega_{\sigma} 
  (\ud{\sigma} + \sigma)\,.
\end{equation}
%
Replacing the Hamiltonian in Eq~(\ref{eq:WedFeb28144028CET2018}) into
Eq.~(\ref{eq:coeffeqs}), we have that the differential equation for
the relevant coefficients are as follows,
\begin{subequations}
  \label{eq:WedFeb28144227CET2018}
  \begin{align}
    i \partial_t \,\mathcal{C}_{10}&= e^{i \phi} \Omega_a  + \left(\Delta_a - i
          \frac{\gamma_a}{2}\right) \mathcal{C}_{10} + g\,
                                   \mathcal{C}_{01} + \Omega_{\sigma} \,
                                   \mathcal{C}_{11}\ + 
          \sqrt{2} e^{-i \phi} \Omega_a \, \mathcal{C}_{20} \,, \\ 
	i \partial_t \,\mathcal{C}_{01}&= \Omega_{\sigma} +  g \, \mathcal{C}_{10} +
                                         \left(\Delta_\sigma - i 
	\frac{\gamma_\sigma}{2}\right) \mathcal{C}_{01} + e^{-i \phi} \Omega_a \,
                                       \mathcal{C}_{11}\,, \\ 
	i \partial_t \,\mathcal{C}_{11}&=  e^{i \phi} \Omega_a \,  \mathcal{C}_{01} + 
                                          \Omega_{\sigma} \, \mathcal{C}_{10}+\left(\Delta_a
                                          +\Delta_\sigma - i 
	\frac{\gamma_a + \gamma_\sigma}{2}\right) \mathcal{C}_{11} +
                                       \sqrt{2} g\, 
          \mathcal{C}_{20}\,, \\  
    i \partial_t \,\mathcal{C}_{20}&=  \sqrt{2}  e^{i \phi}  \Omega_a \, \mathcal{C}_{10} + 
                                      \sqrt{2} g \,\mathcal{C}_{11}+
                                      \left(2 \Delta_a - i \gamma_a
                                   \right) \mathcal{C}_{20}\,, 
  \end{align}
\end{subequations}
where we have assumed that the driving is low enough for the states
containing three or more photons to be neglected. The steady-state
solution for the coefficients Eq.~(\ref{eq:WedFeb28144227CET2018}) is
obtained when the derivatives on the left-hand side of the equation
vanish. Thus, assuming that the coefficient of the vacuum dominates
over all the others, i.e.,~$\mathcal{C}_{00}\approx 1$, and to leading
order in the driving of the cavity, the coefficients are
\begin{subequations}
  \label{eq:WedFeb28145309CET2018}
	\begin{align}
	 \mathcal{C}_{10}  &=  \Omega_a \, \frac{2 e^{i \phi} \left( 2 \Delta_\sigma 
	 	- i \gamma _{\sigma } \right) - 4 \chi g}{4 g^2+\left(\gamma _a+2 i
          \Delta_a\right) \left(\gamma _{\sigma }+2 i \Delta_\sigma \right)}, \\ 
	\mathcal{C}_{01} &= \Omega_a \, \frac{2 \chi \left( 2 \Delta_a
		- i 
		\gamma _a \right) - 4 e^{i \phi} g}{4 g^2+\left(\gamma _a+2 i
		\Delta_a\right) \left(\gamma _{\sigma }+2 i \Delta_\sigma \right)}, \\ 
	\mathcal{C}_{11} & = 4 \Omega_a^2 \, \frac{[2 e^{i \phi} g - \chi (2 \Delta_a - i \gamma_a)]
	[2 g \chi + i e^{i \phi} (\tilde{\gamma}_{11} + 2 i \tilde{\Delta}_{11})]}
	{ [4 g^2+\left(\gamma _a+2 i
      \Delta_a\right) \left(\gamma _{\sigma }+2 i \Delta_\sigma \right)]
       [4 g^2+\left(\gamma _a+2 i
      \Delta_a\right) (\tilde{\gamma}_{11} +2 i \tilde{\Delta}_{11})]},\\   
	\mathcal{C}_{20}  &= \sqrt{8} \Omega _a^2 \,
	\frac{4 g^2 \chi^2
	+ 4 i e^{i \phi} g \chi (\tilde{\gamma}_{11} +2 i \tilde{\Delta}_{11})
+e^{i 2 \phi} [4 g^2 - (\gamma_\sigma + 2 i \Delta_\sigma)
(\tilde{\gamma}_{11} +2 i \tilde{\Delta}_{11})] }
	{ [4 g^2+\left(\gamma _a+2 i
      \Delta_a\right) \left(\gamma _{\sigma }+2 i \Delta_\sigma \right)]
  [4 g^2+\left(\gamma _a+2 i
\Delta_a\right) (\tilde{\gamma}_{11} +2 i \tilde{\Delta}_{11})]}, 
	\end{align}
\end{subequations}
where $\Delta_c = \omega_c - \omega_\mathrm{L}$ (for $c = a, \sigma$), 
$\chi = \Omega_{\sigma}/ \Omega_a$ is the ratio between dot and cavity driving
and $\tilde{\Delta}_{ij} = i \Delta_a + j \Delta_{\sigma}$ (the same notation for
$\tilde{\gamma}_{ij}$). 

The population of both the 2LS
and the cavity, and the~$\g{2}_a$ can be obtained from the
coefficients in Eq.~(\ref{eq:WedFeb28145309CET2018}) as~$ n_a =
|\mathcal{C}_{10}|^2$, $\mean{n_\sigma}= |\mathcal{C}_{01}|^2$ and~$\g{2}_a
= 2|\mathcal{C}_{20}|^2/ |\mathcal{C}_{10}|^4$, respectively; which
coincide with the expressions given in
Eq.~\eqref{eq:Tue29May182623CEST2018} and Eq.~\eqref{eq:Tue29May184632CEST2018} of the
main text.
%
%


\subsection{Exciton-polaritons blockade}
\label{sec:FriMar2181711CET2018}

The Hamiltonian describing exciton-polaritons is given in
Eq.~(\ref{eq:FriMar2182347CET2018}), and the dynamics is complemented
with a master equation that takes into account the decay rate of the
cavity with rate~$\gamma_a$ and of the excitons with
rate~$\gamma_b$. Thus, the effective Hamiltonian that described the
dynamics in the wavefunction approximation is
\begin{equation}
  \label{eq:FriMar2182643CET2018}
H_{\mathrm{eff}}
= \left(\Delta_a 
- i \frac{\gamma_a}{2}\right) \ud{a} a + \left(\Delta_{b}- i
\frac{\gamma_b}{2}\right) \ud{b} b + g \left(\ud{a} b 
+ \ud{b} a \right) + \Omega_a \left(e^{i\phi} \ud{a} + e^{-i \phi} a\right) + \Omega_b (\ud{b} + b) + \frac{U}{2}
\ud{b} \ud{b} b b\,.
\end{equation}
Replacing the Hamiltonian in Eq.~(\ref{eq:FriMar2182643CET2018}) in
Eq.~(\ref{eq:coeffeqs}), we find that the differential equations for
the relevant coefficients are
\begin{subequations}
	\label{eq:polaritonSteadyState_wavefunc}
	\begin{align}
	i \partial_t\, \mathcal{C}_{10} &= e^{i\phi} \Omega_a + \left(\Delta_a - i
          \frac{\gamma_a}{2}\right) \mathcal{C}_{10} + g \ \mathcal{C}_{01} +
          \Omega_b \, \mathcal{C}_{11} +
          \sqrt{2} e^{-i\phi} \Omega_a \, \mathcal{C}_{20}\,, \\
	i \partial_t \,\mathcal{C}_{01} &= \Omega_b +  g \ \mathcal{C}_{10} +
                                        \left(\Delta_\sigma - i 
	\frac{\gamma_\sigma}{2}\right) \mathcal{C}_{01} + e^{-i \phi} \Omega_a \,
                                        \mathcal{C}_{11} + \sqrt{2} \Omega_b \, \mathcal{C}_{02} \,, \\ 
	i \partial_t \,\mathcal{C}_{11} &= e^{i \phi} \Omega_a \, \mathcal{C}_{01} +
	                                  \Omega_b \, \mathcal{C}_{01} +
                                      \left(\Delta_a +\Delta_\sigma -
                                      i 
	\frac{\gamma_a + \gamma_\sigma}{2}\right) \mathcal{C}_{11} +
                                        \sqrt{2} g 
          \left( \mathcal{C}_{20} + \mathcal{C}_{02} \right)\,, \\ 
	i \partial_t \,\mathcal{C}_{20} &= \sqrt{2} e^{i \phi} \Omega_a \,
                                        \mathcal{C}_{10} +  
	\sqrt{2} g \, \mathcal{C}_{11}+
	\left(2 \Delta_a - i \gamma_a \right) \mathcal{C}_{20}\,, \\
	i \partial_t\, \mathcal{C}_{02} &= \sqrt{2} \Omega_b \, \mathcal{C}_{01} + \sqrt{2} g \,
                                        \mathcal{C}_{11}+ 
	\left(2 \Delta_b + U - i \gamma_b \right) \mathcal{C}_{02}\,,
	\end{align}
\end{subequations}
where we have assumed that the driving is low enough for the states
with three or more photons to be neglected. The steady-state solution
of the coefficients in Eq.~(\ref{eq:polaritonSteadyState_wavefunc}) is
obtained when the derivatives in the left-hand side of the equation
vanish.  Thus, assuming that the coefficient of the vacuum dominates
over all the others, i.e.,~$\mathcal{C}_{00}\approx 1$, and to leading
order in the driving of the cavity, the coefficients are
\begin{subequations}
\label{eq:FriMar2190022CET2018}
	\begin{align}
	 \mathcal{C}_{10}  &= 2 \Omega_a \, \frac{e^{i \phi} (2 \Delta_b - i \gamma_b) - 2 g \chi}
	 {4 g^2+(\gamma _a + 2 i \Delta_{a}) (\gamma_b + 2 i \Delta_{b})}\,, \\ 
	\mathcal{C}_{01} &=2 \Omega_a \, \frac{\chi (2 \Delta_a - i \gamma_a) - 2 e^{i \phi} g }
	{4 g^2+(\gamma _a + 2 i \Delta_{a}) (\gamma_b + 2 i \Delta_{b})}\,, \\ 
	 \mathcal{C}_{11}  &= 4 \Omega_a^2 \, [-2 i e^{i \phi} g + (\gamma_a + 2 i \Delta_a)]
	 [e^{i \phi} (U + 2 \Delta_b - i \gamma_b)(\tilde{\gamma}_{11} + 2 i \tilde{\Delta}_{11})
	 - i \chi g (U + 2 \tilde{\Delta}_{11} - i \tilde{\gamma}_{11})] 
	/\mathcal{N} \,,  \\
		\mathcal{C}_{20} & =
	\begin{aligned}[t]
     & i \sqrt{8} \Omega_a^2 \, \lbrace 4 \chi^2 g^2 (U + \tilde{\Delta}_{11} - i
	 \tilde{\gamma}_{11})+ 4 i e^{i \phi} \chi g (U + 2 \Delta_b - i \gamma_b)
	 (U + \tilde{\Delta}_{11} - i \tilde{\gamma}_{11}) + \\ 
	 & \, e^{i 2 \phi}
	 [4 g^2 U - (\gamma_b + 2 i \Delta_b)(\tilde{\gamma}_{11} + 2 i \tilde{\Delta}_{11})
	 (U + \Delta_b - i \gamma_b)]\rbrace / 
	\ \mathcal{N},
    \end{aligned} 
	\\
	 \mathcal{C}_{02}& = i \sqrt{8} \Omega_a^2 \, (\tilde{\gamma}_{11} + 2 i \tilde{\Delta}_{11})
	 (2 e^{i \phi} + i \chi \gamma_a - 2 \chi \Delta_a)^2 / 
	\ \mathcal{N}\,,	 
	\end{align}
\end{subequations}	
where we have used
\begin{equation}
\mathcal{N}= [4 g^2 + (\gamma_a + i \Delta_a)(\gamma_b + \Delta_b)] 
[ (\gamma_a + i \Delta_a) (\tilde{\gamma}_{11} + i \tilde{\Delta}_{11})
(U + 2 \Delta_b - i \gamma_b) + 4 g^2 (U + 2 \tilde{\Delta}_{11} - i \tilde{\gamma}_{11})],
\end{equation}
and $\chi$, $\tilde{\gamma}_{ij}$ and $\tilde{\Delta}_{ij}$ share the
same definition as described above changing $\sigma$ by $b$.

The population of both the cavity and the excitons, and the~$\g{2}_{a,b}$
can be obtained from the coefficients in
Eq.~(\ref{eq:FriMar2190022CET2018}) as~$ n_a = |\mathcal{C}_{10}|^2$,
$\mean{n_b}= |\mathcal{C}_{01}|^2$,
$\g{2}_a = 2|\mathcal{C}_{20}|^2/ |\mathcal{C}_{10}|^4$
and~$\g{2}_b = 2|\mathcal{C}_{02}|^2/ |\mathcal{C}_{01}|^4$,
respectively; which coincide with the expressions given in
Eq.~(\ref{eq:Thu31May102314CEST2018}) of the main text. 

\section{$\g{2}$ decomposition}
\label{app:4}
\subsection{Jaynes--Cummings model}
Given the expressions for $\g{2}$ decomposition from the equations
\eqref{eq:decompositiontermswhole} and steady-state correlators
obtained using the methods described above, we find the following
expressions for the JC model when the cavity is driven ($\chi = 0$):
\begin{subequations}
\begin{align}
\mathcal{I}_0 = & \ 256 g^8  \bigm/f_1\left(g, \Delta_a, \Delta_{\sigma}, \gamma_a,\gamma_\sigma\right)\,,\\
\mathcal{I}_1 = & \ 0\,,\\
\begin{split}
\mathcal{I}_2 = & \ 32 g^4 \bigg[-\gamma _{\sigma }^2 \Big(4 g^2 +
\gamma_a \left(\gamma_a +\gamma_{\sigma }\right) - 4 \Delta_{a}^2\Big)+
4 \gamma_{\sigma } \left(4 \gamma_a + 3 \gamma_{\sigma }\right) \Delta_a \Delta_{\sigma} + 4 \Delta_{\sigma}^2 \Big(4 g^2 + \gamma_a 
\left( \gamma_a + \gamma_{\sigma }\right) - 4 \Delta_{a}^2\Big) - \\ 
& \ 16 \Delta_a \Delta_{\sigma}^3\bigg] \biggm/ 
f_1\left(g, \Delta_a, \Delta_{\sigma}, \gamma_a,\gamma_\sigma\right),
\end{split}
\end{align}
\end{subequations}
where the function $f_1\left(g,\Delta_a, \Delta_{\sigma}, 
\gamma_a,\gamma_\sigma\right)$ is defined as:
\begin{multline}
f_1\left(g, \omega_a, \omega_{\mathrm{L}},
\gamma_a,\gamma_\sigma\right) = 
\Big(\gamma_{\sigma }^2 + 4 \Delta_{\sigma}^2\Big)^2 \Big(16 g^4 +8 g^2 \big[\gamma_a \left(\gamma_a + \gamma_{\sigma }\right)
- 4 \Delta_a \left(\Delta_{a}+\Delta_{\sigma}\right)\big] +\\{} \big[\gamma_a^2 +4 \Delta_{a}^2\big] \big[\left(\gamma_a + 
\gamma_\sigma\right)^2 + 4 \left(\Delta_a + \Delta_{\sigma}\right)^2\big]\Big) ,
\end{multline}
\subsection{Microcavity polariton}
For the polariton system (also for $\chi = 0$), we can do the same
decomposition:
\begin{subequations}
\begin{align}
\mathcal{I}_0 = & 256 U^2 \, g^8 /
 f_2\left(g, \Delta_a, \Delta_b,\gamma_a,\gamma_b\right)\,,\\
\mathcal{I}_1 = & 0,\\
\begin{split}
\mathcal{I}_2= & 
-32 g^4 U \bigg[ \gamma_b^2 \Big(U \gamma_a \left[\gamma_a + \gamma_b\right]
 + 2 \gamma_b \left[2 \gamma_a + \gamma_b \Delta_a\right]
  - 4 U \Delta_a^2 + 4 g^2 \left[U+2 \Delta_2 \right]\Big) + \\
& 2 \gamma_b \Delta_b \Big( 3 \gamma_a^2 \gamma_b + 4 g^2 \left[2 \gamma_a + 3 
\gamma_b\right] - 6 \gamma_b \Delta_a \left[U + 2 \Delta_a\right] + 4 \gamma_a
\left[\gamma_b^2 - 2 U \Delta_a \right]\Big)
- 4 \Delta_b^2 \Big(U \gamma_a \left[\gamma_a + 3 \gamma_b \right]+ 12 
\gamma_b\left[\gamma_a + \gamma_b\right] - \\ &
4 U \Delta_a^2 + 4 g^2 \left[U + 2 \Delta_a\right]\Big) -
\Delta_b^3 \Big(4 g^2+ \gamma_a^2 + 4 \gamma_a \gamma_b - 2 \Delta_a
\left[ U + \Delta_a\right] + 32 \Delta_a \Delta_b^4\Big) 
 \bigg] \biggm/ f_2\left(g, \Delta_a, \Delta_b,
\gamma_a,\gamma_b\right)\,,
\end{split} 
\end{align}
\end{subequations}
where the auxiliary function $f_2$ has the following form:
\begin{multline}
f_2 \left(g, \Delta_a, \Delta_b,\gamma_a,\gamma_b\right) = 
\Big[\gamma _b^2+4 \Delta_b^2\Big]^2 
\bigg[ \left(\gamma_a^2 + 4 \Delta_a^2\right) \left(\left(\gamma_a+\gamma_b\right)^2 
+ 4 \left(\Delta_a + \Delta_b\right)^2\right)
\left(\gamma_b^2 + \left(U + 2\Delta_b\right)^2\right) + \\
 16 g^4 \left(\left(\gamma_a + \gamma_b\right)^2 + \left(U + 2 \left[ \Delta_a + \Delta_b\right]\right)^2\right)
+ 8 g^2 \Big(U^2 \left(\gamma_a \left[\gamma_a + \gamma_b\right]-
4 \Delta_a \left[\Delta_a + \Delta_b\right] \right) + \\
\big(\gamma_a \gamma_b - 4 \Delta_a \Delta_b\big)
\big(\left[\gamma_a + \gamma_b\right]^2 + 4 \left[\Delta_a + \Delta_b\right]^2\big)
- 2 U \left(\gamma_a^2 \left[\Delta_a -\Delta_b\right]- 2 \gamma_a \gamma_b \Delta_b
+ 4 \Delta_a \left[\Delta_a + \Delta_b\right] \left[\Delta_a + 2 \Delta_b\right] \right)
\Big)
\bigg]\,.
\end{multline}

\section{Unconventional antibunching conditions}
\label{app:perfectantibunching}

\subsection{JC Model}

Starting from the equation $\mathcal{C}_{20} =0$ (which is directly
linked with $\g{2}_a = 0$ in the vanishing driving limit):

\begin{equation}
\Delta_a = \frac{i \big(\gamma_\sigma + 2 i \Delta_\sigma \big) \big(\tilde{\gamma}_{11} + 2 i 
	\Delta_\sigma \big) + 4 e^{-i \phi} g \, \chi \big(\tilde{\gamma}_{11} + 2 i \Delta_\sigma \big)
	- 4 i g^2 \big(1 + e^{-2 i \phi} \chi^2\big)}
{2 \gamma_\sigma + 4 i \Delta_\sigma - 8 i e^{-i \phi} g \, \chi }
\end{equation}
Taking the real part gives the expression for UA curve:
\begin{equation}
\Delta_a = \frac{4 g \chi \big\lbrace 2 \cos \phi \big[2 \Delta_\sigma^2 + g^2 \big(1+ \chi^2 \big)\big]
	- g \chi \Delta_\sigma \cos 2 \phi - \gamma_\sigma \sin  \phi \big(g \chi \cos \phi - 2 \Delta_\sigma \big)\big\rbrace - \big[\tilde{\Gamma}^2_\sigma + 4 g^2 \big(1+ 4 \chi^2 \big)\big]}
{\gamma_\sigma^2 + 4 \big(\Delta_\sigma^2 + 4 g^2 \chi^2 \big) - 8 g \chi \big( 2 \Delta_\sigma \cos \phi
	+ \gamma_\sigma \sin \phi \big)}
\end{equation}
While imposing the imaginary part to be zero gives the second contraint:
\begin{equation}
\Delta_\sigma = \frac{4 g \chi \cos \phi \, (\tilde{\gamma}_{11} - g \chi \sin \phi)
	\pm \sqrt{-\tilde{\gamma}_{11} (\gamma_\sigma - 4 g \chi \sin \phi) [-4 g^2 +
		\gamma_\sigma \tilde{\gamma}_{11}- 4 g \chi (g \chi \cos 2 \phi + \tilde{\gamma}_{11} \sin \phi)	] + 4 g^2\chi^2 \sin^2 2 \phi
} }{2 \tilde{\gamma}_{11}},
\end{equation}
which must return a real quantity so that the radicand has to be positive
perforce.

\subsection{Microcavity polariton}

The perfect antibunching conditions can be derived straightaway 
from the equation $\g{2}_a=0$, where $\g{2}_a$ is given by the expression \eqref{eq:PBg2}. Then, we clear $\Delta_a$ from the previous equation, which leads to:
\begin{multline}
\label{eq:UApbcond}
\Delta_a = \big\lbrace
e^{i \phi} \big[4 g^2U - \big(\gamma_b + 2 i \Delta_b\big)
	\big(\tilde{\gamma}_{11} + 2 i \Delta_b\big) \big(U + 2 \Delta_b - i \gamma_b\big)
	\big] \\
	 + 4 i  g \chi \big(U + 2 \Delta_b - i \gamma_b\big) \big(\tilde{\gamma}_{11} + 2 i \Delta_b\big)
	+ 4 e^{-i \phi} g^2 \chi^2 \big(U + 2 \Delta_b - i \tilde{\gamma}_{11}\big)
\big\rbrace
/ \mathcal{N},
\end{multline}
where $\mathcal{N}$ is defined as:
\begin{equation}
\mathcal{N} =
2 \big[ e^{i \phi} \big(\gamma_b + 2 i \Delta_b\big) \big(\gamma_b + i U + 2 i \Delta_b\big)
+4 g \chi \big(U + 2 \Delta_b - i \gamma_b \big) - 4 g^2 \chi^2 e^{-i \phi}\big].
\end{equation}
Nevertheless, by definition $\Delta_a$ must a real quantity.  Taking
the real part of this last expression leads to the equation for the
curve which follows the UA effect. Moreover, the cancellation of its
imaginary part (same as the JC model) turns out to be a second
condition to exactly reach $\g{2}_a = 0$, which can be cleared for any
chosen parameter. Further analysis shows that more restrictions emerge
from the fact of selecting only real-valued (physical) parameters.  As
an illustration, we present here the case of cavity excitation
($\chi = 0$).  Splitting both real and imaginary parts from
Eq.~\eqref{eq:UApbcond}:
\begin{subequations}
\begin{align}
&\Delta_a  = - \Delta_b - \frac{4 g^2 \Delta_b}{\gamma_b^2 + 4 \Delta_b^2}
           + \frac{2 g^2 (U + 2 \Delta_b)}{\gamma_b^2 + (U + 2 \Delta_b)^2}, \\
& 0 = \gamma_a + \gamma_b + 4 g^2 \gamma_b \left( - \frac{1}{\gamma_b^2 + 4 \Delta_b^2}
+ \frac{1}{\gamma_b^2 + (U + 2 \Delta_b)^2} \right) .
\end{align}
\end{subequations}
First expression provides an implicit equation for the 3 distinct curves of
UA shown in Fig.~\ref{fig:08} whereas the latter gives the exact location
where $\g{2}_a$ vanishes.

\section{Particular cases of special interest}
\label{sec:Wed30May110352CEST2018}
In this appendix, we list some expressions which could be easily
derived from the general cases given in the text, but whose importance
and popularity could result a convenience for many readers to be
available explicitely.

This is the two-photon coherence function in presence of cavity
pumping only (the case most discussed in the literature) for
the JC system:
\begin{multline}
\label{eq:JCg2}
\g{2}_a = \Big\lbrace
\big[(16 g^4 + 8 g^2 \left(\gamma_{\sigma } \gamma_a - 4 \Delta_a
\Delta_{\sigma} \right) + 
\Gamma_a^2 \Gamma_\sigma^2\big]  
\big[ 16 g^4 -8 g^2 \big(\gamma_\sigma \gamma_{+} 
- 4 \Delta_\sigma \Delta_{+} \big) +
\Gamma_{\sigma}^2 \Gamma_{+}^2 \big]  \Big\rbrace \biggm/  \\ 
\bigg\lbrace
\Gamma_\sigma ^4 \Big[ 16 g^4 +8
g^2 \big(\gamma_a \gamma_{+} 
- 4 \Delta_a \Delta_{+}\big) +
\Gamma_a^2 \Gamma_{+}^2 \Big] \bigg\rbrace, 
\end{multline}
and for the microcavity polaritons:
\begin{multline}
\label{eq:PBg2}
\g{2}_a = [16g^4 + 8g^2( \gamma_a + \gamma_b- 4\Delta_a
\Delta_b)+\Gamma_a^2 \Gamma_b^2 ]\times{}\\ {}\times \lbrace
16g^4U^2 + \Gamma_b^2 \Gamma_{+}^2 [\gamma_b^2 + (U+2\Delta_b)^2]
-8g^2U [4\gamma_a\gamma_b \Delta_b -8 \Delta_b^2 \Delta_{+} +
2\gamma_b^2 (\Delta_{+}+2\Delta_b) + U(\gamma_b \gamma_{+}-
4\Delta_b \Delta_{+})]\rbrace \bigm/ \\\big[ 8g^2 \Gamma_b^4\{ U^2
(\gamma_a\gamma_{+}- 4 \Delta_a \Delta_{+}) + \Gamma_{+}^2
(\gamma_a\gamma_b -4 \Delta_a \Delta_b) -2 U [ \gamma_a^2 \Delta_{-}
- 2 \gamma_a \gamma_b \Delta_b + 4 \Delta_a \Delta_{+} ( \Delta_{+}
+ \Delta_b)]\} \times {}\\ \Gamma_b^4 \{\Gamma_a^2 \Gamma_{+}^2
[\gamma_b^2 +(U+2\Delta_b)^2] + 16g^4 [\gamma_{+}+(U+2\Delta_{+})^2]
\} \big]\,.
\end{multline}
This is the coupling strength between the cavity and the 2LS which,
for given parameters, results in $\g{2}_a =1$:
\begin{multline}
  \label{eq:FriMar2104433CET2018}
  g_{P} = \frac{1}{2} \Big \lbrace \big [ 16 \Delta_\sigma^4
  + 32 \Delta_a \Delta_\sigma^3 - 8(\gamma_a^2 +3\gamma_a
  \gamma_\sigma + \gamma_\sigma^2 - 4\Delta_a^2)\Delta_\sigma^2 -
  {}\\ {}-8 \gamma_\sigma (4\gamma_a+3\gamma_\sigma)
  \Delta_a \Delta_\sigma + \gamma_\sigma^2 (2\gamma_a^2 +
  2\gamma_a \gamma_\sigma + \gamma_\sigma^2 - 8\Delta_a^2 )
  \big]^{1/2} + \gamma_\sigma^2 - 4\Delta_\sigma^2 \Big
  \rbrace^{1/2}\,.
\end{multline}
A smaller
coupling~$g<g_\mathrm{P}$ produces antibunched light while a larger
coupling~$g>g_\mathrm{P}$ produces bunched light. 

\end{document}